\documentclass[a4paper,11pt]{article}
\usepackage{jheppub} 
\usepackage{hyperref}
\usepackage{amsmath}
\usepackage{amssymb}

\usepackage{slashed}

\long\def\symbolfootnote[#1]#2{\begingroup%
\def\thefootnote{\fnsymbol{footnote}}\footnote[#1]{#2}\endgroup}

\newcommand{\newc}{\newcommand}
\newc{\gsim}{\lower.7ex\hbox{$\;\stackrel{\textstyle>}{\sim}\;$}}
\newc{\lsim}{\lower.7ex\hbox{$\;\stackrel{\textstyle<}{\sim}\;$}}
\newc{\gev}{\,{\rm GeV}}
\newc{\mev}{\,{\rm MeV}}
\newc{\ev}{\,{\rm eV}}
\newc{\kev}{\,{\rm keV}}
\newc{\tev}{\,{\rm TeV}}
\def\egt{{\it e.g.}}

\newc{\mz}{M_Z}
\newc{\mpl}{M_*}
\newc{\mw}{m_{\rm weak}}
\newc{\nr}[1]{N^c_R{}_{#1}}
\usepackage{amsmath}

\newcommand{\cO}{\mathcal{O}}

\newcommand{\ztwo}{\mathbb{Z}_2}
\newcommand{\LSUSY}{\Lambda_{{ SUSY}}}

\def\beq{\begin{equation}}
\def\eeq{\end{equation}}
\newcommand{\bea}{\begin{eqnarray}\begin{aligned}}
\newcommand{\eea}{\end{aligned}\end{eqnarray}}
\def\bitem{\begin{itemize}}
\def\eitem{\end{itemize}}
\newc{\ie}{{\it i.e.}}          \newc{\etal}{{\it et al.}}
\newc{\eg}{{\it e.g.}}          \newc{\etc}{{\it etc.}}
\newc{\cf}{{\it c.f.}}

 \numberwithin{equation}{section}

\newcommand\fverb{\setbox\fverbbox=\hbox\bgroup\verb}
\newcommand\fverbdo{\egroup\medskip\noindent%
            \fbox{\unhbox\fverbbox}\ }
\newcommand\fverbit{\egroup\item[\fbox{\unhbox\fverbbox}]}
\newbox\fverbbox

\newcommand{\GeV}{{\rm GeV}}

\newcommand{\be}{\begin{equation}}
\newcommand{\ee}{\end{equation}}

\usepackage[usenames,dvipsnames,svgnames,table]{xcolor}

\usepackage{soul}

\subheader{\footnotesize
\vspace*{-3.7em}
\begin{flushright}
CERN-TH-2016-240 \\
TTP16-054 
\end{flushright}
}
\title{SUSY Meets Her Twin}

\author[a,b]{Andrey Katz,}
\author[c]{Alberto Mariotti,}
\author[d]{Stefan Pokorski,}
\author[e,f]{Diego Redigolo,}
\author[g]{\\ and Robert Ziegler}

\affiliation[a]{Theory Division, CERN, CH-1211 Geneva 23, Switzerland}
\affiliation[b]{
D\'epartement de Physique Th\'eorique and Center for Astroparticle Physics (CAP), \\ 
Universit\'e de Gen\`eve, 24 quai Ansermet, CH-1211 Gen\`eve 4, Switzerland}
\affiliation[c]{Theoretische Natuurkunde and IIHE/ELEM, Vrije Universiteit Brussel, and International
Solvay Institutes, Pleinlaan 2, B-1050 Brussels, Belgium}
\affiliation[d]{Institute of Theoretical Physics, Faculty of Physics, University of Warsaw
ul. Pasteura 5, PL-02-093 Warsaw, Poland}
\affiliation[e]{Raymond and Beverly Sackler School of Physics and Astronomy, Tel-Aviv University, Tel-Aviv 69978, Israel}
\affiliation[f]{Department of Particle Physics and Astrophysics,
Weizmann Institute of Science, Rehovot 7610001, Israel}
\affiliation[g]{Institute for Theoretical Particle Physics (TTP), Karlsruhe Institute of Technology, \\ Engesserstra\ss e 7,
D-76128 Karlsruhe, Germany}
\abstract{We investigate the general structure of mirror symmetry breaking in the Twin Higgs scenario. We show, using the IR effective theory, that a significant gain in fine tuning can be achieved if the symmetry is broken hardly. We emphasize that weakly coupled UV completions can naturally accommodate this scenario. We analyze SUSY UV completions and present a simple Twin SUSY model with a tuning of around $10\%$ and colored superpartners as heavy as 2~TeV. The collider signatures of general Twin SUSY models are discussed with a focus on the extended Higgs sectors.} 

\begin{document}
\maketitle

\section{Introduction}
Supersymmetry (SUSY) and Compositeness are the leading 
new physics (NP) candidates 
to solve the hierarchy problem of the Standard Model
(SM). However, the null results of the  LHC searches for new colored
particles already put these ideas under pressure. While it is plausible that one of these theories indeed solves the
\emph{big} 
hierarchy problem, it is not easy to see how one can avoid the \emph{little hierarchy problem},
i.e. the mismatch between the electroweak (EW) scale and the scale of top-partners. It is certainly possible that this residual hierarchy is not resolved in nature, and naturalness might
not be the only criterion for physics beyond the SM (BSM). Nevertheless,
it is interesting to explore
alternatives that can solve the little hierarchy
problem and 
circumvent the LHC bounds.
     
The Neutral Naturalness (NN) paradigm is an attractive idea 
in this context. It provides a scenario where the top partners are
charged under an $SU(3)$ group that is  \emph{different from the SM
color group}. As a consequence they can be as
light as required by naturalness while 
evading LHC bounds, because of reduced production cross
sections compared to the usual colored top-partners. Known realizations of NN
include models where the top 
partners are fermions (twin tops~\cite{Chacko:2005pe})
and models where they are bosons 
(\emph{e.g.} the folded stops in the Folded SUSY model~\cite{Burdman:2006tz}).

Our work focuses on the Twin Higgs scenario, 
in which the SM Higgs is a pseudo-Goldstone boson (PGB) of 
an accidental $SU(4)$ symmetry emerging from
a $\ztwo$ symmetry that exchanges the SM with a mirror
SM.\footnote{Having in mind perturbative UV completions of the Twin
Higgs, the $SU(4)$ accidental symmetry is automatically enhanced to
$SO(8)$ and we do not need to distinguish between the two symmetry
groups. We refer to
\cite{Batra:2008jy,Geller:2014kta,Barbieri:2015lqa,Low:2015nqa} for a
careful discussion of the importance of this difference in strongly
coupled UV completions.} Generalizations of the Twin Higgs mechanism
that involves more 
general discrete symmetries
\cite{Craig:2014roa,Craig:2014aea,Craig:2016kue} also share some
basic features with the simplest model.

In spite of its simplicity, the Twin Higgs introduces important 
theoretical challenges that we try to address in this paper. The
first one has to do with the actual gain in fine tuning (FT) of this setup. 
Even though the Twin Higgs was originally suggested to  
ameliorate the little hierarchy problem, all its known
realizations feature an upper bound on their parametric gain in 
FT with respect to colored naturalness models. This gain cannot
be more than 
$\lambda/ \lambda_{\rm SM}$, where $\lambda$ is a perturbative
$SU(4)$-symmetric Higgs quartic coupling and $\lambda_{\rm SM}$ is the
quartic coupling of the SM potential (see for example Ref.~\cite{RiccardoSearch} for a discussion of this bound). A similar bound
has been observed in double protection models, not necessarily
related to the 
NN proposal~\cite{Birkedal:2004xi,Chankowski:2004mq,Berezhiani:2005pb,
Roy:2005hg,Csaki:2005fc,Bellazzini:2008zy,Bellazzini:2009ix}.

The second theoretical challenge concerns the
fact that any NN construction only solves the little hierarchy problem. 
Above the scale of $SU(4)$ symmetry breaking the Twin Higgs models should
be 
embedded into a framework that solves the big hierarchy problem. 
Embeddings in
SUSY~\cite{Falkowski:2006qq,Chang:2006ra,Craig:2013fga} and in
Composite Higgs~\cite{Geller:2014kta,Barbieri:2015lqa,Low:2015nqa}
feature the parametric gain in FT we described
above. In SUSY UV completions, $\lambda/ \lambda_{\rm SM}$ is
severely constrained by the perturbativity bound on $\lambda$
\cite{Craig:2013fga}. This bound can be relaxed in Composite Twin
Higgs scenarios. The latter, however, are  far from achieving $\lambda\sim 4\pi$ \cite{RiccardoSearch,TorreICTP}.

The mirror symmetry in the Higgs potential can be \emph{softly-broken}
by a mass term and/or \emph{hardly-broken} by a quartic coupling.
In this
paper we give a systematic overview of these
breaking patterns and discuss their relations to the
FT.\footnote{A viable 
spontaneous $\ztwo$-breaking necessarily requires extra non-decoupled
Higgses and will not be discussed in this paper, see however
Refs.~\cite{Beauchesne:2015lva,Harnik:2016koz,Yu:2016bku}.}  
The main result of our analysis is that the above mentioned parametric
bound on the FT improvement is 
\emph{not a generic feature of the Twin Higgs construction but rather
an artifact of the soft mirror-symmetry breaking}. This bound can be circumvented if the breaking is dominantly hard, which is however difficult to realize in
composite UV completions of the Twin Higgs and therefore has been mostly
disregarded in the literature (notable exceptions are the models
presented in \cite{Chang:2006ra,Chacko:2005vw}).

We present viable UV completions of the Twin Higgs in
Supersymmetry, where both hard and soft breaking of the mirror
symmetry can be easily achieved, and study their LHC phenomenology. We
show that the FT gain of hard  $\ztwo$-breaking models is substantial and only limited
by the Higgs mass constraint.

Our paper is structured as follows. We review the Twin Higgs mechanism
with an emphasis on its symmetries and the FT gain in
Sec.~\ref{sec:2}, where we work exclusively in the Twin
Higgs IR effective theory and do not ask any questions about possible UV completions. We reproduce the bound on the FT for soft
$\ztwo$-breaking and show how this bound can be relaxed by adding hard
$\ztwo$-breaking. We analyze in detail the parameter 
space of the IR effective theory, emphasize the role of the SM-like Higgs mass constraint,  and discuss how this
requirement determines the final gain in FT of hard $\ztwo$-breaking vs soft
$\ztwo$-breaking.

In Sec.~\ref{sec:3} we discuss SUSY UV completions of the Twin
 Higgs. First, we review the SUSY Twin Higgs with soft
 $\ztwo$-breaking focusing on the simplest scenario of
 Ref.~\cite{Craig:2013fga}. We explore the parameter space of this
 model showing that its FT gain scales as $\lambda/\lambda_{\rm SM}$ and
 the effective $SU(4)$-invariant quartic $\lambda$ is far from being
 maximized. Second, we construct  
 SUSY Twin Higgs models with hard $\ztwo$-breaking. The minimal model
 fails to decouple the colored states due the Higgs mass
 constraint. We show that in next to minimal realizations we can
 comply with such a constraint, decouple the colored states and
 achieve a FT of around $10\%$, a factor of about 5 better
 than in the soft $\ztwo$-breaking scenario.

In Sec.~\ref{Higgscoupl} we present a phenomenological analysis restricted to the
extended Higgs sectors of Twin SUSY theories.  We leave for future
investigations an analysis including the full matter sector, which can lead to many other interesting
signatures both at the LHC and in cosmological observables (see 
\egt~Ref.~\cite{Craig:2015pha,Burdman:2014zta,Chacko:2015fbc,Curtin:2015fna,Cheng:2015buv} 
for further LHC studies and~\cite{Craig:2015xla,Garcia:2015toa,Garcia:2015loa,Farina:2015uea, 
Farina:2016ndq,Barbieri:2016zxn} for the exploration of viable
cosmological scenarios). The best probes of Twin SUSY constructions are certainly direct
searches of the Twin Higgs, which looks very much like a singlet
mixing with the SM Higgs \cite{Buttazzo:2015bka,Bertolini:2012gu,
Robens:2015gla, Falkowski:2015iwa, Gorbahn:2015gxa} and for this reason can be
hunted for in direct searches of a resonance decaying into di-bosons.
Such searches are always better suited for weakly coupled UV
completions of the Twin Higgs rather than indirect searches from Higgs coupling measurements.  
We also find that if the twin Higgs becomes too heavy to be abundantly
produced at the 
LHC, the MSSM-like Higgses become sensibly
lighter than their mirror states. In this case we can hunt for them in
direct searches for MSSM Higgses in the low $\tan\beta$
region~\cite{Craig:2015jba}, via their 
indirect effects on SM Higgs couplings and on the $b\to
s\gamma$ transition rate~\cite{Katz:2014mba}.

Sec.~\ref{discussion} is devoted to our conclusions and future directions. The technical details, such as   
RGE formulae, detailed calculations of the Higgs mass spectrum and
branching ratios are relegated to appendices.

\section{EWSB and fine tuning in Twin Higgs Models}\label{sec:2}
In this section we discuss the general structure of the Twin Higgs potential, 
and study the fine tuning associated to soft and hard $\ztwo$-breaking.
We suggest that the hard breaking can lead to a substantial improvement in FT.

\subsection{Setup}
The basic idea of the Twin Higgs mechanism is that the SM-like Higgs
is light compared to the new physics scale, 
 because it is a pseudo-Goldstone boson (PGB) of an approximate global
 $SU(4)$ symmetry, spontaneously  
broken down to $SU(3)$. This $SU(4)$ is merely an accidental symmetry
that holds at one-loop at  
the level of quadratic terms, as a result of a $\ztwo$ mirror symmetry
between the SM and the ``twin'' sector. As we will  
see later, this $\ztwo$ has to be broken explicitly in the Higgs
potential, in order to allow for a realistic Higgs sector. 
By construction the Twin Higgs model is an 
effective theory that can resolve the
hierarchy problem only up to a scale of around $\sim 5$~TeV. Above
this scale it should be UV-completed by a theory which solves the big
hierarchy  
problem.    

The gauge group of the Twin Higgs is extended to two copies of the SM,
which we denote by $G_A$ and $G_B$, where here and in the following
$A$ refers to the visible sector and $B$ to the twin sector. The Higgs
sector of the original Twin Higgs model~\cite{Chacko:2005pe} consists
of two copies of the SM  
Higgs potential, with two Higgs doublets $H_A$ and $H_B$ transforming
under $G_A$ and $G_B$ respectively. The full matter content of the SM
is also doubled in the $A$ and $B$ sector. 
However, for the purpose of analyzing the FT, we focus on
the top quarks and their twin   
partners that have the largest couplings to the Higgs sector, in the 
spirit of Ref.~\cite{Craig:2015pha}. Although the contribution of the
gauge sector to the little hierarchy problem is not negligible
numerically, we do not address it  
explicitly, since the discussion closely follows the one
of the top sector. We also assume for simplicity a perfect
mirror $\ztwo$ symmetry between the visible and the twin sector,  
which we break explicitly only in the Higgs sector. In agreement with
the original model we imagine that the low energy effective theory has
no light states  
that carry both visible and mirror quantum numbers.\footnote{States
  carrying both $G_A$ and $G_B$ quantum numbers can  
be present 
in other implementations of the Twin Higgs mechanism, for example if
$SU(3)_A\times SU(3)_B$ is embedded in  
$SU(6)$~\cite{Chacko:2005pe,Geller:2014kta}. See Ref.~\cite{Cheng:2015buv} for a 
discussion of the phenomenology of these states.} 

Let us now write down the most general renormalizable Higgs potential
for the visible and the twin Higgs: 
\beq\label{Vtwin}
V = \lambda(|H_A|^2 + |H_B|^2)^2 + m^2 (|H_A|^2 + |H_B|^2)  + \kappa
(|H_A|^4 + |H_B|^4) + \tilde \mu^2 |H_A|^2  
+ \rho |H_A|^4~.
\eeq
Other terms one might be tempted to write down lead to an
equivalent potential up to redefinitions of the existing couplings.  
We now analyze this potential
dividing all the terms into three different categories:
$SU(4)$-preserving, $SU(4)$-breaking but $\ztwo$-preserving,  
and $\ztwo$-breaking.  

\begin{enumerate}
\item {\it $SU(4)$-preserving terms.} These are the first two in
  the potential, to $\lambda $ and $m^2$.  
If $m^2 < 0$ and $\lambda > 0 $, the global $SU(4)$ symmetry is
spontaneously broken down to $SU(3)$. We denote  
the scale of the $SU(4)$ breaking by $f$, and consequently $f^2
\equiv v_A^2 + v_B^2$.  
If we disregard the rest of the potential, after 
the breaking  of the $SU(4)$ global symmetry down to $SU(3)$
we get one real scalar (the so-called radial mode or the ``twin Higgs'') with mass 
 $m_{h_T}=2 \sqrt{ \lambda} f$ and seven 
Goldstone bosons. After we gauge the $SU(2) \times SU(2)$ subgroup of the $SU(4)$, and identify it with the 
visible and dark\footnote{We use the words dark, twin, mirror interchangeably in this paper.} EW gauge groups, 6 of the Goldstone bosons are
``eaten'' by the visible and the mirror $W$ and $Z$-bosons,  
while the remaining one 
is identified with the SM Higgs boson. 

\item {\it $SU(4)$-breaking but $\ztwo$-preserving.} This is the term
  proportional to $\kappa$. If this coefficient is smaller than zero,  
the mirror symmetry is spontaneously broken in the vacuum, and only
one of the Higgses, either $H_A$ or $H_B$,  gets a VEV.  
This possibility does not lead to a viable 
phenomenology in our setup and we will not discuss it further (see
however Ref.~\cite{Beauchesne:2015lva}).  
Instead, if $\kappa$ is larger than zero, the $\ztwo$ symmetry
is preserved  by the vacuum, and $H_A$ and $H_B$ get equal VEVs $v_A =
v_B = f/\sqrt{2}$.  
We further consider $0<\kappa \ll \lambda$, such
that the $SU(4)$-breaking term gives a  
sub-leading contribution to the mass of the radial mode,
but generates a small SM-like Higgs mass $m_h = \sqrt{2 \kappa} f$. 
The hierarchy between the two quartics in the potential is \emph{technically natural} 
and so is the hierarchy between the radial mode and the SM Higgs, because the latter  
is a PGB in the limit $\kappa \ll \lambda$. The main problem with the model at the present stage is the
unbroken $\ztwo$ symmetry which implies that the SM 
Higgs is an equal superposition 
of the visible and the mirror Higgs and it couples with equal strength to the mirror and 
the visible gauge bosons and matter. The couplings of the SM-like Higgs to the visible gauge 
bosons and fermions would be then reduced by a factor $1/\sqrt{2}$. 
This scenario is excluded both by LEP EWPM and the LHC Higgs coupling measurements. Therefore 
we must include also explicit $\ztwo$-breaking terms. 

\item {\it $\ztwo$-breaking terms.} There are two ways to break the mirror symmetry within a renormalizable theory: via
a relevant operator proportional to $\tilde \mu$ in Eq.~\eqref{Vtwin}, or via a marginal one -  the $\rho$ term.    
We define 
$\tilde \mu^2 \equiv \sigma f^2$ and work with the dimensionless parameters $\sigma$ and $\rho$, 
which are responsible for the 
Higgs VEV misalignment and the mixing angle between the SM-like Higgs and the twin Higgs. 
Maximal mixing 
is already excluded and the largest possible 
misalignment allowed by data translates into the bound $f/v \gtrsim 2.3$ or $f \gtrsim 400 \, \GeV$ (see for example Ref.~\cite{Buttazzo:2015bka}). 
\end{enumerate}

\subsection{Electroweak Symmetry Breaking and Radiative Corrections}
In order to analyze EWSB and the Higgs mass in the Twin Higgs scenario it is instructive to integrate out the heavy 
radial mode and switch to an effective Higgs theory, where we can write   the Higgs fields in a non-linear realization as 
\beq
H_A = f \sin \frac{\phi}{\sqrt 2 f},\ \ \ \   H_B  = f \cos \frac{\phi}{\sqrt 2 f}~.
\eeq
Hereafter we identify $\phi$ with the SM-like Higgs.  It is straightforward to plug these expressions into Eq.~\eqref{Vtwin}
and obtain the effective SM Higgs potential in the low energy effective theory. Minimization of this potential with 
respect to $\phi$ yields the following expressions for the VEV and the mass of the SM-like 
Higgs:\footnote{The same expressions can alternatively be obtained at the level of the linear sigma model \eqref{Vtwin}
by solving the EWSB conditions and expanding them at leading order in 
$\kappa,\sigma \ll \lambda$. We refer to Appendix \ref{app:twin} for a discussion of the sub-leading corrections in this expansion.} 
\begin{gather}
\frac{2v^2}{f^2}  = \frac{2\kappa - \sigma}{ 2\kappa + \rho}  \label{vh1}~, \\ 
m_h^2  =  4 v^2 \left(2\kappa +
  \rho\right)\left(1-\frac{v^2}{f^2}\right)=  2f^2 \left( 2\kappa -
  \sigma \right) 
\left(1-\frac{v^2}{f^2}\right)~.
\label{vh2}
\end{gather} 
Most of previous work on Twin Higgs has concentrated
 on \emph{soft} $\ztwo$-breaking, i.e
  $\sigma\gg\rho$.
In this paper we go further and consider \emph{hard}
$\ztwo$-breaking, i.e. the presence of a tree-level $\rho$, such
  that  $\rho\gg\sigma$. A hard breaking of the  
mirror symmetry was already introduced in some specific Twin Higgs models
(see for example Refs.~\cite{Chacko:2005vw, Chang:2006ra}). 
Note that, in models where $\ztwo$-breaking effects are generated by loops of the matter/gauge
sector, $\rho$ is unavoidably generated. In 
that case, however, it is typically smaller than
$\sigma$.\footnote{A notable example are models where the ``twin''  
hypercharge $U(1)_B$ is ungauged~\cite{Craig:2015pha} and
$\ztwo$-breaking effects are generated by $U(1)_A$ gauge loops.  
In this case $\sigma\sim g_1^2/16\pi^2$ and $\rho\sim
g_1^4/16\pi^2$ and the latter can safely be neglected.}  Here
for the first time we attempt to provide a systematic understanding of the
breaking of $\ztwo$ mirror symmetry in the Higgs sector from the  
EFT point of view.

\begin{figure}[t]
\begin{center}
\includegraphics[scale=0.7]{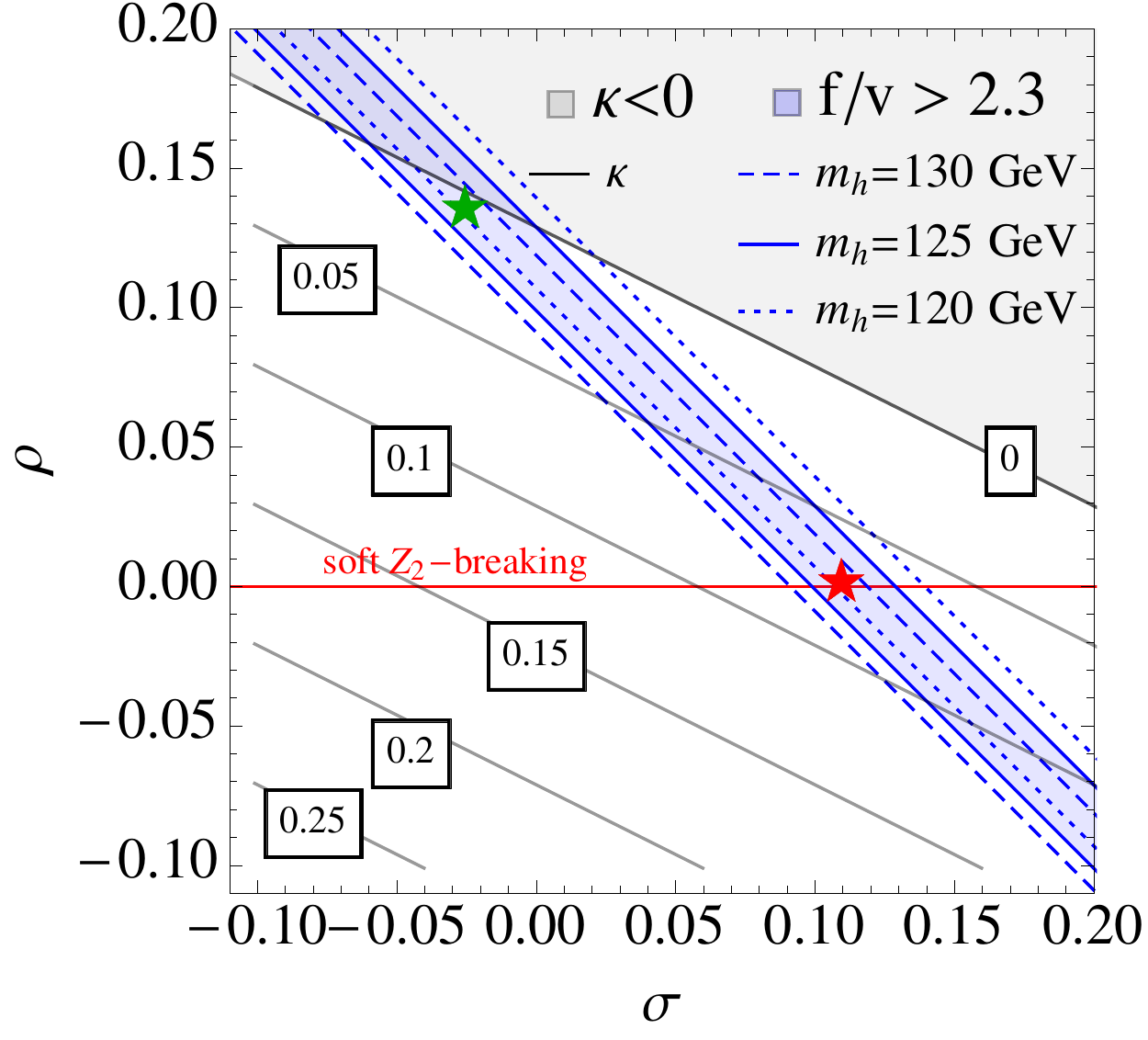}
\caption{The blue region inside the solid blue lines corresponds to the allowed parameter space with $f/v>2.3$ 
and $m_h=125$~GeV. The dashed (dotted) blue lines shows how the allowed region shifts if one 
assumes $m_h=120$~GeV ($m_h=130$~GeV). The black lines correspond to the values of $\kappa$. The 
non-viable region with negative $\kappa$ is shaded in grey. The green/red star indicates the region of hard/soft breaking.}
\label{PGB1}
\end{center} 
\end{figure}

As a guideline for model building, it is very helpful to visualize the
parameter space of the Twin Higgs model when both soft and hard $\ztwo
$-breaking terms are present.   
In order to do this we solve the equations for the measured values of
the Higgs mass~\eqref{vh2} and the VEV~\eqref{vh1}  
and remain with only two free parameters,\footnote{We use $m_h =
  125$ GeV and $v = 174$ GeV in all our calculations.   
Here we allow for a $\pm 5$~GeV Higgs mass shift because 
we do not take into account higher order corrections.}
which we choose to be the  $\ztwo$-breaking ones, namely  $\sigma$ and $\rho$. 
The values of the $\ztwo$-preserving quartic $\kappa$  and of the $SU(4)$-breaking scale $f$ are 
then fixed at each point in the $(\sigma, \rho)$
plane. 
We show the contours of  $\kappa$ and the region that satisfy the constraint $f/v > 2.3$ in Fig.~\ref{PGB1}.   
The first striking message of this figure is that, given the values
of the Higgs boson mass and the electroweak VEV, the acceptable values of the  $\ztwo$-odd parameters  are confined 
to a  narrow band in the $\ztwo$-breaking parameter space. 
 It is also instructive to consider the two limiting cases of dominant soft and dominant hard $\ztwo$-breaking. 
 The corresponding regions in 
the parameter space are marked in Fig.~\ref{PGB1} with red  and green stars, respectively. The values of $\kappa$ in 
both regions  can also be inferred from the figure: in the hard $\ztwo$-breaking region $\kappa$ is forced to be roughly a factor of 5 smaller than in the soft  $\ztwo$-breaking region as a consequence of the Higgs mass constraint.

Now we are ready to include the radiative corrections to the various parameters. These are coming from top loops and EW gauge boson loops (we discuss only the former in detail) but also from Higgs quartic loops, because the hard $\ztwo$-breaking quartic $\rho$ reintroduces for $\sigma$ a quadratic sensitivity to the mass threshold at which $\ztwo$-breaking  is generated. 

For the radiative top corrections, we should a priori consider the contributions of both visible and mirror tops and 
assume that the numerical value of the top-Higgs coupling in the visible and the twin sector can be different from 
one another. However, as was already pointed out 
in Ref.~\cite{Craig:2015pha}, the top Yukawas in the visible and the
hidden sector should agree to a level of better than 1\% in order 
to avoid an unacceptable FT. To simplify our discussion we enforce an \emph{exact} $\ztwo$ symmetry between the top sectors (and the gauge sectors also).

The dominant radiative corrections to the dimensionless parameters read
\begin{align}\label{kappaRC}
\Delta \kappa & =  \frac{3 y_t^4}{16 \pi^2} \log \frac{\Lambda_t^2}{m_{t_B}^2} + 
\frac{3 \lambda \rho}{32 \pi^2} \left( \log \frac{\Lambda_\rho^2}{m_{h_T}^2}+ 
\log \frac{\Lambda_\rho^2}{m_h^2}\right) \, , \\ \label{rhoRC}
\Delta \rho & = \frac{3 y_t^4}{16 \pi^2} \log \frac{f^2}{v^2} \, , \\\label{sigmaRC}
\Delta \sigma & = \frac{3 \rho }{16 \pi^2} \left( \epsilon \frac{\Lambda_\rho^2}{f^2} + 2 \lambda \log 
\frac{\Lambda_\rho^2}{m_h^2}\right) \, ,
\end{align}
where we kept frozen the Higgs dependence in the
logarithms and expanded to first
order in $\kappa$ and $v/f$. The running top Yukawa coupling, $y_t$, is evaluated at the 
UV cutoff scale $\Lambda_t$, fixing the scheme dependence of our one-loop computation. 

\begin{figure}[t]
\begin{center}
\includegraphics[scale=0.6]{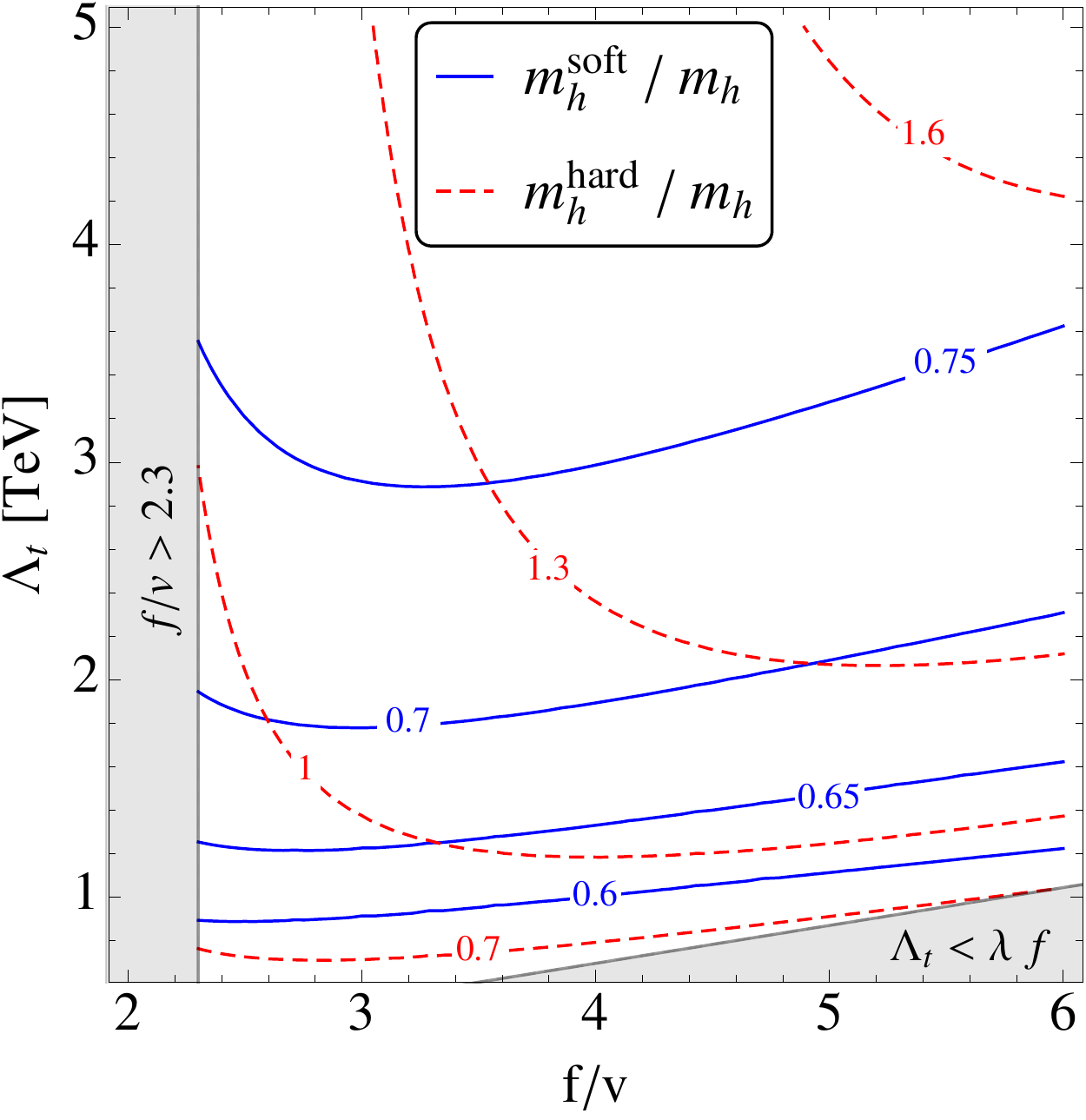}
\caption{The blue contours indicates the irreducible contributions to the Higgs mass in the 
pure soft breaking scenario $(\rho_0=0)$, coming from top loops normalized to $m_h=125\text{ GeV}$. 
The red dashed contours indicate the irreducible contributions to the Higgs mass in the pure 
hard breaking case ($\sigma_0 = 0$), where we fix $\lambda = 1$,
$\Lambda_\rho=1\text{ TeV}$ and $\epsilon=+1$ in Eqs.~\eqref{kappaRC}
and~\eqref{sigmaRC}. 
 The grey shaded region at the left edge with $ f < 2.3 \, v$ is excluded by Higgs coupling measurements and 
 the grey region at the bottom right of the plot has $\Lambda_t< \lambda f$.} 
\label{PGB2}
\end{center} 
\end{figure}

Several clarifications are in order. First, we introduced in these expressions two different mass thresholds: 
$\Lambda_t$ and $\Lambda_\rho$, cutting-off the top loops and the Higgs loops respectively.  Loosely speaking, in a UV complete natural theory $\Lambda_t$  will be identified with the mass scale of new colored states, while the states associated with $\Lambda_\rho$ can be complete SM singlets. While in strongly coupled UV completions  it is hard to imagine a wide separation between $\Lambda_t$  and $\Lambda_\rho$,  in weakly coupled UV completions ( {\it e.g.} SUSY) there could be appreciable differences between the two. Therefore, we keep track of these scales separately and we will
see in the concrete SUSY models of Sec.~\ref{sec:3} that $\Lambda_t$  and $\Lambda_\rho$
correspond to different SUSY mass thresholds. Of course
we should keep in mind that these scales cannot be arbitrarily
separated from one another, because of higher-order quantum
corrections.  

Second, in Eq.~\eqref{sigmaRC} we introduced a new parameter $\epsilon$. This parameter stand for the 
\emph{sign} of the UV mass threshold corrections and a priori $\epsilon = \pm 1$. Since one cannot calculate the sign of these radiative corrections within the IR effective theory, we remain agnostic and consider
both positive and negative threshold corrections. As we will see in the next section, within a full UV complete theory this sign is 
determined 
unambiguously.

Having at hand all the radiative contributions to $\rho,\ \kappa$ and
$\sigma $, we can estimate how big are the radiative contributions to the Higgs mass. 
We concentrate on two extreme cases: pure soft breaking, defined as $\rho_0 = 0$, and pure hard breaking, 
defined as $\sigma_0 = 0$. From Eq.~\eqref{vh2} we can infer that the
radiative contributions to the Higgs mass squared, up to  
$\cO(v^2/f^2)$ corrections, are given by $\Delta m_h^2  = 4v^2 (2 \Delta \kappa + \Delta \rho)$. We show 
the size of these corrections in Fig.~\ref{PGB2}. 
The first important conclusion that we draw from this figure is that while radiative corrections contribute at least $\sim 60\%$ of the 
Higgs mass for pure soft $\ztwo$-breaking, the radiative corrections
typically overshoot the Higgs mass for pure hard $\ztwo$-breaking ,  
though not by orders of magnitude.   
In what follows we will see that in the hard breaking case 
the FT $\Delta_{v/f}$ can be ameliorated compared to the soft case at the price of 
adjusting the Higgs mass (with either a large $\Lambda_\rho$ or a negative tree-level contribution to $\kappa$).

\subsection{Fine Tuning in the Low Energy Effective Theory and Beyond}\label{sec:FT}
We are now ready to discuss the FT quantitatively.  For illustration purposes let us start
from the pure soft case, where the IR FT is well-known to scale as $\sim f^2/v^2$.  More precisely, quantifying 
the fine-tuning \`a la Barbieri-Giudice~\cite{Barbieri:1987fn}, one gets 
\begin{align}
\Delta^{\rm soft}_{v/f} \equiv \left| \frac{\partial \log v^2}{\partial \log \sigma} \right|= \frac{f^2 - 2 v^2}{2v^2} \, .
\end{align} 
This is only a part of the total fine-tuning that one can estimate in the IR effective theory. On top of that one should also consider 
the fine-tuning of the scale $f$, which one can think of as the
inverse Fermi constant of the spontaneously broken $SU(4)$, with respect to the top
cutoff scale $\Lambda_t$.  Strictly speaking this fine-tuning should
be  
computed in the 
full UV theory, but one can get a reasonable estimate by analyzing the threshold 
corrections to the scale $f$ in the effective theory.
  These radiative corrections are given by
\beq
\Delta f^2 = \frac{1}{32 \pi^2 } \left( \frac{3 y_t^2}{\lambda}
  \Lambda_t^2 - 5 \, \Lambda_\lambda^2 \right)~.\label{eq:radiativef} 
\eeq
We will see in the next section how this expression reproduces the
dominant RGE effects of the UV theory, once the ``cut-off'' scales  
are identified with physical mass thresholds. A priori the sensitivities to the thresholds $\Lambda_t$ and $\Lambda_\lambda$ 
are equally 
dangerous and, unlike in the SM, it is not clear that the dominant sensitivity comes from the tops rather than the Higgses themselves.
However, at the threshold $\Lambda_\lambda$ we expect to find \emph{colorless} particles (singlet scalars and singlinos), 
which are weakly constrained by the LHC. Instead, one expects
the colored top partners at the scale $\Lambda_t$, which  
are more tightly bounded from direct searches. Therefore we will
assume that $\Lambda_t > \Lambda_\lambda$ and will mostly worry about
  the top threshold as main source of the fine-tuning.

From Eq.~\eqref{eq:radiativef} we can estimate the fine-tuning $\Delta_{f/\Lambda}$ 
of the scale $f^2$ with respect to the scale $\Lambda_t$ 
as:\footnote{One  
should be careful at this point. $\Lambda_t$ ``cut-off'' is not a parameter of the full UV theory and therefore one technically need a full UV theory to render this calculation reliable. However, we expect this calculation to be a good estimate of the fine tuning after the cut-off is properly mapped onto the parameters of the full theory.} 
\begin{align}
\Delta_{f/\Lambda_t}^{\rm soft} \equiv  \left| \frac{\partial \log f^2}{\partial \log \Lambda_t^2} \right| =  
\frac{3 y_t^2}{32 \pi^2 \lambda} \frac{\Lambda_t^2}{f^2} \, .
\end{align}
What is now the total fine-tuning of the soft Twin Higgs model? It is
tempting to say that  
$\Delta^{\rm soft} = \Delta_{v/f}^{\rm soft} \times
\Delta_{f/\Lambda_t}^{\rm soft}$.  
From the point of view of the effective field theory 
this factorization is definitely correct and  we can estimate the
total fine-tuning as  
\beq\label{eq:softFTIR}
\Delta^{\rm soft} \approx \frac{3 y_t^2}{32 \pi^2\lambda} \left( \frac{\Lambda_t^2}{2v^2}\right)~.
\eeq
We will see later that this result is an excellent approximation also well beyond the effective IR theory. 

Interestingly, Eq.~\eqref{eq:softFTIR} reproduces the familiar
expression for the fine-tuning in common SUSY models up to the crucial change  
$\lambda \leftrightarrow \lambda_{\rm SM}$, where the effective SM
quartic is defined via $m_h^2 = 4 \lambda_{\rm SM} v^2$, such that
  $\lambda_{\rm SM} \approx 0.13$ (as we will see later we can identify $\Lambda_t$ with the average stop soft mass $M_s$ times a $\log$ encoding the dependence on the high SUSY scale). 
We conclude that the fine-tuning in the Twin Higgs with a softly
broken $\ztwo$  
can be reduced with respect to the garden-variety SUSY by a factor
$\lambda_{\rm SM}/\lambda$ at best, 
in agreement with the results of Ref.~\cite{Craig:2013fga} and with later claims 
about the generic nature of the fine-tuning in Twin Higgs models (see
{\it e.g.} Ref.~\cite{RiccardoSearch}).   
This parametric structure of the fine-tuning measure is
characteristic of so-called double-protection  
models~\cite{Birkedal:2004xi,Chankowski:2004mq,Berezhiani:2005pb,Roy:2005hg,Csaki:2005fc,Bellazzini:2008zy,Bellazzini:2009ix}, 
in which the Higgs is a PGB,  and the scale of the spontaneous global
symmetry breaking is protected by SUSY.  
 
Therefore, the fine-tuning gain is at best moderate in weakly coupled theories
where perturbativity implies that $\lambda$  cannot exceed  
$\sim 1-1.5$. In practice, 
the gain is even more modest, because getting such high values of
$\lambda$ is not easy in full SUSY UV complete models, where $\lambda$
is a function of other parameters of the underlying theory. This difficulty has
motivated  model  
building in the direction of strongly-coupled Twin Higgs UV
completions~\cite{Barbieri:2015lqa,Low:2015nqa}, with the hope  
to saturate the parametric gain by taking $\lambda \approx  4\pi$. 
This kind of constructions have their own problems, most notably the
large number of states below the cut-off scale and  
the tensions with EW precision measurements, which again give an upper
bound on the  
value of $\lambda$~\cite{RiccardoSearch,TorreICTP}.

\begin{figure}[t]
\begin{center}
\includegraphics[scale=0.6]{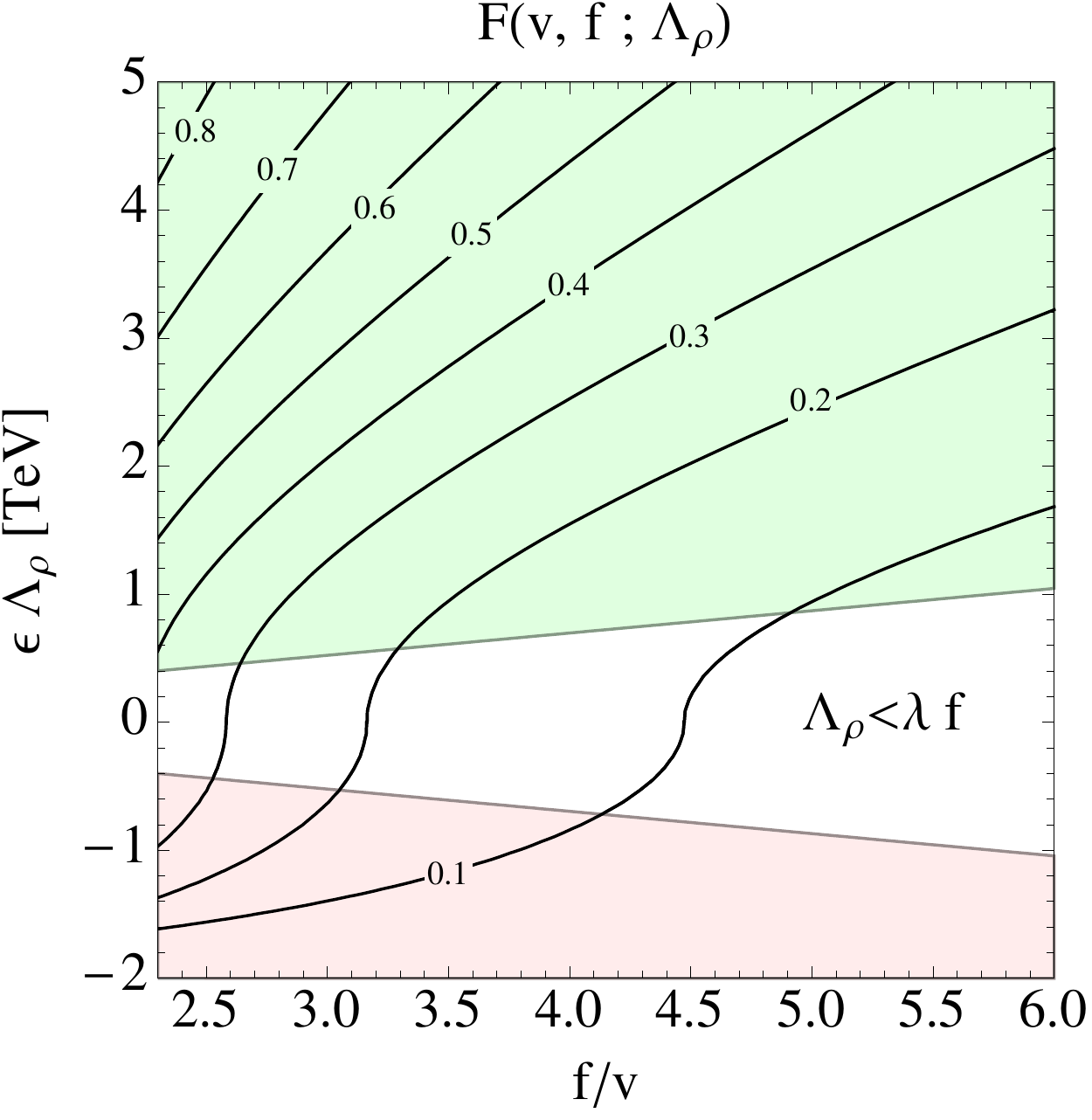}
\caption{Contours of $F(v, f; \Lambda_\rho)$ as it is defined in Eq.~\eqref{eq:Fdef}. The two different colors 
(light green and light pink) correspond to the two different signs of the one-loop threshold in Eq.~\eqref{sigmaRC} 
($\epsilon=+1$ and $\epsilon=-1$). In the white cone $\Lambda_\rho<\lambda f$ with $\lambda=1$.}
\label{PGB4}
\end{center} 
\end{figure}
 
 What happens if we consider pure hard $\ztwo$ breaking, i.e. $\sigma_0 = 0$? In this case $\sigma$ is 
 entirely given by the radiative corrections in Eq.~\eqref{sigmaRC}. In this case the fine-tuning in the IR effective theory 
 is given by the logarithmic variation of $v^2$ with respect to the parameter $\rho $ and reads
 \beq\label{eq:HardFT}
 \Delta^{\rm hard}_{v/f} = \frac{f^2 - 2v^2}{2 v^2} \times F(v, f; \Lambda_\rho)~,
 \eeq
 where we eliminated $\rho$ using the EWSB condition in Eq.~\eqref{vh1}
 and we dropped the $\log \Lambda_\rho$ contributions to $\sigma$ because they are largely subdominant 
to the quadratic contributions in most portions of the parameter space (but not all). 
Here we have introduced for further convenience 
 \beq\label{eq:Fdef}
 F(v, f ; \Lambda_\rho) \equiv \frac{3 \epsilon \Lambda_\rho^2 + 32 \pi^2 v^2}{3 \epsilon\Lambda_\rho^2 + 16 \pi^2 f^2}~. 
 \eeq
 Interestingly the first piece in~\eqref{eq:HardFT} is precisely the fine-tuning in the pure soft breaking scenario, but in the hard 
 breaking case 
 it is multiplied by a function $F(v, f; \Lambda_\rho)$.  
 We illustrate the behavior of this function in Fig.~\ref{PGB4}.\footnote{Note that on all our plots, including this figure, 
 we single out the region where the mass thresholds are below $\lambda
 f$, i.e. roughly the twin Higgs mass, because we do not expect our
 low energy EFT 
 to be valid there.} 
 This function is always smaller than 1 and in the limit 
 $3 \Lambda_\rho^2  \ll 32 \pi^2 v^2 $ reduces to $2v^2 /f^2$, while in the limit 
 $3 \Lambda_\rho^2  \gg 16 \pi^2 f^2 $ it saturates the upper bound of 1. If we assume that the fine-tuning of the scale $f^2$ with respect to the top cutoff scale is not different from the 
 soft case and the factorization works, this tells us (at least naively) that the reduction of the fine-tuning with respect to common
 SUSY scenarios should be $(\lambda_{\rm SM}/\lambda) \times F(v, f; \Lambda_\rho)$, which can be a significant improvement compared to the Twin Higgs with softly broken $\ztwo$.  
    \begin{figure}[t]
\begin{center}
\includegraphics[scale=0.55]{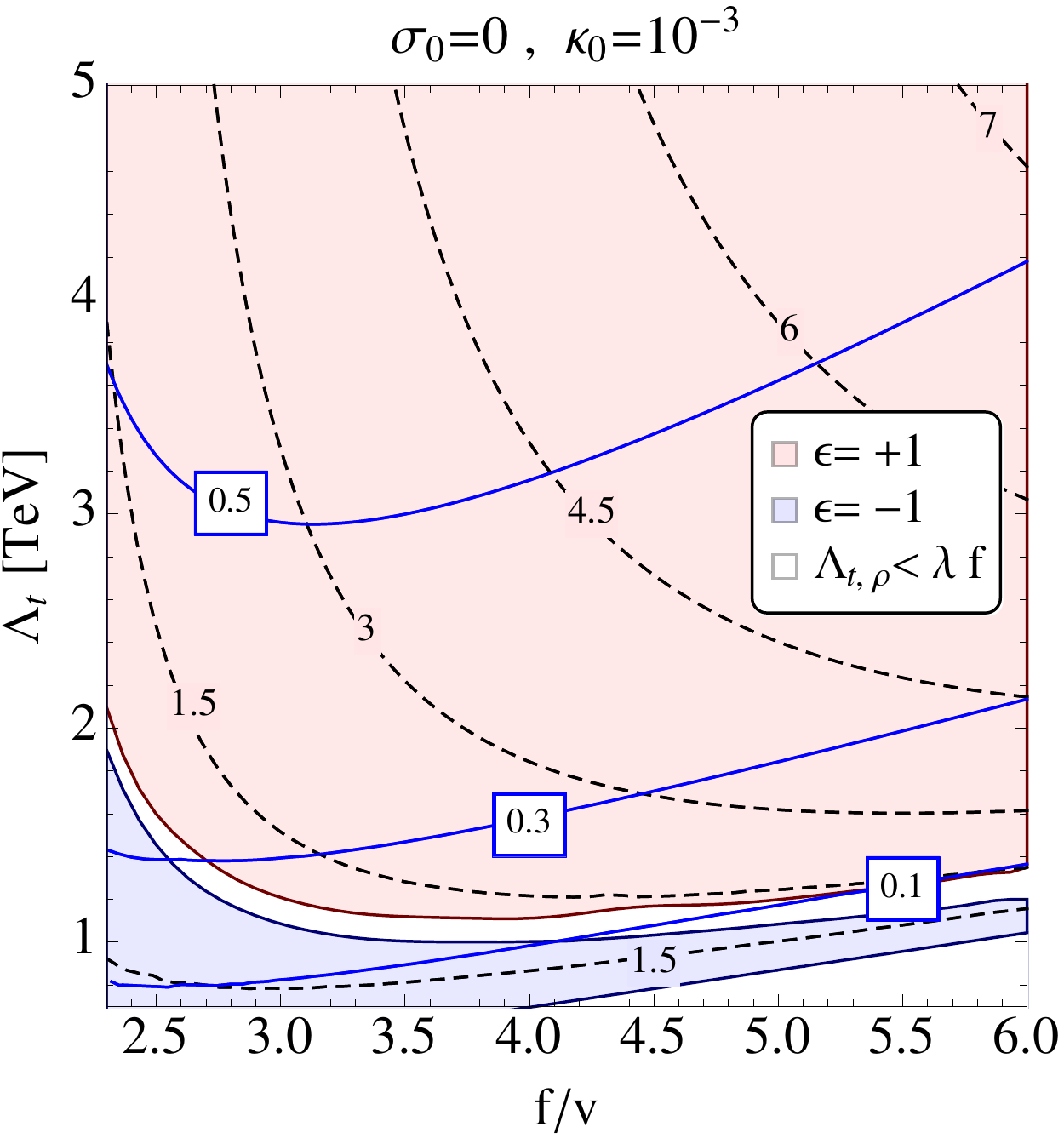}\hfill
\includegraphics[scale=0.55]{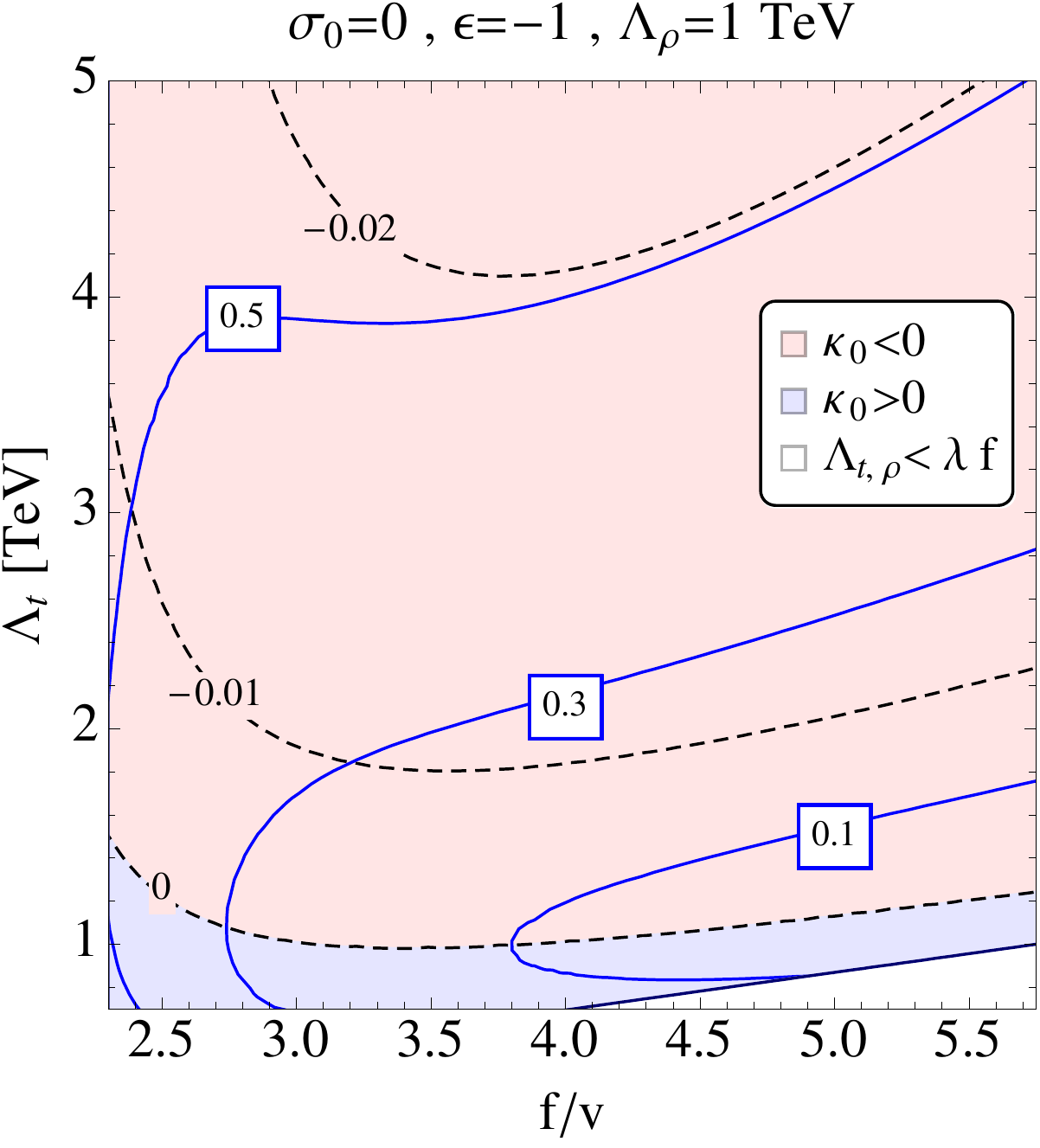}
\caption{The solid blue lines in both plots indicate the gain in
  fine-tuning of the hard-breaking model with respect to the soft one.  
In the white region $\Lambda_t<\lambda f$ or $\Lambda_\rho<\lambda f$ with $\lambda=1$. 
{\bf Left:} we take $\kappa_0=10^{-3}$ and 
the black dashed contours show the $\epsilon \times \Lambda_\rho$ (in
TeV units) adjusted to obtain $m_h = 125$~GeV.  
The light red/blue region is the allowed parameters space with
$\epsilon=\pm 1$. {\bf Right:} we fix $\epsilon=-1$ and
$\Lambda_\rho=1\text{ TeV}$ and fix  $\kappa_0$ (black dashed
contours) to get $m_h=125\text{ GeV}$. The light red/blue region is
the allowed parameters space with $\kappa_0\lessgtr0$. 
}
\label{PGB5}
\end{center} 
\end{figure}

Alternatively one can understand the parametric dependence of the
fine-tuning measure in the hard breaking scenario by fixing the  
 cut-off scale $\Lambda_\rho$ using the EWSB condition in Eq.~\eqref{vh1}, so
 that the fine-tuning in Eq.~\eqref{eq:HardFT}  
 can be rewritten as
  \beq\label{eq:HardFT2}
 \Delta^{\rm hard}_{v/f} = \frac{f^2 - 2v^2}{2 v^2} \times
 \frac{2\kappa}{2\kappa+\rho}= \frac{f^2 - 2v^2}{2 v^2}\times ~
   \frac{8 \kappa v^2}{m_h^2}\left(1-\frac{v^2}{f^2}\right). 
 \eeq
This expression makes manifest that the gain in fine-tuning of the hard-breaking scenario with respect to the 
soft breaking one 
is roughly proportional to the relative size of quartic couplings $\kappa$ and $\rho$. In the second equality we have
used that the combination of the two quartics in the denominator is fixed by the Higgs mass constraint~\eqref{vh2}, so that the 
gain in fine-tuning depends just on the value of $\kappa$, 
which has to be a small (see Fig.~\ref{PGB1}), in agreement with our requirement to minimize the fine-tuning. Clearly 
$\kappa$ cannot be arbitrarily small and its natural value is indicated by Eq.~\eqref{kappaRC}, numerically 
$\cO(10^{-2} \ldots 10^{-3})$. As we 
see in Fig.~\ref{PGB2} this slightly overshoots the Higgs mass, but the overshooting is $\cO(1)$ rather than order of 
magnitude, therefore the fine-tuning associated  with adjusting the value of $\kappa$ is also merely of order~$\cO(1)$.

In  Fig.~\ref{PGB5} we show the improvement in the fine-tuning of two different hard-breaking scenarios. Both these scenarios will have 
 direct analogs in the SUSY UV completions to be discussed in Sec.~\ref{sec:hard}. On the left panel we consider a purely hard 
 $\ztwo$ breaking with $\sigma_0 = 0$ and a small $\kappa_0$ that we take $10^{-3}$ having in mind its typical value in 
 SUSY UV completions at small $\tan\beta$.  The value of $\epsilon\times\Lambda_\rho$ is indicated by the black 
 dashed isolines on the left panel of Fig.~\ref{PGB5}. Since $\kappa_0$ is fixed, this quantity is determined in order 
 to get the measured Higgs mass value $m_h=125~\text{GeV}$.
  Interestingly, in the light blue region of Fig.~\ref{PGB5} where the threshold correction proportional to 
 $\Lambda_\rho$ is \emph{negative} (i.e. for $\epsilon=-1$), the Higgs mass induces a 
 strong upper bound on $\Lambda_t$. In the opposite case where the threshold correction is 
 \emph{positive},  one can always adjust the Higgs mass in  Eq.~(\ref{vh2}) via a large radiatively induced 
 $\sigma$, increasing $\Lambda_\rho$. 
Of course this comes at the price of reducing the gain in
fine-tuning of the  
 hard-breaking model, as it is shown by the blue isolines on  the left
 panel of Fig.~\ref{PGB5}. 

On the right panel of Fig.~\ref{PGB5} we consider a purely hard $\ztwo$-breaking case with $\sigma_0 = 0$, $\Lambda_
\rho=1 \, \text{TeV}$ and we fix $\epsilon=-1$ (i.e. the sign of the
threshold correction in Eq.~\eqref{sigmaRC}). If we require $\kappa_0$ to compensate for the overshooting of the
Higgs mass, than a \emph{negative} $\kappa_0$ is needed in order to
avoid an upper bound on the scale of the colored states $\Lambda_t
\lesssim 1$~TeV.  
This leads to a new contribution to the fine-tuning from the
adjustment of $\kappa_0$ against $\Delta \kappa$ that we sum up in
quadrature with Eq.~\eqref{eq:HardFT}, and the final gain in fine-tuning
with respect to the soft breaking scenario is indicated  by the blue
isolines of Fig.~\ref{PGB5} (right). 

We now comment on the caveats of our effective analysis. 
First, the factorization of the fine-tuning measure \emph{very often fails} in UV complete models. 
This failure is related to RGE effects that introduce a dependence of $f$ on the  $\ztwo$-breaking parameters. 
These effects could be important in UV complete hard breaking models
as we will see in Sec.~\ref{sec:3}. Second, there is the
$\log$-dependence of both $\kappa$ and  $\sigma$ on $\Lambda_\rho$,
which we have neglected in order to have a simple analytical
understanding. Third there might be extra threshold corrections coming
from the  
UV theory which sensibly affect the Higgs mass estimate. All these
effects can slightly weaken our naive  
estimates of the reduction in fine-tuning.   

In the next section we will show how all these caveats are taken under control in explicit SUSY UV completions, which indeed follow the parametrical intuition developed here in most of the parameter space. In particular we will show examples of hard breaking models that sensibly ameliorate the fine-tuning of the Twin Higgs.

\section{SUSY UV completions}\label{sec:3}
In this section we discuss simple SUSY UV completions of the Twin
Higgs. In the minimal setup the  
Twin SUSY UV completion consists of  two copies of the MSSM
$(\text{MSSM}_A\times\text{MSSM}_B)$ that are  
symmetric under the action of the $Z_2$ exchange symmetry
$A\leftrightarrow B$.  
The Higgs sector of the model consists of two replicas of the two
Higgs doublet model charged under the $A$ and the $B$ EW group respectively.     

Even if the most general SUSY potential is quite involved, we can
develop an analytical understanding by going in the decoupling limit
of the SUSY Higgses and matching it to the non-SUSY Twin Higgs
potential in Eq.~\eqref{Vtwin} (with 2 doublets $H_{A,B}$) by replacing 
\bea
h_u^{A} & = H_A s_A\, ,  &h_u^{B} & = H_B s_B \, , \\
h_d^{A} & = H_A^{\dagger} c_A\, ,  &h_d^{B} & = H_B^{\dagger} c_B \, ,
\label{decouplinglim}
\eea
where $s_{A,B} = \sin \beta_{A,B}, c_{A,B} = \cos \beta_{A,B}$. Of
course this matching gives a good description of the SUSY Twin
Higgs EWSB as long as the SUSY Higgses are heavier than the Twin
Higgs. We give a treatment of the full four Higgs doublet model beyond
this approximation in Appendix~\ref{app:2}. 

We first review Twin SUSY constructions with only soft
$\ztwo$-breaking, which exemplifies several generic features  
of the SUSY UV completions of the Twin Higgs.  
We pay particular attention to the simplest soft Twin SUSY model
proposed in Ref.~\cite{Craig:2013fga} and validate  
our analytical understanding against a full numerical treatment.  We
show how the parameter space of this model can  
be completely solved once the Higgs mass and EWSB constraints are imposed
and comment 
on possible directions to ameliorate  
the fine tuning.  

Then we move to Twin SUSY models with hard $\ztwo$-breaking, which
UV-complete some previously described setups
and feature similar improvement in the FT.  
We first discuss the
prominent role of the Higgs  
mass constraint in these models showing that the simplest
implementation of hard $\ztwo$-breaking in SUSY has a strong upper
bound 
on the stop mass scale $M_s$\footnote{Hereafter we define the stop
  mass scale as a geometric mean of the stop
masses $M_s \equiv \sqrt{M_{\tilde t_1} M_{\tilde t_2}}$. Note also
that $M_{\tilde t_i}$ stands for the physical  
mass of the stop, rather than for its soft mass.} once the constraint
$m_h=125\text{ GeV}$ is imposed. 
Finally, we present an explicit model where the measured Higgs mass is
obtained via a \emph{negative} contribution to $\kappa$, very much in
the spirit of the effective theory presented in the right panel of
Fig.~\ref{PGB5}. In this model we get $\sim10\%$ FT with
$M_s\sim 2\text{ TeV}$ and $\lambda_S=1$, ameliorating the fine-tuning
of soft Twin-SUSY by a factor of about $\sim 5$.

\subsection{Soft Twin SUSY}\label{subsec:softTwin}
In this subsection we analyze SUSY UV completions of the
Twin Higgs with \emph{softly} broken mirror symmetry. Even though this
kind of UV completions have already been discussed to some extent in the literature, reviewing them carefully will clarify the basic building blocks of any SUSY UV
completion of the Twin Higgs. Analogously to
Sec.~\ref{sec:2}, we organize our discussion according to the global
symmetries that each term of the scalar potential preserves.  

Generating the $SU(4)$-invariant part of the Twin Higgs potential
in Eq.~\eqref{Vtwin} already introduces some degree of 
 model dependence in SUSY UV completions. Indeed, if we consider the
 matter content of just two copies of the MSSM, the only
 $SU(4)$-preserving operators at the renormalizable level are 
the mass terms. 
Therefore the $SU(4)$-invariant quartics
 require some additional dynamics. The simplest possibility is to use
 a non-decoupling F-term from a heavy singlet $S$ that has NMSSM-like
 couplings with the $A$ and $B$ Higgses, as in Refs.~\cite{Falkowski:2006qq,Craig:2013fga}.   
The superpotential and soft masses of this setup are
\bea
&W_{SU(4)} = \left(\mu+\lambda_{S} S \right)  \mathcal{H}_u
\mathcal{H}_d+ \frac{{\cal M}_S}{2} S^2 \, ,\\ 
\label{SU4super}
&V_{SU(4)}  = m_{H_u}^2 |{\cal H}_u|^2 + m_{H_d}^2 |{\cal H}_d|^2  -b
\left(\mathcal{H}_u\mathcal{H}_d+ {\rm h.c.} \right) + m_S^2 |S|^2~. 
\eea
To make our equations more compact, we have switched here
to manifestly $SU(4)$ invariant notations. We will further  
use $\mathcal{H}_{u,d} =(h_{u,d}^A, h_{u,d}^B)$, wherever the $SU(4)$ 
conventions are appropriate, and with a slight abuse of notation,  
we will use $\mathcal{H}_{u,d}$ both for the Higgs superfields and
their lowest components.  

We also assume that the singlet soft mass is much larger than the
SUSY one and integrate out $S$ in this limit.  
The potential we get is:  
\begin{equation}
V_{SU(4)}^{\rm eff}\approx m^2_u \vert \mathcal{H}_u\vert^2+m^2_d \vert
\mathcal{H}_d\vert^2-b \left(\mathcal{H}_u\mathcal{H}_d+ {\rm h.c}
\right)+\lambda_S^2\vert \mathcal{H}_u\mathcal{H}_d\vert^2  \, ,  
\label{U4eff}
\end{equation}
where we have defined $m^2_{u,d}=\mu^2+m_{H_{u,d}}^2 $. We also kept
just the renormalizable operators and neglected the extra quartics of  
order $\mathcal {O}(\mathcal{M}_S^2/m_S^2)$,  ${\cal O}
(\mu\mathcal{M}_S/m_S^2)$ and  
$\mathcal{O} (\mu^2/m_S^2)$. By construction these
sub-leading quartic terms are also $SU(4)$-invariant. 

Throughout this paper we will often trade the $b$-term for the mass of
the heavy CP-odd Higgs $m_{A_T}=2b/\sin(2 \beta)$.
Note,  
that unlike in the MSSM, where $2b/ \sin(2 \beta)$ is the mass-squared
of the CP-odd Higgs, here, in the SUSY Twin Higgs, it controls the
mass-squared of the {\it mirror CP-odd Higgs}. The mass of the
``visible'' CP-odd Higgs turns out to be always lighter: $m_A^2\approx
m_{A_T}^2-\lambda_S^2 f^2$ (see Appendix \ref{app:2}).  

We further match the $SU(4)$-invariant parameters of the SUSY
potential in Eq.~\eqref{U4eff} to the parameters  
of the Twin Higgs potential in Eq.~\eqref{Vtwin}:
\beq \label{eq:softmatching}
\lambda\approx\frac{\lambda_S^2}{4}s_{2\beta}^2~, \  \  \  \ \ \ \  
 m^2\approx  m_u^2 s_\beta^2+m_d^2 c_\beta^2-b s_{2\beta}~. 
\eeq
In these expressions we have disregarded the terms that depend on the
difference between the $\beta$ angles in the different sectors.
We will later analyze in detail the role of the
$\beta$-misalignment. So far we are taking the SUSY
decoupling limit, which is defined by Eq.~\eqref{decouplinglim}, while
we are keeping the twin Higgs in the spectrum treating the Twin Higgs
model as a full linear sigma model. The effective PGB theory discussed
in Sec.~\ref{sec:2} can be obtained integrating out the twin Higgs and
expanding at the first non-trivial order in $\lambda \gg
\kappa,\sigma$ (see Appendix~\ref{app:2} for details on this point).   

Eq.~\eqref{eq:softmatching} immediately explains the origin of the
problem already mentioned in the previous section: 
SUSY UV completions have a hard time to maximize $\lambda$ and therefore
also 
the fine-tuning gain, which is  $\lambda_{\rm SM}/\lambda$.  
In SUSY UV completions $\lambda_S$, rather than $\lambda$, should be
perturbative. Moreover, as we will shortly see, in order 
to get the right Higgs mass we will have to stick to moderate values
of $\tan \beta$, thus further suppressing the effective $\lambda$. As a
result, we will generally get $\lambda < 1$,  
such that the gain in fine tuning will never be
large. In passing, we notice that the fine tuning in the
soft models can be slightly ameliorated by changing the functional
dependence of $\lambda$ on $\tan \beta$ and other fundamental parameters,
as we comment later on. 

We now proceed to discus the leading $\ztwo$-even but
$SU(4)$-breaking operators in the Twin SUSY potential. At  leading
order  the  $SU(4)$-breaking originates  at tree-level
from the electroweak D-terms and at one-loop from the top Yukawa
sector: 
 \begin{align}
V_{\slashed{U}(4)}^D&= \frac{g_{\text{ew}}^2}{8} \left[ \left(
    |h^A_u|^2 - |h^A_d|^2 \right)^2 +  \left( |h^B_u|^2 - |h^B_d|^2
  \right)^2  \right] \, , \label{Dterm}\\ 
V_{\slashed{U}(4)}^{\rm{top}}  &\approx \frac{3y_t^4 }{16 \pi^2}
\left[ (|h^A_u|^4+|h^B_u|^4) \log \frac{M_s^2}{ m_{t_B}^2}
  +|h^A_u|^4\log \frac{f^2 }{ v^2}\right] \, ,\label{stoptop} 
\end{align}
where we defined $g_{\text{ew}}^2=g^2+g^{\prime 2}$, and $y_t$ is the
SUSY superpotential coupling, related to 
the top mass as $m_t= y_t v \sin\beta$. In order to 
get Eq.~\eqref{stoptop} we compute the  
CW potential and 
and set the dynamical Higgses to their VEVs in the non-polynomial
terms. The RH and LH stops mass are assumed to be equal, as well as the soft masses in the $A$ and $B$
sectors.\footnote{We also assume the trilinear $A$-terms to be negligible  
and expand these expressions at the  leading order in
$m_{t_{A,B}}^2/M_s^2$.}

Matching again the effective $SU(4)$-breaking potential
to Eq.~\eqref{Vtwin} we identify  
\begin{equation}
\kappa\approx \frac{g_{\text{ew}}^2}{8} c_{2\beta}^2+\frac{3 m_t^4
}{16\pi^2 v^4} \log\left( \frac{M_s^2}{m_t^2} \frac{v^2}{f^2}
\right)~.  
\label{softkappa}
\end{equation}
This expression indicates that the $SU(4)$-breaking $\ztwo$-even
quartic $\kappa$ gets an unavoidable  positive contribution  
from the EW D-terms at tree level in any SUSY UV completion.  
The ballpark of this contribution is $\cO(10^{-2} \ldots 10^{-3})$
depending on $\tan \beta$. 
The contribution from the top-stop sector reproduces the result
obtained in Eq.~\eqref{kappaRC}. Given that top loops alone already set the Higgs mass in the right
ballpark (see the blue contours in Fig.~\ref{PGB2}),  
we expect that for a fixed $M_s$ a Twin SUSY model would have an
\emph{upper} bound on $\tan \beta$. We will later see that 
 because of this
bound only models with small or moderate $\tan
 \beta$ are viable.    

In those Twin SUSY theories, where the mirror symmetry is broken only
softly, the  $\ztwo$-breaking terms can be introduced as soft masses in the
potential:  
\begin{equation}
V_{\slashed{Z}_2}^{\rm soft}\approx \Delta m_u^2 \vert
h_u^A\vert^2+\Delta m_d^2 \vert h_d^A\vert^2+\Delta b
\left(h_u^Ah_d^A+ {\rm h.c} \right) .\label{Z2super} 
\end{equation}
These can easily be matched to the respective term in the IR
effective potential in Eq.~\eqref{Vtwin}  
\begin{equation}
\sigma \approx \sigma_u s_\beta^2+\sigma_d c_\beta^2~,
\label{Z2oddsoft}
\end{equation}
where we defined $\sigma_{u,d}= \Delta m_{u,d}^2/f^2$
and assume  $\Delta b=0$.

Using the matching conditions described so far, we can rephrase in 
SUSY language the two EWSB conditions, which were already captured by
the low energy Twin Higgs potential in Eq.~\eqref{Vtwin}.
The first EWSB
condition is the one we discussed in the PGB limit in Eq.~\eqref{vh1} which fixes the EW scale $v$ in terms of $f$. This is achieved in soft
$\ztwo$-breaking models by balancing $\sigma$ against $\kappa$. These
two parameters are determined in terms of four parameters of the SUSY
theory: $\beta$ and $M_s$ determine $\kappa$ through
Eq.~\eqref{softkappa}, while $\sigma_u$, $\sigma_d$ and $\beta$ give
$\sigma$ through Eq.~\eqref{Z2oddsoft}.  

Solving the EWSB condition in the IR effective theory we can trade
$\sigma$ for $f/v$. The second combination of
$\sigma_u$ and $\sigma_d$ 
is unimportant in the 
$\delta t_\beta = 0$ limit. Imposing the Higgs mass constraint in the
PGB limit then fixes 
$\kappa$ and allows us to predict the value of $\beta$ in the
$(f/v, M_s)$ plane by solving Eq.~\eqref{softkappa} for
$\beta$. This procedure holds only as far as
$\kappa/\lambda\ll1$ and we will later discuss how the Higgs mass
formula gets modified beyond the PGB approximation.  

The second condition is essentially the $SU(4)$ breaking condition,
namely $f^2\approx m^2/2 \lambda$, rephrased in terms of the parameters
of the underlining SUSY theory (note that here $m$ stands for the 
mass parameter in Eq.~\eqref{Vtwin}) using
the matching conditions in Eq.~\eqref{eq:softmatching}. The dependence of the scale $f$ on the stop threshold $M_s$
can then be estimated via the RGE of $m_{H_u}$, which largely
dominates 
the fine-tuning measure. In this approximation we get that $\delta
f^2\approx \frac{2 s_\beta^2}{\lambda_S^2 s_{2\beta}^2}\times\delta
m_{H_u}^2$ . The fine tuning measure derived in Eq.~\eqref{eq:softFTIR} 
in terms of the SUSY parameters is 
\begin{equation}
\Delta^{\rm{soft}}\approx\frac{3y_t^2 s_\beta^2}{2\pi^2\lambda_S^2
  s_{2\beta}^2}\frac{M_s^2\log 
\frac{\LSUSY}{M_s}}{2v^2}\, .\label{softTwinFT}
\end{equation}
 In this estimate we also neglect the contribution of the singlet soft
 mass $m_S$ in the RGE for $m_{H_u}$. This contribution
 turns out to be sub-leading with respect to the stops one as long as
 $m_S\sim 1 \ldots 2\text{ TeV}$.  

Looking back at the low energy fine-tuning~\eqref{eq:softFTIR} we see
that  in the SUSY theory $\Lambda_t^2$ is identified with $2
M_s^2\times \log\frac{\LSUSY^2}{M_s^2}$. The UV cut-off $\LSUSY$ can
be interpreted as a scale where the soft masses are formed, corresponding for example to the
messenger scale in low energy SUSY-breaking scenarios. For the
purpose of numerical calculations we will  further use $\LSUSY=100\, M_s$. 

 All in all  the gain in fine tuning of soft Twin SUSY with
 respect to standard SUSY models like the NMSSM is 
\begin{equation}
\frac{\Delta^{\rm{soft}}}{\Delta^{\rm{NMSSM}}}\approx \frac{m_h^2}{2v^2}\times
\frac{1}{\lambda_S^2s_{2\beta}^2} = \frac{2 \lambda_{\rm SM}}{\lambda_S^2 s_{2 \beta}^2}~,
\label{gainFTsoft}
\end{equation}
in agreement with the analysis performed in Ref.~\cite{Craig:2013fga} as
well as with our expectations from the IR effective theory analysis in
Sec.~\ref{sec:2}.  It is important to remember that
the Higgs mass constraint in this model fixes the value of $\beta$ for
a given $M_s$ and $f/v$ determining the final gain in fine tuning.

\begin{figure}[t]\centering
\includegraphics[scale= 0.49]{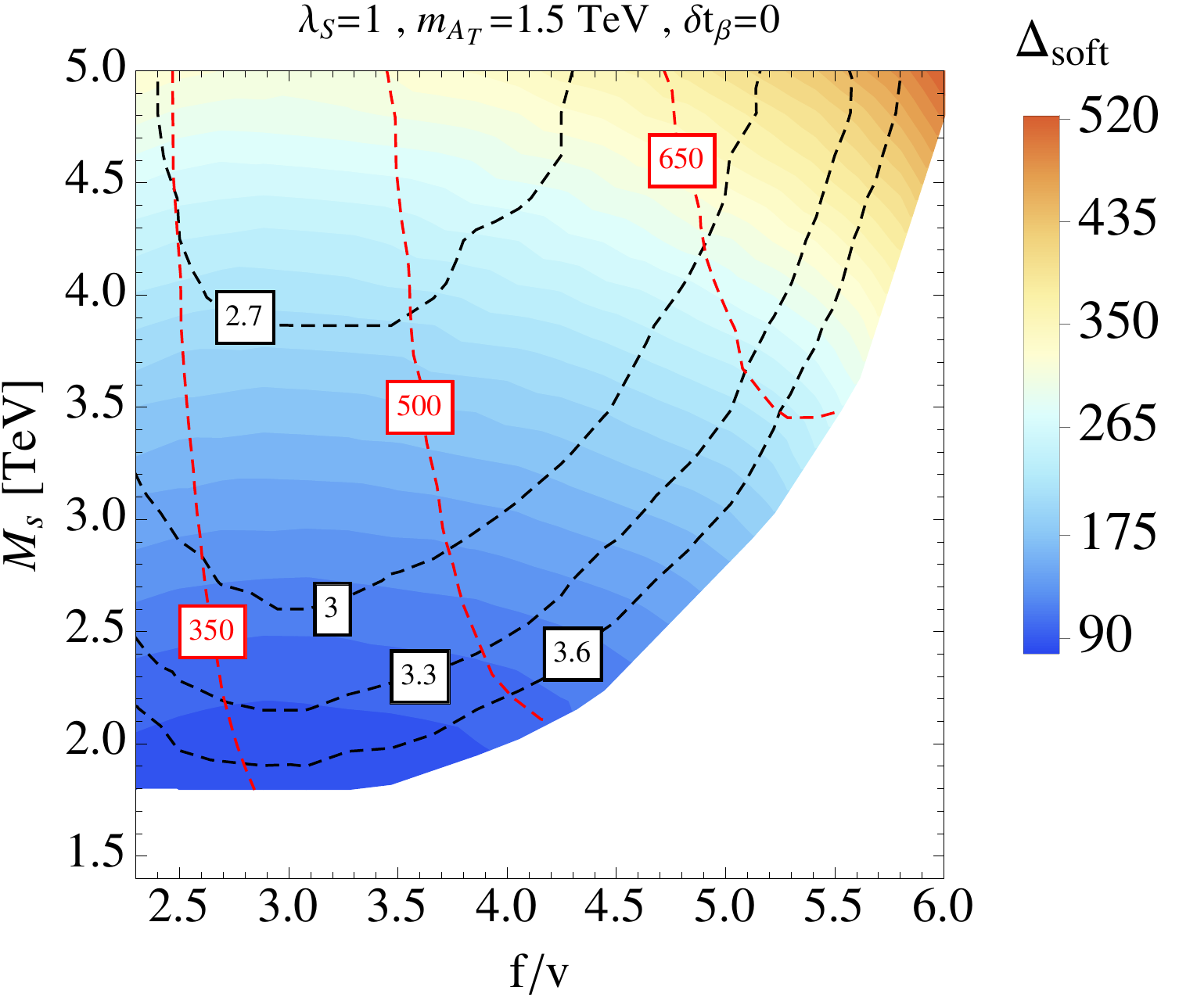}\hfill
\includegraphics[scale= 0.49]{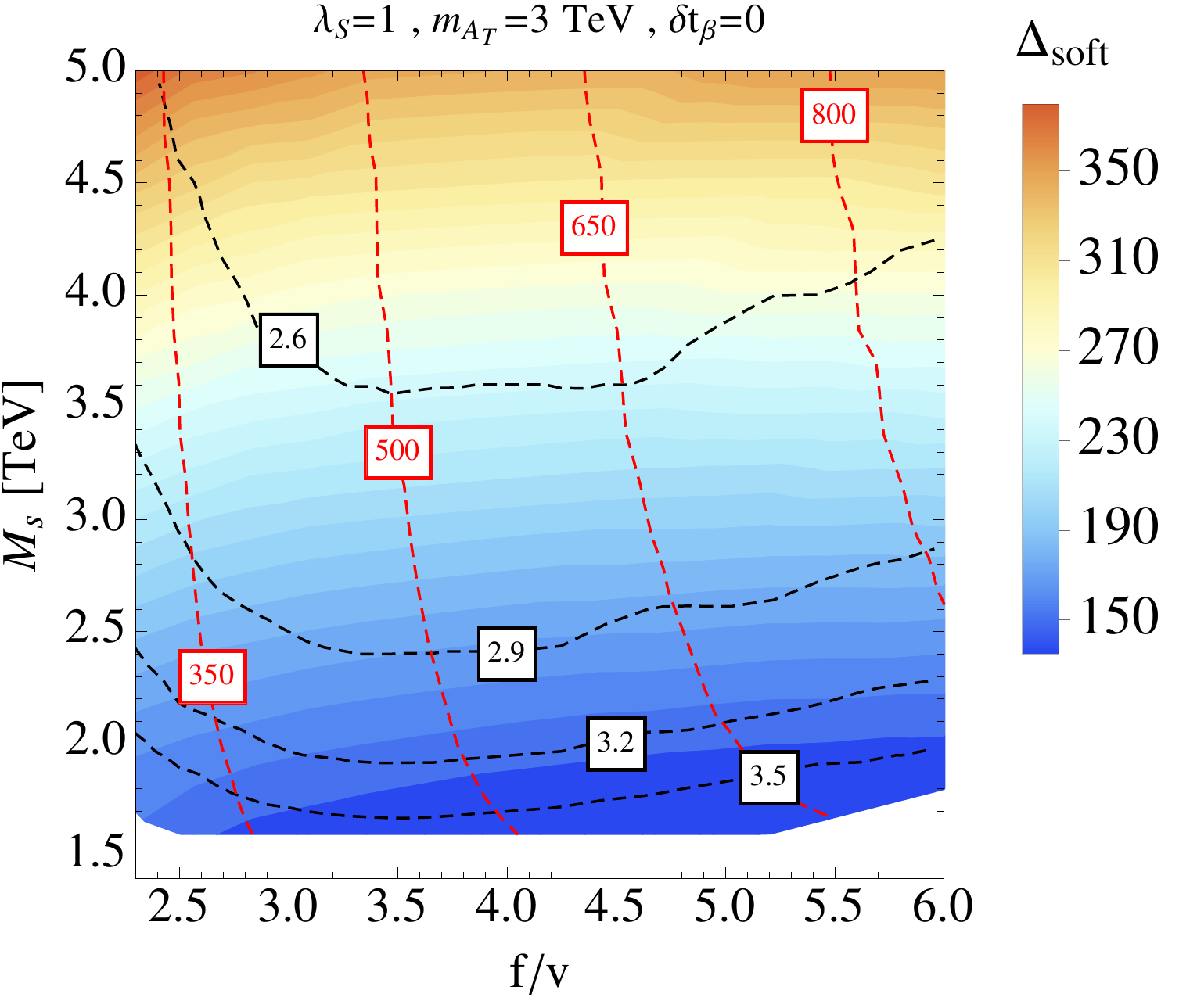}
\caption{Soft Twin SUSY parameter space with $m_{A_T}=1.5\text{ TeV}$
  (left panel) and $m_{A_T}=3\text{ TeV}$ (right panel). In the white
  region $m_h=125\pm 2$~GeV cannot be achieved as explained in the
  text. We also assume here $\delta t_\beta=0$, $\lambda_S=1$ and
  $m_S = 1$~TeV,  
$\mu=500\text{ GeV}$, $M_3=M_s$. Black/red dashed lines are contours 
of $\tan\beta$/next-to-lightest CP-even Higgs mass in GeV. The gradient color
function indicates region of increasing $\Delta_{\text{soft}}$,
calculated in Eq.~\eqref{numerical_def}.  
\label{numeric_FT_soft}}
\end{figure}

We summarize the behavior of the soft
Twin SUSY model in Fig.~\ref{numeric_FT_soft}. We work directly in the
SUSY theory, 
solve numerically the EWSB conditions and the Higgs mass
constraint, and  present the allowed parameter space in the $(f/v,\
M_s)$ plane. 
In this plane we show the behavior of the
fine tuning measure, the minimal value of $\tan\beta$ that satisfies the
Higgs mass constraint \footnote{Since we are allowing for a range of
possible SM Higgs masses we select the minimal $\tan\beta$ within
the range of the allowed ones.} 
and the mass of the next-to-lightest
CP-even Higgs. We show our results for two different choices of the
mirror CP-odd Higgs masses $m_{A_T}=1.5, 3\text{ TeV}$. In the rest of
the section we give an analytical understanding of the results of Fig.~\ref{numeric_FT_soft}.

The fine tuning we show in Fig.~\ref{numeric_FT_soft} is computed \`a
la Giudice-Barbieri by evaluating numerically the logarithmic
derivatives of the EW scale $v$ with respect to \emph{all} the UV
parameters of the full soft Twin SUSY model:  
\begin{equation}
\Delta^{\text{soft}}=\sqrt{\sum_{i=\mathcal{P}_{\text{soft}}}\left(
    \frac{\partial\log v^2}{\partial\log p_i} \right)^2}~. 
\label{numerical_def}
\end{equation}
In this expression $i$ runs over the parameters   
of the Higgs sector $\mathcal{P}_H=\{ m_{H_u}^2\, , m_{H_d}^2\, , 
\mu^2\, , b \}$, of the colored sector  
$\mathcal{P}_Q=\{m_Q^2\, , m_U^2\, , M_3\}$, of the singlet sector
$\mathcal{P}_S=\{m_S^2\, ,\lambda_S\}$, and of  
the $\ztwo$-breaking sector 
$\mathcal{P}_{\slashed{Z}_2}=\{\Delta m_{u}^2\, ,\Delta
m_{d}^2\}$. All these parameters are considered  at the ``messenger
scale'' $\LSUSY$ for the purpose of the fine-tuning computation and   
we use full one-loop RGE (see Appendix~\ref{app:RGE}) to obtain the final fine tuning.

One more comment is appropriate here regarding the total fine tuning
measure in Eq.~\eqref{numerical_def} that we are using.  
This is slightly different from the measure which has been used by
Barbieri and Giudice in~\cite{Barbieri:1987fn}, which assumed the
total fine-tuning to be the \emph{maximum} of the one-parameter fine
tunings, rather than their \emph{sum in  
quadrature}, as we do. We choose this particular measure to give a
numerical expression to an intuition, that if a model  
has a similar fine-tuning with respect to two independent parameters,
it is more fine-tuned, than a model that is one-tuned  
only with respect to a single parameter.  We checked it explicitly,
and in vast majority of the parameter space our measure is  
not very different numerically from the nominal Barbieri-Giudice
measure. Note however, that in case of multiple free parameters and
small fine-tuning our measure overestimates the fine-tuning with
respect to the Barbieri-Giudice measure.  

As expected, our tuning measure can be well approximated by
formula~\eqref{softTwinFT},  
meaning that the gross features of the Twin SUSY model are captured by
the PGB intuition we have developed so far. In particular the factorized
formula for the tuning measure works pretty accurately and we can
 rewrite Eq.~\eqref{numerical_def} as  
\begin{equation}
\Delta^{\text{soft}}\approx
\sqrt{\sum_{i=\mathcal{P}_{\slashed{Z}_2}}\frac{\partial\log
  v^2}{\partial\log p_i}} \times  
\sqrt{\sum_{j=\mathcal{P}_H\cup\mathcal{P}_Q\cup\mathcal{P}_S}\frac{\partial\log
  f^2}{\partial\log p_j}}\approx  
\sqrt{\sum_{i=\mathcal{P}_{\slashed{Z}_2}}\frac{\partial\log
  v^2}{\partial\log p_i}}\times  
\frac{\partial\log f^2}{\partial\log m_{H_u}^2}~,
\label{factorization_soft}
\end{equation}
where the last expression exactly reproduces the approximated
formula in Eq.~\eqref{softTwinFT}.  

The gain in fine tuning with respect to the NMSSM~\eqref{gainFTsoft}
follows essentially the contours of  
$\tan\beta$ in Fig.~\ref{numeric_FT_soft}.  Since $\tan\beta$
decreases at larger $M_s$, the effective $\lambda$ grows (see
Eq.~\eqref{eq:softmatching}), and consequently this particular soft
Twin SUSY model has a bigger gain  
with respect to the NMSSM at higher values of $M_s$. For $\lambda_S=1$
the gain is merely a factor of $0.4 \ldots 0.5$, 
depending on the region of the parameter space. This can be slightly
improved up 
to $0.2 \ldots 0.3$ by taking $\lambda_S=1.4$,  
i.e. very close to its perturbativity bound.

The fine tuning measure is a relatively flat function of $f/v$ as
expected from the factorized formula ~\eqref{softTwinFT}. This is
especially true in the right panel where we take $m_{A_T}=3\text{ TeV}$
and the SUSY states are completely decoupled in the full range of
$f/v$.
The residual dependence on $f/v$ in the left panel is introduced indirectly by the Higgs mass
constraint, which in turn determines the value of $\tan\beta$ at fixed
$f/v$ and $M_s$.  

We now analyze the role of the Higgs mass in the soft Twin SUSY
parameter space. In particular, we would like to understand the
absolute \emph{lower bound} on $M_s$ which is present in both plots and the \emph{upper bound} on  
$f/v$ at fixed $M_s$ in the left panel. These two bounds are
responsible for the white region in Fig.~\ref{numeric_FT_soft}, where
there is no value of $\tan\beta$ for which the Higgs mass constraint
can be satisfied. Assuming that there is an upper bound on $\tan\beta$,
  the lower bound on $M_s$ can be easily understood from the PGB mass
  formula:
\beq
m_h^2 \approx g_{\text{ew}}^2 v^2 c_{2\beta}^2 + \frac{3m_t^4}{2  \pi^2 v^2}
\log \left ( \frac{M_s^2}{m_t^2} \frac{v^2}{f^2}
\right)~,
\eeq
which shows that for most of the moderate $\tan \beta$ we undershoot
$m_h = 125$~GeV unless we enhance the radiative corrections, yielding
a lower bound on the $M_s$. 

\begin{figure}[t]\centering
\includegraphics[scale= 0.55]{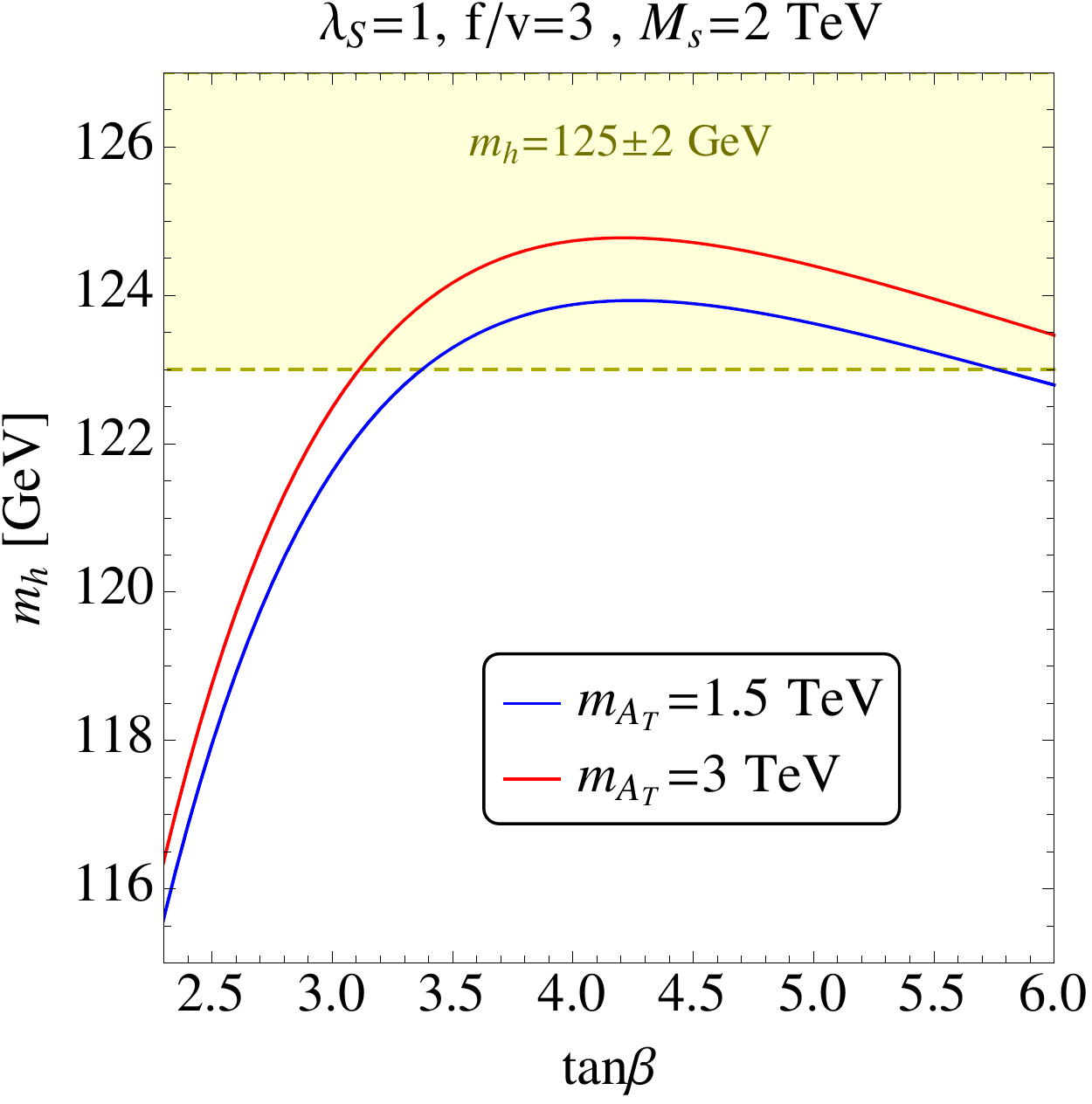}\hfill
\includegraphics[scale= 0.55]{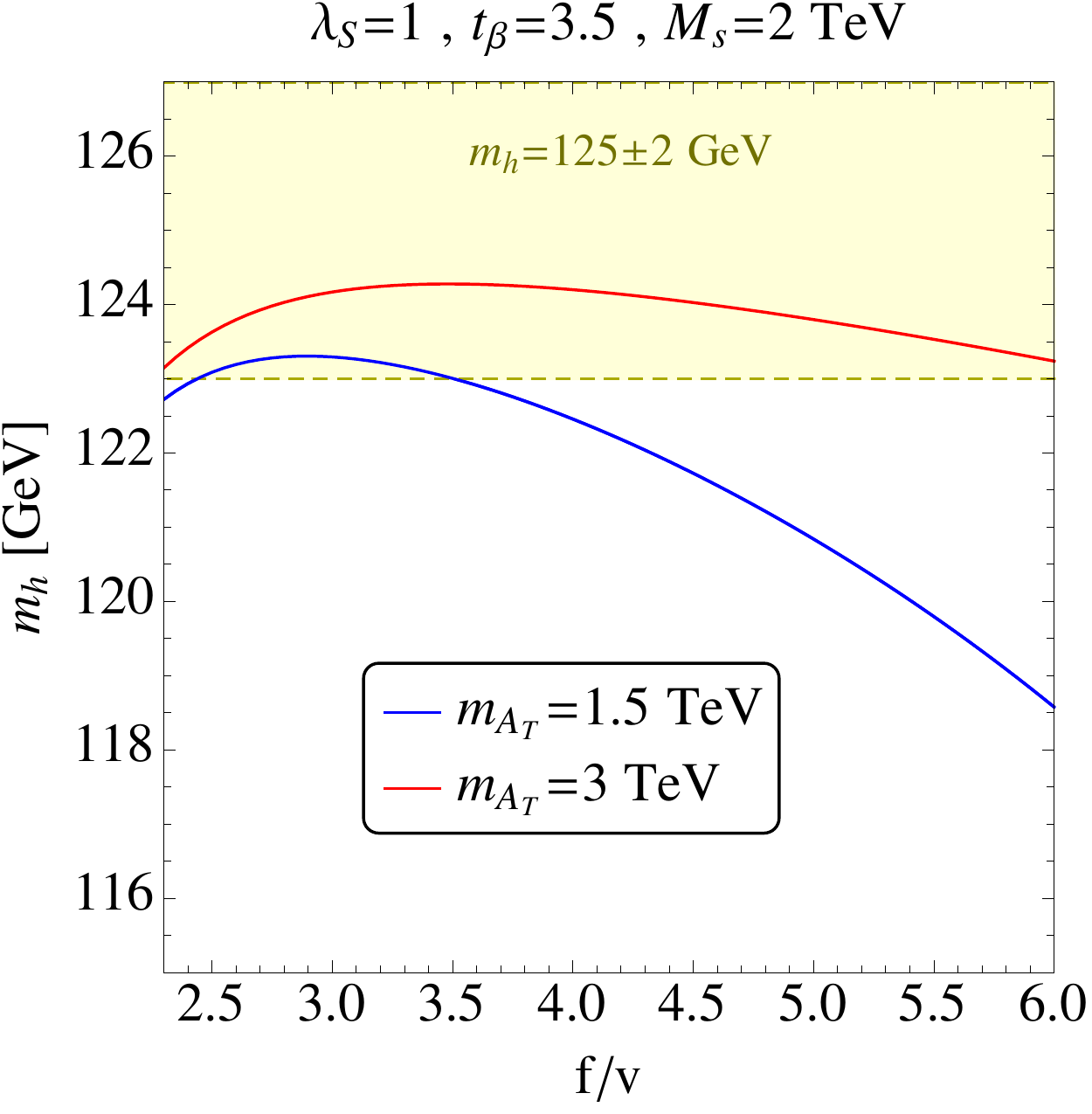}
\caption{The Higgs mass for $M_s=2\text{ TeV}$ and $\lambda_S=1$ as a
  function of $\tan \beta$(left) and $f/v$ (right). Blue and red line
  correspond to 
  $m_{A_T}=1.5,\, 3\text{ TeV}$ respectively. In the yellow band we
  get  $m_h=125\pm2\text{ GeV}$. 
\label{mh_function}}
\end{figure}
 
To understand the upper bound on $\tan\beta$ we have to 
go beyond the PGB approximation, which breaks down for large $\tan
\beta$, because the PGB approximation requires 
\beq
\frac{\kappa }{2 \lambda } \ll 1 \ \ \ \ \ \Longrightarrow  \ \ \ \ \
\left( \frac{g_{\text{ew}}}{2 \lambda_S} \right)^2 \cot^2 2 \beta \ll 1~.
\eeq 

Beyond the PGB approximation we should take into account all the
$\kappa/\lambda$ corrections, which is not easy to do
analytically (see however Appendix~\ref{app:2}). Nonetheless, we can 
arrive to the right conclusions analyzing the first non-vanishing 
correction to the PGB approximation:
\begin{equation}\label{eq:hmassLM}
m_h^2\approx 8
\kappa v^2
\left(1-\frac{v^2}{f^2}\right)\left(1-\frac{\kappa}{2\lambda}+\dots\right) 
\approx g_{\text{ew}}^2 v^2 \cos^22 \beta \left( 1 -
  \frac{g_{\text{ew}}^2}{4 \lambda_s^2} \cot^2 2 \beta \right)~.
\end{equation}
As $\tan \beta$ grows, the term in the brackets becomes important, and
drives the Higgs mass down, signalling that in the full Higgs mass we
have a ``sweet spot'' for the $\beta$ angle at which we can maximize
the Higgs mass, and therefore the minimal allowed value for $M_s$.

We show the value of the Higgs mass as a function of $\tan \beta $ and
$f/v$ using the full numerical
calculation at all orders in $\kappa/\lambda$ on the left panel 
Fig.~\ref{mh_function}. This confirms that the full function of the
Higgs mass has a sweet spot for $\tan \beta$, in qualitative agreement with
the simplified formula in Eq.~\eqref{eq:hmassLM}.

The upper bound on $f/v$ at fixed $M_s$ instead strongly depends on
the value of $m_{A_T}$ which controls the masses of the SUSY Higgses
(see Table \ref{tab:spectrum}  and Appendix~\ref{app:2} for 
details). In particular we see in 
Fig.~\ref{numeric_FT_soft} that the   upper bound on $f/v$
disappears for $m_{A_T}=3\text{ TeV}$. This feature is also
illustrated on the right panel of Fig.~\ref{mh_function} and can be understood as
a consequence of level splitting between the SM Higgs and the CP-even
SUSY Higgs whose mass squared is proportional to $\propto
(m_{A_T}^2-\lambda_S^2 f^2)$. For fixed $m_{A_T}$ increasing $f/v$ 
reduces the SUSY Higgs, thus enhancing the level splitting  and decreasing the SM Higgs mass.
This effect can always be compensated by taking a larger $m_{A_T}$
(see Fig.~\ref{mh_function}),
however this comes at the price of an $\mathcal{O}(1)$ increasing of the overall
fine-tuning for $M_s\lesssim 3\text{ TeV}$.   

Interestingly, an indirect bound on $f/v$ at fixed
$M_s$ implies an upper bound on the twin Higgs mass. 
Since the twin Higgs is the radial mode of the $SU(4)$-charged Higgs,
its mass is roughly $\sim\lambda_S f$ up to the $\kappa/\lambda$
corrections.  For example, for $M_s=2\text{
  TeV}$ and $m_{A_T}=1.5 \text{ TeV}$ the twin Higgs cannot exceed 
$500\text{ GeV}$ and can be probed at the LHC as we will discuss in more detail in
Sec.~\ref{Higgscoupl}. 

Up to now we completely ignored the role of $\delta t_\beta$ which we
set to zero in Fig.~\ref{numeric_FT_soft} for simplicity. However, in
the full SUSY theory there are two more relations which determine $\tan \beta$ and the misalignment between the $\beta$ angles in terms
of the fundamental parameters (at the leading order in $\delta t_\beta$):
\begin{align}
t_\beta^2  & \approx \frac{m_d^2}{m_u^2} \, , & \delta
t_\beta & \approx \frac{2 f^2}{m_A^2
  s_{2\beta}}\left[2\left(1-2\frac{v^2}{f^2}\right)
  \frac{g_{\text{ew}}^2}{8}\vert c_{2\beta}\vert
  +\sigma_d\right] \, , \label{extraEWSBsoft} 
\end{align}
where  $m_A^2=m_{A_T}^2-\lambda_S^2 f^2$ is the mass of the MSSM-like
CP-odd Higgs. 

\begin{figure}[t]\centering
\includegraphics[scale= 0.49]{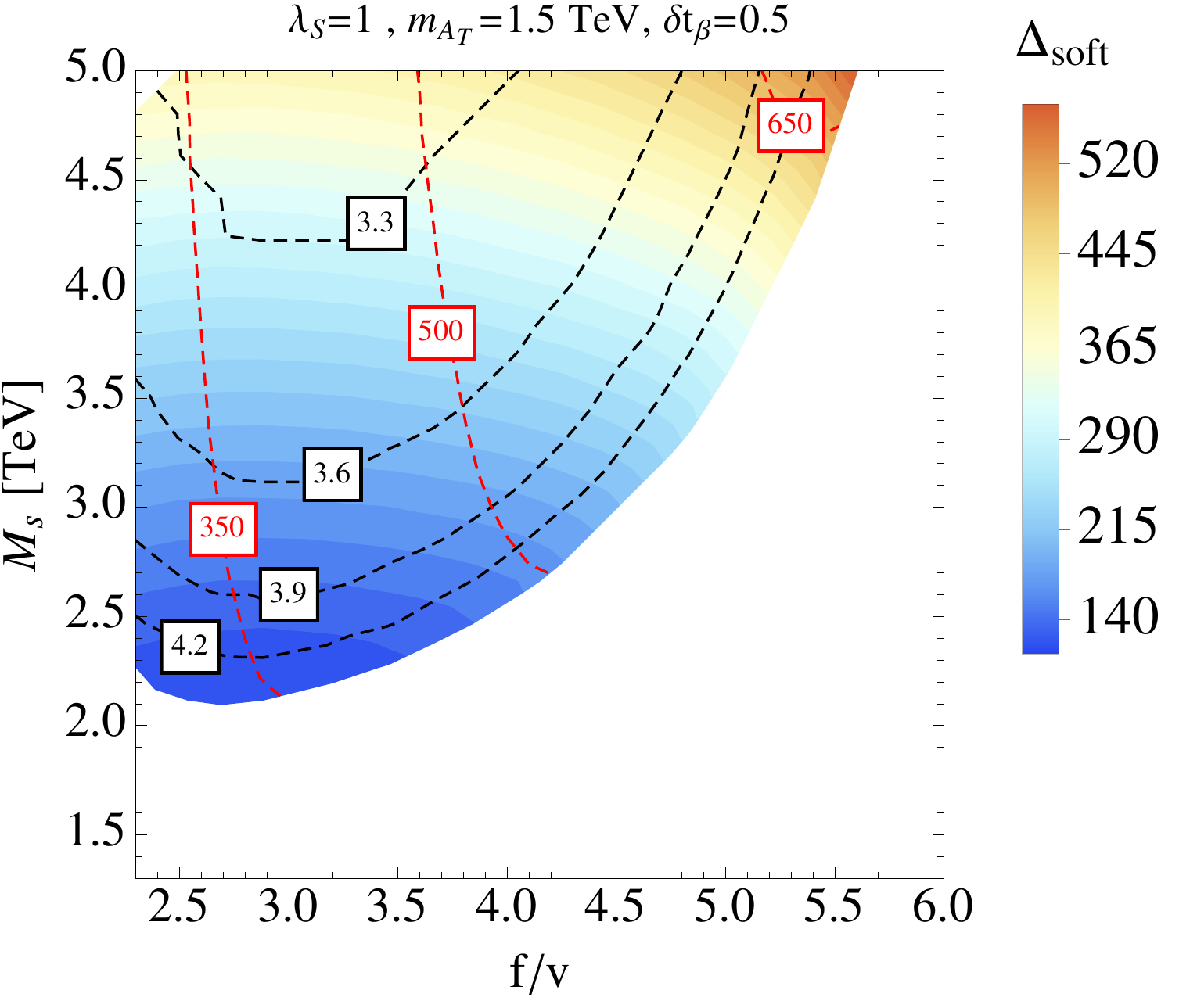}\hfill
\includegraphics[scale= 0.49]{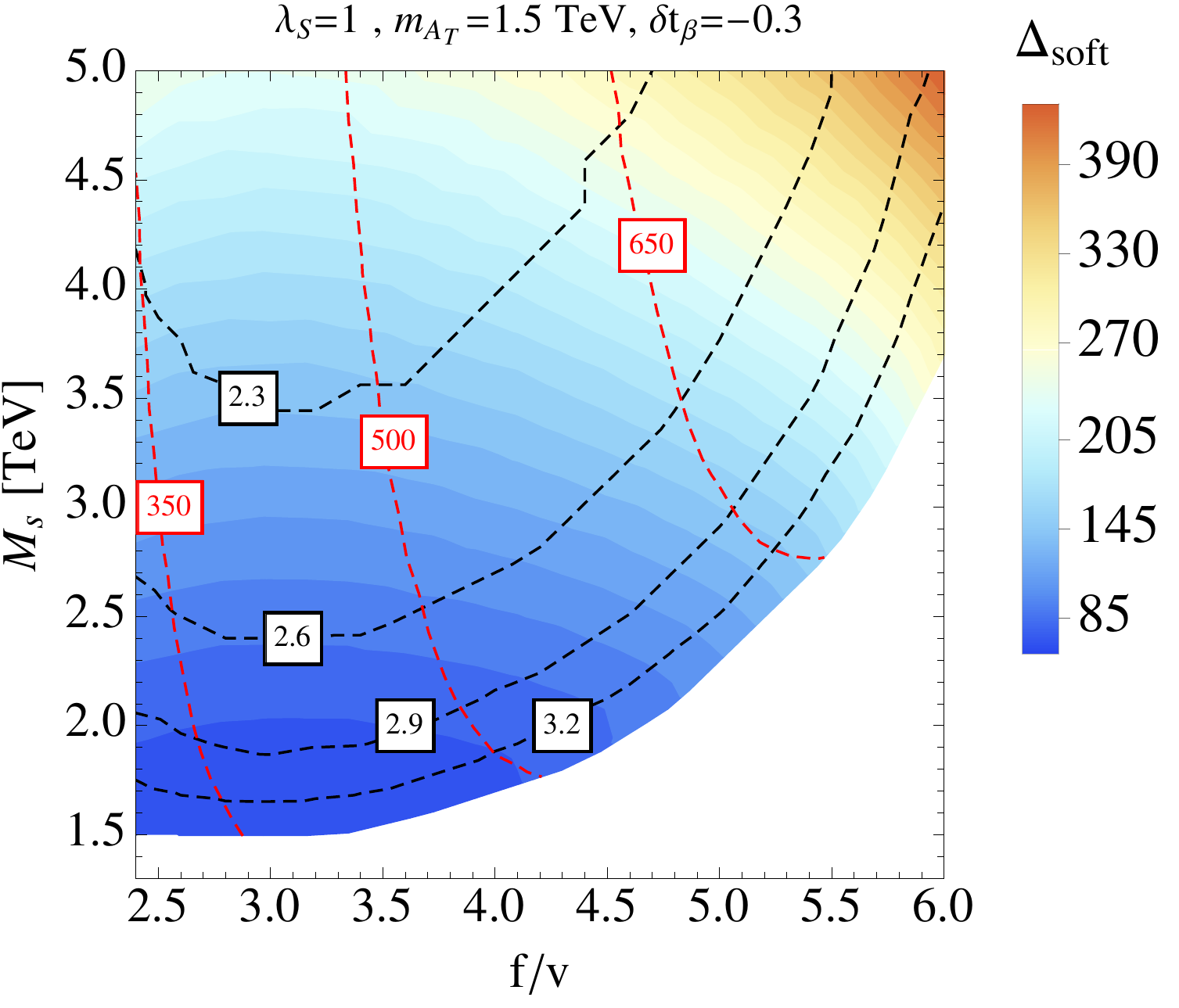}\hfill
\caption{Soft Twin SUSY parameter space with $m_{A_T}=1.5\text{ TeV}$
  and $\delta t_\beta=0.5$ (left panel) and $\delta t_\beta=-0.3$
  (right panel). In the white region $m_h=125\pm 2$ cannot be achieved
  as explained in the text. We also fix $\lambda_S=1$ and $m_S=1\text{
    TeV}$, $\mu=500\text{ GeV}$, $M_3=M_s$. Black/red dashed lines are
  contours of $\tan\beta$/next-to-lightest CP-even Higgs mass in GeV. The gradient
  color function indicates region of increasing
  $\Delta_{\text{soft}}$, which is defined in Eq.~\eqref{numerical_def}.  
\label{numeric_FT_soft_2}}
\end{figure}

From these expressions one can estimate what would be the natural
value for the $\beta $ angle misalignment between the different
sectors. If we require no unnatural cancellations and the ``sweet
spot'' values for $\tan \beta $ (namely of order 2...3), we
expect that $\vert\delta t_\beta\vert \lesssim 1$, but usually not smaller than
0.1.  Because these natural values are not necessarily small, we show in
Fig.~\ref{numeric_FT_soft_2} how the non-trivial $\beta$-angle
misalignment changes the allowed parameter space and the fine-tuning.  
Negative values of $\delta t_\beta$ are allowed, because $\sigma_d$
can be either positive or negative,
allowing to extend the parameter space at larger $f/v$. A positive
$\delta t_\beta$ 
makes instead the parameter space shrink. The qualitative features of the
plots are however left unchanged with respect to the $\delta
t_\beta=0$ case. 

We conclude that this particular soft Twin SUSY model has a
number of features which crucially depend on the fact 
that $\lambda$ is generated via the non-decoupling F-term of a
NMSSM-like singlet and hence suppressed even for moderate $\tan\beta$. In  
particular the $\tan\beta$-dependence of $\lambda$ renders the behavior
of the tuning measure sub-optimal and certainly motivates
further exploration in the theory space of soft $\ztwo$-breaking
models. For example, one can consider models along the lines
  of~\cite{Batra:2003nj} where an extra $U(1)_x$ is gauged and both
  $A$ and $B$ Higgses carry the same $U(1)_x$ charge. If the $U(1)_x$
  is spontaneously broken by fields with large  
soft masses, one would obtain a non-decoupling $SU(4)$-symmetric
quartic from the D-terms of the $U(1)_x$, getting a very different
dependence of the effective $\lambda$ on the $\beta$-angle. We leave
an exploration of soft Twin SUSY models for future work.

Let us finally mention some caveats of our analysis. First, the allowed
parameter space presented in Fig.~\ref{numeric_FT_soft} depends
crucially on our requirement $m_h=125\pm 2$~GeV. This range is
probably 
too conservative since our one-loop computation is subject to
large theoretical uncertainties dominated by two-loop QCD
effects. It is known that these effects can sometimes shift the one-loop result
by  
more than 2~GeV in SUSY theories where they have been already
computed like in the MSSM. In particular a large positive shift from
two loop effect might quantitatively change our conclusions on the
allowed parameter space of the model. Nevertheless our analysis illustrates
some important features of tree level effects to the Higgs mass that
will certainly be present also after the inclusion of two loops
corrections.  

Second, an interesting option to enlarge the parameter space of this
particular soft Twin SUSY model and possibly ameliorate its fine
tuning would be to introduce a sizeable negative $\delta
t_\beta$. This value does not come out naturally, but might be
justified in a UV complete model where the origin of both $\ztwo$-even
and $\ztwo$-odd soft masses is specified. We leave these questions open for
future work.

\subsection{Hard Twin SUSY}\label{sec:hard}

In this section we discuss models of Twin SUSY where a
$\ztwo$-breaking quartic is generated at the tree-level. The easiest
implementation of this class of models borrows the $SU(4)$-invariant
sector    from the soft model described above and just
extends the singlet sector adding a new singlet $S_A$ which couples to
the $A$-sector Higgses \emph{only} 
\begin{align}
W_{\slashed{Z}_2}^{\rm hard} & = \lambda_A S_A H_u^A H_d^A +
\frac{{\cal M}_{S_A}}{2 } S_A^2 \, , & V^{\rm
  hard}_{\slashed{Z}_2} & = m_{S_A}^2 |S_A|^2 \,
.\label{Z2SA} 
\end{align}
Integrating out the singlet in the limit $m_{S_A} \gg {\cal
  M}_{S_A}\, , \mu$ we generate a $\ztwo$-breaking quartic at
treel level while the $\ztwo$-breaking soft masses in Eq.~\eqref{Z2super} are
generated at 1-loop: 
\begin{align}
V_{\slashed{Z}_2}^{\rm hard} & \approx \lambda_A^2 |h_u^A h_d^A
|^2-\frac{3\lambda_A^2 m_{S_A}^2}{32\pi^2}(|h_u^A|^2+|h_d^A|^2)~,
\label{Z2superhard} 
\end{align}
with the same approximations that we have used before and neglecting the log-pieces in the CW potential.  Matching Eq.~\eqref{Z2superhard} to the $\ztwo$-odd parameters of the Twin
Higgs potential in Eq.~\eqref{Vtwin} we get  
\begin{align}
&\rho\approx \frac{\lambda_A^2}{4} s_{2 \beta}^2\,  ,
&\sigma \approx -\frac{3\lambda_A^2}{32 \pi^2} \frac{m_{S_A}^2}
{f^2} \equiv -\lambda_A^2\sigma_{S_A}  
\ . 
\label{Z2oddhard} 
\end{align}
After comparing these matching conditions with the
results of Sec.~\ref{sec:2} we see that the required values
of $\ztwo$-breaking quartic $\rho\gtrsim0.1$ to get $f/v\gtrsim 2.3$
are easy to achieve with a moderate coupling
$\lambda_{A}\gtrsim0.6$ as long as $\tan\beta\lesssim2$. 

The biggest problem of this model is that the one-loop threshold
corrections to $\sigma$ are negative and their size is controlled by the singlet mass.
A negative $\sigma$ is subject to a strong upper
bound coming from the stability of the EWSB vacuum ({\it c.f.}
Fig.~\ref{PGB1} and the discussion in Sec.~\ref{sec:2}). 
In a SUSY UV completion 
we get a similar bound  
by solving the EWSB condition for $f/v$ in terms of
$\lambda_A$ 
\begin{equation}
\lambda_A^2\approx-\frac{1- \frac{2 v^2}{f^2}}{\frac{\sigma_{S_A}}{2}-
  \frac{v^2}{4f^2}s_{2\beta}^2}\kappa\ . 
\label{stability} 
\end{equation}
Since $\lambda^2_A>0$ we need
$\frac{\sigma_{S_A}}{2}<\frac{v^2}{4f^2}s_{2\beta}^2$ as long as
$\kappa>0$. This poses an \emph{upper} bound on the mass of the
singlet $m_{S_A}$ around 1 TeV.  This bound was expected from Sec.~\ref{sec:2}
where for negative $\sigma$ $(\epsilon= -1)$ we got a strong constraint on $\Lambda_\rho$ 
which is here identified with the singlet mass up to numerical factors. 

\begin{figure}[t]\centering
\includegraphics[scale =.6]{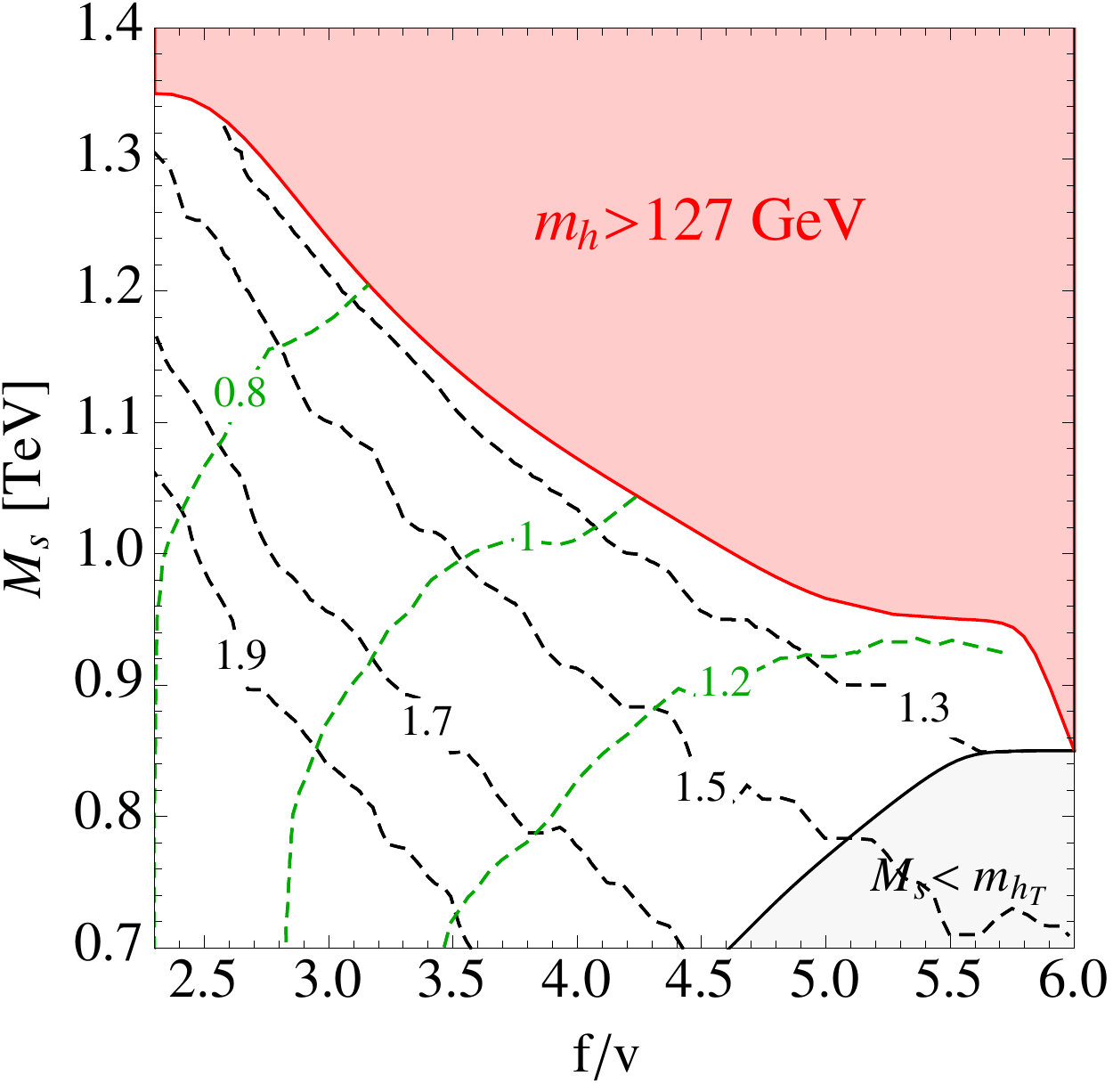}\hfill
\caption{Allowed parameter space of the simplest hard model, described
  in Eq.~\eqref{Z2SA}. We assume $\delta t_\beta=0.1$ and
  $\lambda_S=1$, $\mu=500\text{ GeV}$, $M_3=M_s$. A strong upper bound
  on $M_s$ 
  comes from the Higgs mass constraint $m_h<127$~GeV and the red 
  shaded area is inaccessible. The black dashed contours stand for the
  maximal allowed value of $\tan \beta$  and the green dashed contours
  for the mass of the heaviest CP-odd Higgs in TeV.   
\label{numeric_FT_hard}} 
\end{figure}

We illustrate the parameter space of the model in
Fig.~\ref{numeric_FT_hard}. The figure also shows that the upper
bound on the singlet mass we derived in Eq.~\eqref{stability} gives indirectly a bound on the
maximal allowed $\tan \beta$, which is even stronger constrained than in
the soft Twin SUSY.  An important difference between the hard $\ztwo$-breaking scenario
compared and the soft  one is that the masses of the
MSSM Higgses controlled by $m_{A_T}=2b/s_{2\beta}$ cannot be
arbitrarily decoupled if $\delta t_\beta$ is fixed. The $\beta$-angle
misalignment in this scenario is 
\begin{equation}
\delta t_\beta\approx\frac{2}{c_\beta^2 s_{2\beta}}\frac{f^2
}{m_A^2}\left[\left(1-\frac{2v^2}{f^2}\right)\left(\frac{g_{\text{ew}}^2}{8}\vert
    c_{2\beta}\vert+\delta \lambda_u
    s_\beta^4\right)+\frac{\lambda_A^2\sigma_{S_A}\vert
      c_{2\beta}\vert}{2}\right]\ , 
\label{dtb2}
\end{equation}
where  $\delta\lambda_u=\frac{3m_t^4 }{16\pi^2 s_\beta^4 v^4}\log\left(
  \frac{M_s^2}{m_t^2} \frac{v^2}{f^2} \right)$ encodes the top-stop
contribution to $\kappa$. The 
equation for $\tan\beta$ however stays intact  up to 
sub-leading corrections.  The expression~\eqref{dtb2} shows that the mass
of the CP-odd Higgs is linear in $f$ and drops at lower $\tan \beta$,
exactly what we see in Fig.~\ref{numeric_FT_hard}. Note also that
the range of natural 
values of $\delta t_\beta$ is of order $\cO(10^{-2} \ldots 10^{-1})$
and it can only be positive. Completely
neglecting the $\beta$-misalignment is often not a bad approximation in this
scenario.

Although light SUSY Higgses can help with the Higgs mass
constraint 
because level splitting among CP-even Higgses can decrease the mass of
the lightest eigenstate (i.e. the SM-Higgs), the masses of the SUSY
Higgses  are subject
to bounds both from direct and indirect searches, {\it e.g.} $b \to s
\gamma$ and the SM Higgs couplings fit. 
 Even though
these bounds are somewhat  model dependent, we conservatively
impose $m_{A}\gtrsim400\text{ GeV}$ in our numerical scan, which
correspond to $m_{H^{\pm}}\gtrsim350\text{ GeV}$ using the formulas in
Appendix \ref{app:2}. We will further discuss these bounds in detail in
Sec.~\ref{Higgscoupl}. 

At the end of the day the simplest model of
Twin SUSY has a fine tuning of around $\sim10\%$ but is so severely constrained by
$m_h=125\text{ GeV}$ that the SUSY colored states are always forced to be within the reach of the LHC. This traces back to the fact that with a negative $\sigma $ and an irreducible tree level contribution to $\kappa$ coming from the EW D-terms, we always overshoot the Higgs mass. The only
way out of this is to take $\tan\beta\sim 1$ and render the log in
Eq.~\eqref{softkappa} extremely small. The upper bound on $M_s$ was again expected from the discussion in Sec.~\ref{sec:2} when for $\epsilon=-1$ we got a strong upper bound on $\Lambda_t$. Interestingly, the Higgs mass constraint is acting on
the parameter space of hard Twin SUSY models in the opposite direction 
with respect to the standard MSSM-like SUSY constructions where it
often gives a lower bound on $M_s$. 

 A crucial question remains then
to be addressed:    
\emph{Is it possible to find a model that decouples the colored
  states?} Here we present an existence proof for such a model. 
Notice that in the IR effective theory for a given $\Lambda_\rho$
threshold one can access higher values
of $\Lambda_t$ by introducing a \emph{negative} contribution to $\kappa$ at
the tree level, compensating for the unavoidable positive contribution
from the D-terms.  
Of course this is not
the only way one can explore, but we focus on this option for 
concreteness and comment later on possible alternative solutions.

In order to get a negative contribution to
$\kappa$ we  introduce a pair of bi-doublets $B+\bar{B}$
under $SU(2)_A \times SU(2)_B$ with hypercharges
$\pm 1/2$ and the 
following potential 
\begin{align}
\Delta W_{Z_2} & = \lambda_{BD}  B h_u^A h_u^B + 
M_B B \bar{B}\, , & V_{Z_2} & =  m_{B}^2 \left( |B|^2 + |\bar{B}|^2
\right) \, . \label{bidoublets} 
\end{align}
Integrating out the bidoublet in the limit of $m_{B} \gg M_B\, ,\mu $
and matching it to the Twin Higgs potential we obtain: 
\begin{align}
\Delta V_{Z_2} & =  s_\beta^4|\lambda_{BD}|^2  |H_A|^2  |H_B|^2~.
\end{align}
In the Twin Higgs parametrization in Eq.~\eqref{Vtwin}
this gives us a negative contribution to $\kappa$ and a
positive one to $\lambda$: 
\begin{align}
\Delta \kappa & = - \Delta \lambda = - \frac{1}{2} s_\beta^4 | \lambda_{BD}|^2  \, .
\end{align}
Therefore we expect that for a given point in the $(f/v\, ,M_s)$ plane
where the $\ztwo$-breaking quartic $\lambda_A$ is fixed to get the
desired value of $f/v$, one can always choose an appropriate $
\lambda_{B}$ in order to satisfy the Higgs mass constraint, hence
rescuing the hard-breaking $\ztwo$-breaking model. An
analytical understanding of the structure of the  EWSB conditions
can be easily derived from our previous discussion
performing the shift:
$\delta\lambda_u\to\delta\lambda_u-\frac{\lambda_{BD}^2}{2}$.  

Of course this solution for correcting the Higgs mass comes
at a price. The main problem is the presence of states charged
under both $A$ and $B$ gauge group that  generically generate a
mixing between the photon and the dark photon at one-loop, which is
  strongly constrained by experiments
  \cite{Davidson:2000hf,Hook:2010tw}. However, 
this problem can be solved either by ungauging the mirror hypercharge or
by breaking it, giving a mass to the mirror photon. These effects would generate extra
$\ztwo$ breaking contributions in the Higgs potential which however do not change
the structure of our model drastically. In particular the ungauging of the twin hypercharge 
leads to a $\ztwo$ breaking quartic via the D-terms, which however is too small to dominate the mirror symmetry breaking in a realistic scenario.

\begin{figure}[t]\centering
\includegraphics[scale= 0.6]{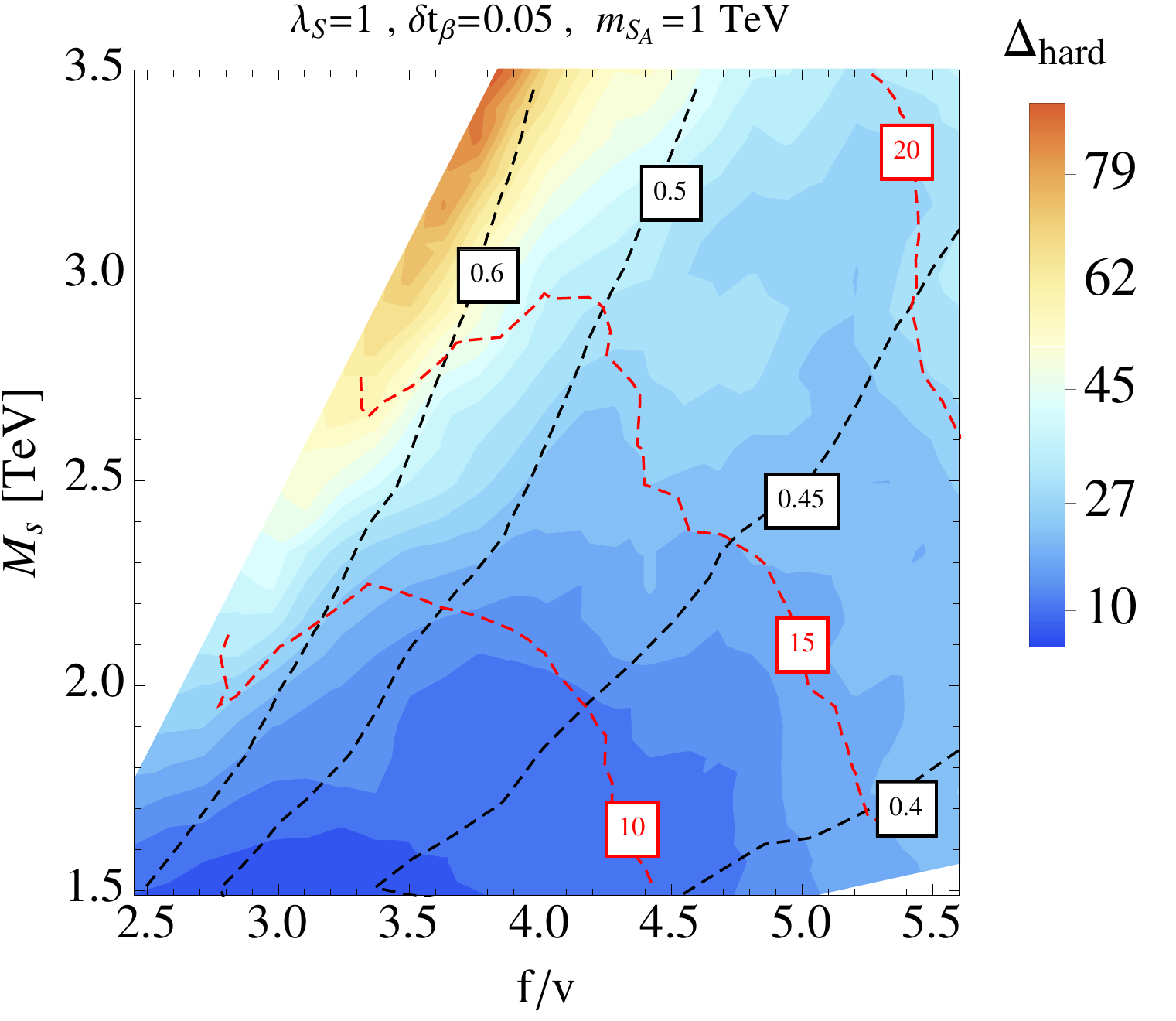}\hfill
\caption{Allowed parameter space of hard Twin SUSY with 
  the heavy bi-doublets. Dashed black contours stand 
  for the minimal allowed couplings of 
  $\lambda_{BD}$.  The gradient color function indicates region of
  increasing fine tuning. The red dashed contours indicate the
  fine tuning of the EW scale with respect to $\lambda_{BD}$. We set
  $\delta t_\beta=0.05$, $m_{S_A}=1 \text{ TeV}$ and $m_B=1
  \text{ TeV}$. In the white region a stable vacuum cannot be
  achieved. We also fix $\lambda_S=1$ and $m_S=1\text{ TeV}$,
  $\mu=500\text{ GeV}$ and $M_3=M_s$. 
\label{numeric_FT_BD_1}} 
\end{figure}

We show the parameter space of the bidoublet model in 
Fig.~\ref{numeric_FT_BD_1}. The bidoublet coupling $\lambda_{BD}$ ranges between $0.4 \ldots 0.6$
depending on the region of the parameter space. This corresponds to the   
negative shift in $\kappa$ of order $0.05 \ldots 0.1$. Among other parameters we scan
here on $\tan\beta$, which however stays around $\tan\beta\lesssim2$ in
almost the entire parameter space. 
We calculate the fine tuning as in the previous
section, varying the EW breaking scale $v$ with respect to all the
parameters of the hard model.\footnote{In this case the hard sector
  parameters include the parameters of the Higgs sector
  $\mathcal{P}_H=\{ m_{H_u}^2\, , m_{H_d}^2\, , 
\mu^2\, , b \}$, of the colored sector  
$\mathcal{P}_Q=\{m_Q^2\, , m_U^2\, , M_3\}$, of the $SU(4)$-preserving 
singlet sector $\mathcal{P}_S=\{m_S^2\, ,\lambda_S\}$ of the
$\ztwo$-breaking singlet sector 
$\mathcal{P}_{\slashed{Z}_2}=\{\lambda_A\, ,m_{S_A}^2\}$ and of 
the bi-doublet sector $\mathcal{P}_{B}=\{\lambda_B\, ,m_{B}^2\}$.  As 
in the soft Twin SUSY case all these parameters are taken to be  at
the ``messenger scale''  $\LSUSY$ for the purposes of the fine-tuning
computation and   
we use full one-loop RGE (see Appendix~\ref{app:RGE}) to obtain the
final fine tuning.  }

The behavior of the fine tuning isolines in Fig.~\ref{numeric_FT_BD_1}
can be intuitively understood by means of the following approximate
formula   
\begin{equation}
\Delta^{\text{hard}}\approx 
\sqrt{\left( \frac{\partial\log v^2}{\partial\log \lambda_A}
  \right)^2+\left( \frac{\partial\log v^2}{\partial\log \lambda_{BD}}
  \right)^2}\times  
\frac{\partial\log f^2}{\partial\log m_{H_u}^2}\ .\label{bidoubletFT}
\end{equation}
The idea behind this approximation is that the fine-tuning of $f$ with
respect to the SUSY scale is dominated by the running of $m_{H_u}$
as in soft Twin SUSY. The fine-tuning of the EW scale with respect
to $f$ is instead accounted by the log-derivative with respect to the
hard $\ztwo$-breaking quartic and the $\ztwo$-preserving quartic
$\lambda_{BD}$ which is tuned to satisfy the Higgs mass constraint at
fixed $f/v$ and $M_s$. This second contribution to the fine-tuning is
shown on 
Fig.~\ref{numeric_FT_BD_1} with dashed red contours and it increases
at 
larger $M_s$ as expected. The approximate formula in
Eq.~\eqref{bidoubletFT} is the SUSY analogue of the FT measure we
discussed from the low energy perspective in the right panel
of Fig.~\ref{PGB5}. For $M_s=2\text{ TeV}$ the theory is tuned at $\sim
3-10\%$, depending on $f/v$ which is roughly a factor of $\sim 3-10$ better than the soft Twin SUSY
model. Going to larger
$M_s$, the fine tuning needed to satisfy $m_h=125\text{ GeV}$ becomes
larger and introducing hard $\ztwo$-breaking is not beneficial
anymore. That is why we cut the parameter space of
Fig.~\ref{numeric_FT_BD_1} at $M_s\sim 3.5 \text{ TeV}$.

Of course the approximate formula in Eq.~\eqref{bidoubletFT}  relies on a
number of assumptions which are not always satisfied in the full
parameter space of the model. The major caveat is for sure the
factorization of the fine tuning measure which is broken in the
presence of extra quartics by the one-loop corrections to the soft
masses proportional to both $\lambda_A$ and $\lambda_{BD}$ (see for
example the RGE of $m_{H_u}$ in Appendix~\ref{app:RGE}). These
corrections are however always sub-leading compared to the top-stop
contribution to the RGE of $m_{H_u}$ as long as both the couplings are
smaller or equal to $y_t$ and both the singlet and the bi-doublet soft
masses are not too large. For this reason we set
$m_{S_A}=m_{B}=1\text{ TeV}$ in our numerical results.  

The white region at low $f/v$ region  can be understood by further
analyzing the expressions in Eqs.~\eqref{stability} and~\eqref{dtb2}. First
of all we see from Eq.~\eqref{dtb2} that $m_A$ decreases with $f/v$
and hence we expect the lower bound $m_A>400\text{ GeV}$ to give a
\emph{lower} bound on $f/v$.  
It is then clear
that small values of $\delta t_\beta$ are favored because they can
easily raise the mass of $m_A$ as one can see again from
Eq.~\eqref{dtb2}. 
In this regime we can approximate $\delta t_\beta\approx0$ and
solve~\eqref{dtb2}  for $\sigma_{S_A}$. Plugging back this solution
into~\eqref{stability} we get  
\begin{equation}
\lambda_A^2=-\frac{2(f^2-2v^2)}{v^2}\left(\frac{(\delta
    \lambda_u-\frac{\lambda_{BD}^2}{2}) s_\beta^2}{\vert
    c_{2\beta}\vert}+\frac{g_{\text{ew}}^2}{4} \right)~, 
\end{equation} 
from which we see that the lower bound on $f/v$ in order to get
$\lambda_A>0$ becomes stronger at larger $M_s$ (which correspond to a larger $\delta\lambda_u$), explaining the
diagonal shape of the boundary of the parameter space on
Fig.~\ref{numeric_FT_BD_1}.

This lower bound on $f/v$ is of course not a feature of hard Twin SUSY
by itself but of our specific model. It is interesting however that it 
provides a lower bound on the Twin Higgs mass, which was 
instead bounded from above in the soft Twin SUSY model of
Sec.~\ref{subsec:softTwin}. Conversely the mass scale $m_A$ of the
SUSY Higgses, which was a free parameter in the soft Twin SUSY
model, is now fixed by Eq.~\eqref{dtb2} and possibly becomes a
promising signature at the LHC as we will discuss in
Sec.~\ref{Higgscoupl}.   

In conclusion, besides being sub-optimal for the reasons explained
above, we believe that the SUSY  model described in this section
provides a simple existence proof of the hard $\ztwo$-breaking
mechanism that can ameliorate the fine-tuning in perturbative UV
completions of the Twin Higgs. 

Of course many other constructions in the model space of hard Twin
SUSY are still left unexplored. For example a possible alternative
solution to the overshooting of the Higgs mass would  
be to flip the sign of the one-loop $\ztwo$-breaking 
threshold to $\sigma$ via a non  minimal singlet sector. In such a
model the constraint $m_h=125\text{ 
  GeV}$ can be achieved by tuning the positive $\sigma$. These
solutions would be the UV completion of the effective analysis in the
left panel of Fig.~\ref{PGB5} and is left for future studies.

\section{Twin SUSY Higgs phenomenology}\label{Higgscoupl}
In this section we discuss the phenomenological signatures of the SUSY
Twin Higgs.  
In particular we focus  on the Higgs sector of the theory, which leads
to distinctive and model independent signatures at the LHC. 
The Twin Higgs mechanism implies the existence of an extra scalar in
the spectrum, the ``twin'' Higgs, which is the radial mode of the
spontaneously broken $SU(4)$-symmetry. The phenomenology of the twin
Higgs at LHC essentially resembles the one of a scalar singlet which
mixes with the SM Higgs and has been already studied 
e.g. in Refs.~\cite{Buttazzo:2015bka,Bertolini:2012gu, Robens:2015gla, 
  Falkowski:2015iwa, Gorbahn:2015gxa}. Here we recast present and
future bounds from LHC and Higgs coupling measurements on the twin
Higgs focusing on the particular parameter space of Twin SUSY models
(see Appendix \ref{app:2} for detailed formulas about the spectrum,
the mixing angles and the decay widths). In addition we show that extra
MSSM-like Higgses, which are generically present in SUSY UV
completions, provide complementary probes on the parameter space of
Twin SUSY both at the LHC and in indirect measurements.  

We are not going to explore possible extra signatures coming from the
twin matter sector. These depend very much on its structure that we
leave unspecified in this paper (see
Refs.~\cite{Craig:2015pha,Burdman:2014zta,Chacko:2015fbc,Curtin:2015fna}
for some studies about LHC signatures in  Twin Higgs
scenarios). Moreover, we assume pair production of colored SUSY states
to be out of reach of LHC with $300\text{ fb}^{-1}$, which roughly
corresponds to $M_s\gtrsim 2\text{ TeV}$. In the previous section
we have shown that this regime is automatically achieved in the
simplest soft Twin SUSY model once $m_h=125\text{ GeV}$ is imposed,
because the Higgs mass constraint bounds $M_s$ from below. For hard
Twin SUSY models instead, the 
situation is reversed and the region with $M_s\gtrsim 2\text{ TeV}$ is
accessible at a price of some model building gymnastics. 

For $M_s\lesssim 2\text{ TeV}$ the LHC bounds on extra Higgses should be in principle compared with the canonical SUSY
searches for pair produced superparticles. The same comparison should
be performed for prospects at HL-LHC with $3000\text{ fb}^{-1}$. We
leave such a detailed comparison for future works and focus on the
physics of the extra Higgses from now on.

\subsection{The Higgs sector of Twin SUSY}
The Higgs sector of the SUSY Twin Higgs exhibits a very rich structure
since it contains \emph{at least} a double copy of the MSSM. 
In Sec.~\ref{sec:3} we have shown that
scalar states beyond the visible and hidden Higgs doublets are needed 
in concrete realizations of Twin SUSY in order to get the required
structure of the quartics in the low energy Twin Higgs potential of
Eq.~(\ref{Vtwin}). Since these states are generically required to have
a large soft mass of order $\sim M_s$, we assume that they are
sufficiently heavy to be decoupled from the LHC phenomenology.  

We summarize the minimal Higgs sector of
Twin SUSY in Table~\ref{tab:spectrum}.
 The PGB formula of the SM-like Higgs mass in Twin SUSY has
 already been discussed in Sec.~\ref{sec:2}, and the expressions beyond
PGB can be found in Appendix \ref{app:2}. The behavior of the heavy
Higgs spectrum is instead described with very good accuracy by the
tree-level formulas in Table \ref{tab:spectrum}. Here we have assumed 
that $\lambda^2_S \gg g_{\text{ew}}^2$, $f^2 \gg v^2$,
$m_{A_T}^2 \gg \lambda_S^2 f^2$,  
neglected subleading corrections proportional to $\lambda_A$ and set $\delta t_\beta= 0$.

\begin{table}[t]
\begin{center}
\begin{tabular}{||c|c|c|c|c||}
\hline
&\multicolumn{4}{|c||}{CP-even Higgses} \\
\hline
 States: & $h$ &  $h_T$ &  $H$ &  $H_T$  \\
 \hline
Masses: &$m_h^2 $ & $  \lambda_S^2 s_{2 \beta}^2  f^2$ & $m_{A_T}^2 -
\lambda_S^2 f^2 $& $m_{A_T}^2- \lambda_S^2 f^2 s_{2 \beta}^2$ \\ 
\hline
\hline
& \multicolumn{2}{c|}{CP-odd Higgses} & \multicolumn{2}{c||}{Charged Higgses}  \\
\hline
 States: &   $A_T$ &  $A$ &  $H^{\pm}$  &  $H_T^{\pm}$  \\
 \hline
Masses: &$ m_{A_T}^2$ &$m_{A_T}^2 - \lambda_S^2 f^2 $&  $m_{A_T}^2 -
\lambda_S^2 f^2 $& $m_{A_T}^2 - \lambda_S^2 f^2 $ \\ 
\hline
\end{tabular}
\end{center}
\caption{\label{tab:spectrum}
Higgs spectrum of the Twin SUSY model.}
\end{table}

From these expressions we first recognize the twin Higgs $h_T$ with a
mass squared given by $m_{h_T}^2=4\lambda f^2$, where the
SU(4)-invariant quartic $\lambda$ is given by the SUSY matching
condition in Eq.~\eqref{eq:softmatching}. We further notice that among
the extra Higgs states there is a fully degenerate $SU(2)$-doublet
($H,A, H^\pm$) with mass set by the CP-odd Higgs mass $m_A^2 \equiv
m_{A_T}^2-\lambda_S^2 f^2$, which becomes light at large $f$.  
Because this mass should be non-tachyonic,
$m_{A_T}^2=2b/s_{2\beta}$ should always be the largest mass
scale of the Higgs sector. The other SUSY $SU(2)$-doublet ($H_T, A_T,
H_T^\pm$) has instead larger splittings: the heaviest state is the
CP-odd Higgs $A_T$ with mass $m_{A_T}^2$ followed by the CP-even $H_T$
and the charged Higgses $H_T^\pm$.  
The charged Higgses $H^{\pm}$ and $H_T^{\pm}$ are mass-degenerate in
this approximation and do not mix because of gauge invariance. As a
consequence the charged Higgses $H^{\pm}$ have exactly the same
couplings as in the MSSM, while $H^{\pm}_T$ are completely dark and
hence uninteresting for collider purposes.    

The mixing of the CP-odd and the CP-even SUSY Higgses is controlled by
$v/f$ exactly like the one of the SM Higgs $h$ and the twin Higgs
$h_T$: $A_T$ and $H_T$ are purely dark in the limit $v/f\to 0$ while
$A$ and $H$ become MSSM-like.\footnote{Any mass term that mixes Higgs
  states in $A$ and $B$ sector must be proportional both to the
  breaking of $SU(2)_A$ and $SU(2)_B$ and therefore vanishes in the
  limit $v \to 0$.}  The full expressions for both mixing angles and
mass eigenvalues can be found in Appendix~\ref{app:2}.  

\begin{figure}[t]
  \centering
 \includegraphics[width=0.49\textwidth]{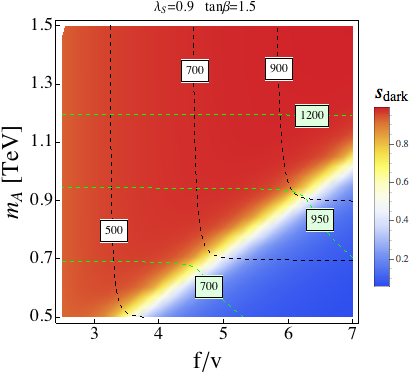}
  \includegraphics[width=0.49\textwidth]{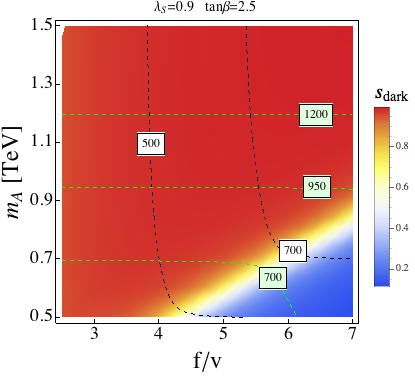}\\
  \vspace{.25cm}
   \includegraphics[width=0.49\textwidth]{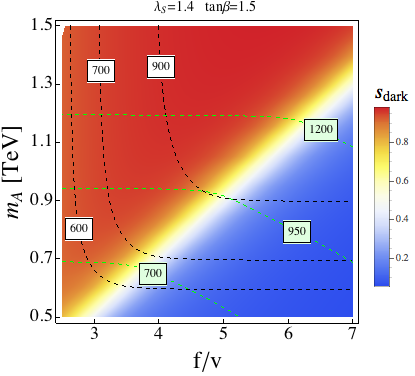}
    \includegraphics[width=0.49\textwidth]{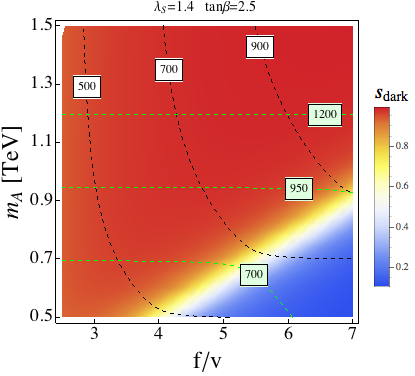}
 \caption{Mass contours in the $m_A$ vs $f/v$ plane. The four plots
   correspond to different values of $(\lambda_S,\tan \beta)$. The black and green
 contours are the second and third lightest CP-even mass state
 respectively. The temperature map denote the B-sector composition of
 the second lightest CP-even mass eigenstate as explained in the text,
 where $s_{\text{dark}}\equiv\sqrt{V_{23}^2+V_{24}^2}$.} 
  \label{fig:masses}
\end{figure}

In most of the allowed parameter space the twin Higgs $h_T$ is lighter
than the visible MSSM-like $SU(2)$-doublet and hence it is the
next-to-lightest CP-even scalar after the SM-like Higgs. However, when
$\lambda_S^2 f^2 \gtrsim m_{A}^2$ (corresponding to $m_{A_T}^2 \gtrsim
\lambda_S^2 f^2)$, the hierarchy can be inverted and a full MSSM-like
$SU(2)$-doublet forms the first level of extra scalars above the
SM-like Higgs. The transition between these two regimes is shown in
Fig.~\ref{fig:masses}, where we plot the composition of the
next-to-lightest  CP-even eigenstate $H_2$ and its mass 
for the benchmark values of $\lambda_S$ and $\tan
\beta$. We define the composition of mass ordered  CP-even states
$H_I$ in terms of the gauge eigenstates of the $A$ and $B$ sector as: 
\begin{align}
H_I & = \sum_i V_{Ii} h_i\, ,  & h_i & = \{h_u^A,h_d^A,h_u^B,h_d^B\} , 
\end{align}
In the temperature plot in Fig.~\ref{fig:masses} we show the
combination of the mixing matrices
$s_{\text{dark}}\equiv\sqrt{V_{23}^2+V_{24}^2}$ that indicate how much
$H_2$ is composed by  $B$-sector Higgses. Both  
mixing angles and the mass eigenvalues are computed numerically taking 
into account all orders in the $\lambda_S^2 f^2/m_A^2$ expansion 
(setting $\lambda_A=\lambda_{BD}=\delta t_{\beta}\approx0$ for
simplicity). 

As expected, $H_2$ is mainly $B$-like (i.e. $h_T$) in the upper left
corner of the plot ($m_{A} \gtrsim \lambda_S f$), while it is mainly
$A$-like (i.e. $H$) in the bottom right corner of the plot
($m_{A}\lesssim \lambda_S f$).  
The four plots in Fig.~\ref{fig:masses} show how the details of the
transition between the two regimes depend on the value of $\lambda_S$
and $\tan\beta$: comparing the plots row-wise we see that a larger
$\tan\beta$ suppresses the radial mode mass hence enlarging the
parameter space in which $H_2$ is mostly $B$-like (red
region). Column-wise, we see that a larger
$\lambda_S$ enlarges the parameter space in which $H_2$ is mostly
$A$-like (blue region), because the twin Higgs mass $m_{h_T}$ gets
enhanced while the MSSM doublet mass $m_A$ gets suppressed.

\begin{figure}[t]
  \centering
 \includegraphics[width=0.49\textwidth]{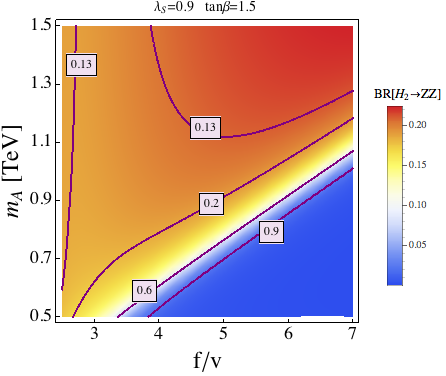}
  \includegraphics[width=0.49\textwidth]{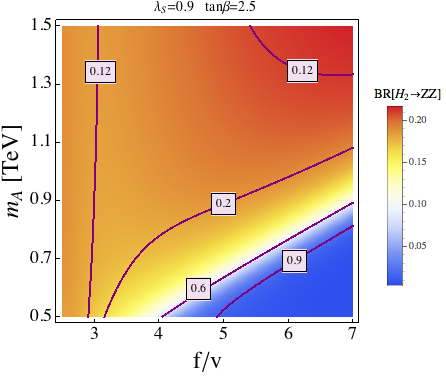}\\
  \vspace{.25cm}
  \includegraphics[width=0.49\textwidth]{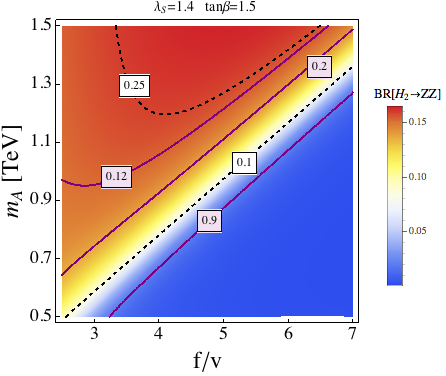}
  \includegraphics[width=0.49\textwidth]{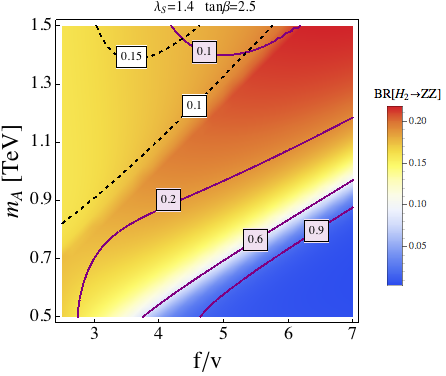}
   \caption{Branching ratio of the second lightest CP-even
     eigenstate. The four plots
   correspond to different values of $(\lambda_S,\tan \beta)$. The
     purple (black dashed) contours denotes the BR into SM $t \bar t$ pairs (the invisible
     BR into $W_D$ and $Z_D$). The temperature map indicates the BR into SM $ZZ$.
 }
  \label{fig:BR}
\end{figure}

We are now ready to understand the parametric dependence of the decays
of the CP-even Higgses in the $(f/v\, , m_A)$ plane, which we can
numerically compute using the decay rates of the CP-even eigenstates
that are collected in Appendix~\ref{app:2}. In Fig.~\ref{fig:BR} we
show the relevant branching ratios for the next-to-lightest CP-even
state~$H_2$.  

Clearly we have two different regions depending if the lightest
CP-even state is the twin Higgs $h_T$ (red region) or the MSSM-like
heavy Higgs $H$ (blue region).  
In the first case the visible decays of the twin Higgs are into pairs
of SM gauge bosons/Higgses and into top pairs. An invisible decay
width can also be present  if the decays into
dark gauge bosons are kinematically open.  
Since we assume equal gauge couplings in the visible and dark sector,
the dark gauge boson masses are fixed and scale approximately linearly
with $f$,  
e.g. $m_{W_B} \approx \frac{g_2}{\sqrt{2}} f$.
Comparing the upper and the lower panel of Fig.~\ref{fig:BR} we see
that a non-zero invisible decay width for the twin Higgs ultimately depends on the value of the $SU(4)$-invariant coupling $\lambda_S$,
which sets the scale of the twin Higgs mass. 
In the upper row we take $\lambda_S=0.9$, such that the invisible decay
channels are kinematically closed while in the second row we take
$\lambda_S=1.4$. Of course the latter value is favored by fine-tuning
arguments as discussed in Sec.~\ref{sec:3}, however since
$\lambda_S\lesssim 1$ makes the phenomenology of the twin Higgs
radically different we decided to include $\lambda_S=0.9$ in our
discussion.  

The most interesting decay channel to hunt for the twin Higgs at the
LHC is certainly the one into $Z$ pairs. In Fig.~\ref{fig:BR} we see
that this branching ratio goes roughly from $25\%$ to $10\%$ in the
red area,  depending on the region of the parameter space. A $25\%$
branching ratio into $Z$ pairs is the  value expected from the
Goldstone equivalence principle if one assumes $\sim100\%$ branching
ratio of the twin Higgs into SM gauge bosons/Higgs like was done in
\cite{Buttazzo:2015bka}. From Fig.~\ref{fig:BR} we see that this
expectation is only (marginally) saturated in the region
where~$m_A>1\text{ TeV}$ and $f/v>5$. In the bulk of the SUSY Twin
Higgs parameter space the branching ratio into $Z$ pairs (like the
ones into $WW$ and $hh$) is depleted because of an irreducible
branching ratio into top pairs (purple contours). The $t\bar{t}$
channel is suppressed only for very large $f/v$ (i.e. very small
mixing) or for very small $f/v$ because of the reduced phase space.  

Using the formulas from Appendix~\ref{app:2},
assuming $m_A\gg \lambda_S f$ and neglecting the phase space
suppressions we get 
\begin{equation}
\frac{\Gamma(h_T\to ZZ)}{\Gamma(h_T\to t\bar{t})}\approx
\frac{m_{h_T}^2}{12 c_W^2 m_t^2} \approx \frac{\lambda_S^2
  s_{2\beta}^2 f^2}{12 c_W^2 m_t^2}\approx 1\times \left(\frac{
    f/v}{3}\right)^2\times \left(\frac{\lambda_S}{1}\right)^2\times
\left(\frac{s_{2\beta}}{0.98}\right)^2 \, , \label{approxttvsZZ} 
\end{equation}
which shows that an irreducible branching ratio into $t\bar{t}$ for
the twin Higgs is somewhat typical for Twin SUSY scenarios, where the
branching ratio into $ZZ$ gets suppressed for $\tan\beta>1$ (where
$\sin2\beta$ is reduced) and cannot be enhanced arbitrarily by taking
a larger $\lambda_S$ because of perturbativity.

In the second row of Fig.~\ref{fig:BR} we set $\lambda_S=1.4$ and the
invisible width of the twin Higgs is different from zero (dashed black
contours). When the decays into dark gauge bosons is kinematically
open, if one would assume $100\%$ decay into (Pseudo)-Goldstones, the
Goldstone equivalence theorem would predict a branching ratio of $\sim
0.14$ into $ZZ$ and an invisible width of $\sim3\times
0.14$. However, this naive estimate is once again modified by the
irreducible decay width into $t\bar{t}$ (though less relevant than
$\lambda_S=0.9$ in agreement with formula~\eqref{approxttvsZZ}) and by
the fact that in most of the Twin SUSY parameter space the twin Higgs
mass is not parametrically larger than the dark gauge boson
masses. Indeed  the invisible width of the radial mode gets sensibly reduced because
 of phase space suppression,  as can be seen from the contours in Fig.~\ref{fig:BR}. 
The phase space suppression of the invisible channels 
becomes more important at larger $\tan\beta$, where the twin Higgs
mass is further reduced.

Finally,
for fixed value of $f/v$, decreasing the value of $m_A$ implies that the mixing
with the MSSM-like CP-even Higgs becomes important and the branching
ratio into $t\bar{t}$ is enhanced with respect to $ZZ$ (and with
respect to invisible ones if present). Decreasing $m_A$ further the
MSSM-like Higgs becomes the next-to-lightest state and as a consequence
the decay width into $ZZ$ is suppressed while the one into $t\bar{t}$
gets rapidly close to $100\%$.

\subsection{Probing the SUSY Twin Higgses}

From the previous discussion one can infer what are the most promising
phenomenological signatures associated to the Higgs sector of the SUSY
Twin model.  
We essentially have two different types of signatures depending on the
region of the parameter space: 
\begin{itemize}
\item In the region where $m_A \gtrsim \lambda_S f$ the twin Higgs is
  the next-to-lightest CP-even state (this is the red region in Figs.~\ref{fig:masses} and \ref{fig:BR}). The radial mode is copiously
  produced at the LHC via its mixing with the SM Higgs and the most
  promising channels to probe it are di-boson final states; in
  particular di-Higgs (four b-jets) or Z-boson pairs
  \cite{Craig:2013fga,Buttazzo:2015bka,Craig:2015jba,Khachatryan:2015cwa}. Indirect
  bounds on the twin Higgs also arise from modifications of the
  SM-like Higgs couplings (in particular the ones to gauge bosons).  

\item In the region where $m_A \lesssim \lambda_S f$ we have a full
  MSSM-like $SU(2)$ doublet which becomes light. Since having
  $m_h=125\text{ GeV}$ always forces $\tan\beta$ to be quite small,
  this region of the parameter space is preferably tested by searches
  for charged MSSM-like Higgses. These can be probed either indirectly
  through their contribution to $b\to s\gamma$, or directly hunted at
  the LHC via $t\bar{b}$ and $b\bar{t}$ final states
  \cite{CMS:2014pea,Craig:2015jba}. Another interesting channel is
  provided by the associated production of a CP-even/odd Higgs with
  $t \bar t$~\cite{Craig:2015jba}. However, searches for charged
  MSSM-like Higgses typically provide a cleaner channel for similar
  mass scales. 
\end{itemize}

We now provide a preliminary study of the LHC reach on the parameter
space of Twin SUSY models, which focus on direct and indirect searches
for 
the twin Higgs and charged MSSM-like Higgses. We present both existing
constraints and the future reach of LHC searches and indirect
measurements. For LHC prospects we consider projections for the final
stage of LHC with $300\ \text{fb}^{-1}$ and for HL-LHC, where $3000\
\text{fb}^{-1}$ are expected~\cite{Gianotti:2002xx}.  

For the  $ZZ$ channel we use the results of
Ref.~\cite{Buttazzo:2015bka}, where the current and future bounds for
a scalar singlet mixing with the SM-Higgs are given as a function of
its mass and signal strength into $ZZ$. We recast these
bounds in the Twin SUSY parameter space by computing the
production cross section of the radial mode from
both gluon fusion and vector boson fusion (VBF) processes and
multiplying by the branching ratio into $ZZ$.\footnote{To estimate the
  8 TeV cross section we use the Higgs production cross section for
  the 
  gluon fusion and the VBF processes for equivalent mass
  \cite{Heinemeyer:2013tqa} and weight them with the appropriate
  mixing angles. For the  $\sqrt{s}=13$ TeV and $\sqrt{s}=14$ TeV
  cross section we just re-weight the $\sqrt{s}=8$ TeV cross sections
  for gluon fusion and VBF 
with the corresponding parton luminosities ratios taken from
Ref.~\cite{Heinemeyer:2013tqa}.}  

For what concerns Higgs coupling deviations, 
a global fit of Higgs couplings and EWPT in Twin SUSY has already been
performed in Ref.~\cite{Craig:2013fga}, where it has been shown that
current data 
cannot probe our region of interest on which we already impose $f/v
\gtrsim 2.3$ and $m_A \gtrsim 400$ GeV. To estimate the HL-LHC
prospects on Higgs couplings, one can focus on the measurement of the
SM Higgs coupling to $Z$-bosons. 
The HL-LHC prospects for the Higgs coupling measurements will
ameliorate the current bound on the mixing angle 
by roughly a factor of 2 (the precise numbers can be found in
Refs.~\cite{Buttazzo:2015bka} and \cite{Dawson:2013bba}), which leads
to a future bound of approximately $f/v\gtrsim4.6$. We will comment on
such a limit in the following discussion.  

The LHC prospects for direct probe of the MSSM charged Higgses are
taken from Ref.~\cite{Craig:2015jba} where the process $p p \to H^{+}
\bar t b \to t \bar b \bar t b $ (and charged conjugate) is
employed. Notice that we can use these bounds without any further
modification since the visible charged Higgses do not mix with the
dark sector ones.

Concerning the indirect searches through $b \to s\gamma$, we
parametrize the BSM contributions as modifications to the Wilson
coefficients $\Delta C_7$ and $\Delta C_8$ with respect to their SM
values~\cite{Misiak:update}: 
\be
B_{s \gamma}^{\text{th}} \times 10^4 = 3.36 \pm 0.23 -8.22 \Delta C_7 -1.99 \Delta C_8\, ,
\ee
where $\pm 0.23$ is the theoretical uncertainty of the Standard Model
prediction. For the experimental measurement we use the combined
result from Ref.~\cite{Trabelsi:2015eps}  
\be
B_{s \gamma}^{\text{exp}}  \times 10^4 = 3.41 \pm 0.16 \, ,
\ee
and we derive the bounds allowing for $2\sigma$ deviations from the
experimental value, summing in quadrature the theoretical and
experimental uncertainties. 
The contribution of the charged MSSM-like Higgs to this observable
implies a lower bound on its mass for a fixed value of $\tan
\beta$. In particular, the existing limits set a lower bound of around
$m_{H^+} \gtrsim 500$ GeV for $\tan \beta \simeq 2$, and hence does
not constraint our parameter space. In order to estimate the future
prospects for this observable, we optimistically assume that both the
theoretical and the experimental uncertainties get reduced by a factor
of $2$.\footnote{M.~Misiak private communication.} The bound on the
charged MSSM-like Higgs mass is then significantly improved and goes
up to $m_{H^+} \geq 775$~GeV for $\tan \beta \simeq 2$.

\begin{figure}[h!]
  \centering
 \includegraphics[height=0.49\textwidth]{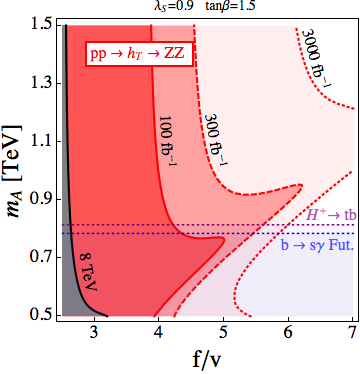}\hfill
  \includegraphics[height=0.49\textwidth]{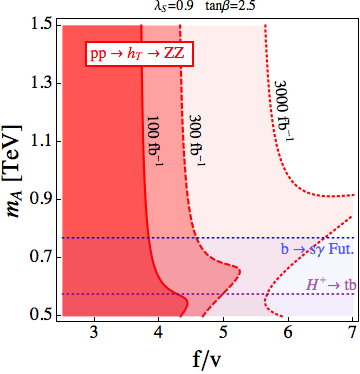}\\
  \vspace{.25cm}
   \includegraphics[height=0.49\textwidth]{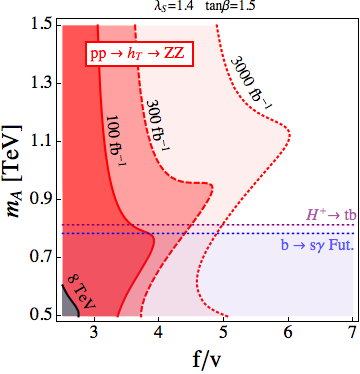}\hfill
    \includegraphics[height=0.49\textwidth]{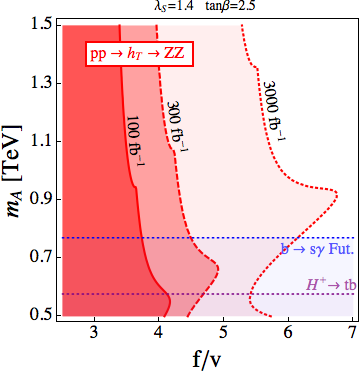}
    \caption{Phenomenology of the SUSY Twin Higgs model in the $(f/v,m_{A})$ plane. The dark grey region denotes the $8$ TeV
      exclusion from di-boson searches. The red regions denote the
      LHC prospects for di-boson signatures associated with the
      twin Higgs state, respectively with 100, 300 and 3000~fb$^{-1}$
      for the 
    solid contour, the dashed contour and the dotted contour.
   Projected bounds from $b \to s \gamma$ are denoted in blue while
   the HL-LHC reach on the charged Higgs state with 3000~fb$^{-1}$ is
   indicated in purple. 
    }
  \label{fig:pheno}
\end{figure}

In Fig.~\ref{fig:pheno} we project the current and future bounds from
the searches described above in the $(f/v\,, m_A)$ plane. Analogously
to Figs.~\ref{fig:masses} and~\ref{fig:BR} we show the results for
four different choices of $\lambda_S=\{1.4\,, 0.9\}$ and
$\tan\beta=\{1.5\,, 2.5\}$. From these plots we can see that direct
searches for the twin Higgs in the $ZZ$ channel  
give a lower bound on $f/v$ as long as $m_A\gtrsim
\lambda_S f$. In contrast direct and indirect searches on the MSSM
charged Higgses set a lower bound on $m_A$
\emph{independently} on the value of~$f$.  
We first discuss the features of the searches targeting the MSSM-like scalars.
Comparing the plots row-wise we see that direct LHC searches on the
charged Higgs depend significantly on $\tan\beta$, which sets the
couplings to $\bar t b$. 
The projected bound from $b\to s\gamma$ is also dependent on
$\tan\beta$ and gets slightly stronger for very low values of
$\tan\beta$. Note that here we used 
the complete expression for the charged Higgs mass which includes also
$\cO(v^2)$ corrections. Conversely, comparing plots column-wise one can see that varying
$\lambda_S$ does not change the physics of the MSSM charged Higgs as
expected.  

Now we focus on the region probed by the di-boson resonance searches.
The shape of the exclusion lines from the $ZZ$ final states presents a
rich structure, which depends on multiple effects controlled by
$\lambda_S$ and $\tan\beta$. 
The overall reach is affected by variations of the coupling to tops
(the main production mechanism is gluon fusion), variations in the
structure of the mass spectrum (cf. Fig.~\ref{fig:masses}) and in the
branching ratio to $ZZ$ (c.f.~Fig.~\ref{fig:BR}). 
However, independently on the values of $\lambda_S$ and $\tan\beta$,
the exclusion regions exhibit a peculiar horn-like shape. The horn is
located where 
the next-to-lightest eigenstate changes from mostly twin-like to
mostly MSSM-like 
In such region,
because of the transition, there is a local enhancement in the
A-sector components of the $H_2$ state and the sensitivity is
increased. 
Then, in the region where the $H_2$ state is very much MSSM-like, the
sensitivity drops because of the reduced branching ratio into gauge
bosons. 
In the low $m_A$ region of the plots (close to $m_A\sim 500\text{
  GeV}$) the reach of HL-LHC is again enhanced because the very small
branching ratio of $H_2$ into gauge bosons is compensated by a
sizeable enhancement of the production cross section for such low
masses.

In the first row of Fig.~\ref{fig:pheno} we fix $\lambda_S= 0.9$ such
that the invisible decay channels are closed.  
Comparing the plots row-wise we see that the bounds from LHC are very
similar. The reason is that for lower $\tan\beta$ the enhancement of
the top coupling, and hence of the gluon fusion channel, is partially
compensated by a larger twin Higgs mass (cf. the contours of
Fig.~\ref{fig:masses}). 
The transition between the twin Higgs regime and the MSSM-like regime
is not the same in the two plots and reflects the discussion of the
previous subsection.  

In the second row of Fig.~\ref{fig:pheno} we fix $\lambda_S= 1.4$ and
the invisible decay channel for the twin Higgs opens up, depleting the
signal strength into $ZZ$ with respect to the $\lambda_S= 0.9$
case. However, in the bottom right plot the larger value of $\tan
\beta$ reduces $m_{h_T}$ and closes the invisible decay into dark
gauge bosons exactly where the small bump in the
exclusion curves is, roughly where $m_{h_T} \sim 2 m_{W_D}$ (compare with
contours in Figure \ref{fig:masses}).

In summary, we see that searches for the twin Higgs into $ZZ$ final
states at the LHC are extremely interesting in Twin SUSY
constructions, because perturbativity of the $SU(4)$-invariant
coupling $\lambda_S$ and $\tan\beta>1$ give an upper bound on the twin
Higgs mass for fixed value of $f/v$ and fixed $m_A$. This has to be
contrasted with strongly coupled UV completions where usually the
strongest constraint comes from Higgs coupling measurements.  
Note that the prospects for Higgs coupling measurements at HL-LHC
reach roughly the value of $f/v \simeq 4.6$ as we mentioned before. 
For the largest $\lambda_S$ we considered, which corresponds to the
largest $m_{h_T}$ compatible with perturbativity, the Twin SUSY
phenomenology starts to be similar to strongly coupled UV completions
with the Higgs coupling measurements having a reach not very far from
the direct searches (see left bottom plot of
Fig.~\ref{fig:pheno}). However, for lower values of $\lambda_S$ Higgs
coupling measurements are never competitive with the direct resonance
search as can be seen from the other plots on Fig.~\ref{fig:pheno}. 
Finally, the region with large $f/v$ features a radial mode too heavy to be probed in $ZZ$ searches at the LHC, even at High Luminosity. 
For small $m_A$, the spectrum here resembles the usual MSSM and 
MSSM-like Higgses searches will explore this region for $m_A \lesssim
1$ TeV. 

Looking back at the parameter space of the soft and the hard Twin SUSY
models of Sec.~\ref{sec:3}, we can roughly compare the sensitivity of
the direct searches at the LHC in these two particular cases. 
In the simplest soft Twin SUSY model of Fig.~\ref{numeric_FT_soft} and
Fig.~\ref{numeric_FT_soft_2} we found a strong \emph{upper} bound on
$f/v$ for $M_s\approx 2\text{ TeV}$, which, together with the
quite large value of $\tan\beta\approx 3.5$, gives a stringent upper
bound on the mass of the twin Higgs $\lesssim 400-700\text{ GeV}$,
depending on the particular choice of $\delta t_\beta$ and
$m_{A_T}$. A considerable amount of parameter space of soft Twin SUSY
is then likely to be probed by $ZZ$ searches already with $100\text{
  fb}^{-1}$ of data. The mass of the MSSM-like Higgses depends instead
on $m_{A_T}$, which is unconstrained in this model and can always be
taken to be heavy.  

The hard Twin SUSY model we presented in Fig.~\ref{numeric_FT_BD_1}
has instead a \emph{lower} bound on the Twin Higgs mass which
increases at large $M_s$. This lower bound is still pretty mild
for $M_s\approx 2\text{ TeV}$ and Twin Higgs searches are certainly promising
to explore these models. Quite interestingly the masses of the
MSSM-like Higgses cannot be arbitrarily decoupled in this model as far
as natural values of $\delta t_\beta$ are considered and the charged
Higgses can easily be below the TeV leaving some hope for indirect
signatures in $b\to s\gamma$ or the direct searches at the LHC. 

\section{Conclusions}\label{discussion}
In this work we have performed a systematic study of perturbative UV
completions of the Twin Higgs mechanism based on Supersymmetry. In
this context we showed that breaking the $\ztwo$ mirror symmetry with
large quartics can be beneficial in terms of fine tuning, leading to
theories with colored states decoupled from the LHC and tuned to
the level of 
$\sim 10\%$.   

In order to explore the role of $\ztwo$-breaking quartics, we
performed a detailed comparative study between hard and soft breaking
of the $\ztwo$-symmetry from the effective field theory point of view
in Sec.~\ref{sec:2}. This study provides a complete picture of the
parameter space of the Twin Higgs and opens up new model building
avenues. The main result is that hard $\ztwo$-breaking models
can lead to a gain in fine tuning with respect to soft
$\ztwo$-breaking at the price increasing the SM-like
Higgs mass, which is usually predicted to be too high. Once the Higgs
mass 
constraint is satisfied, the fine tuning of hard $\ztwo$-breaking
models can still be around a factor of $\sim5$ better compared to soft
$\ztwo$-breaking.  

\begin{figure}[t]
\centering
\includegraphics[width = .55\textwidth]{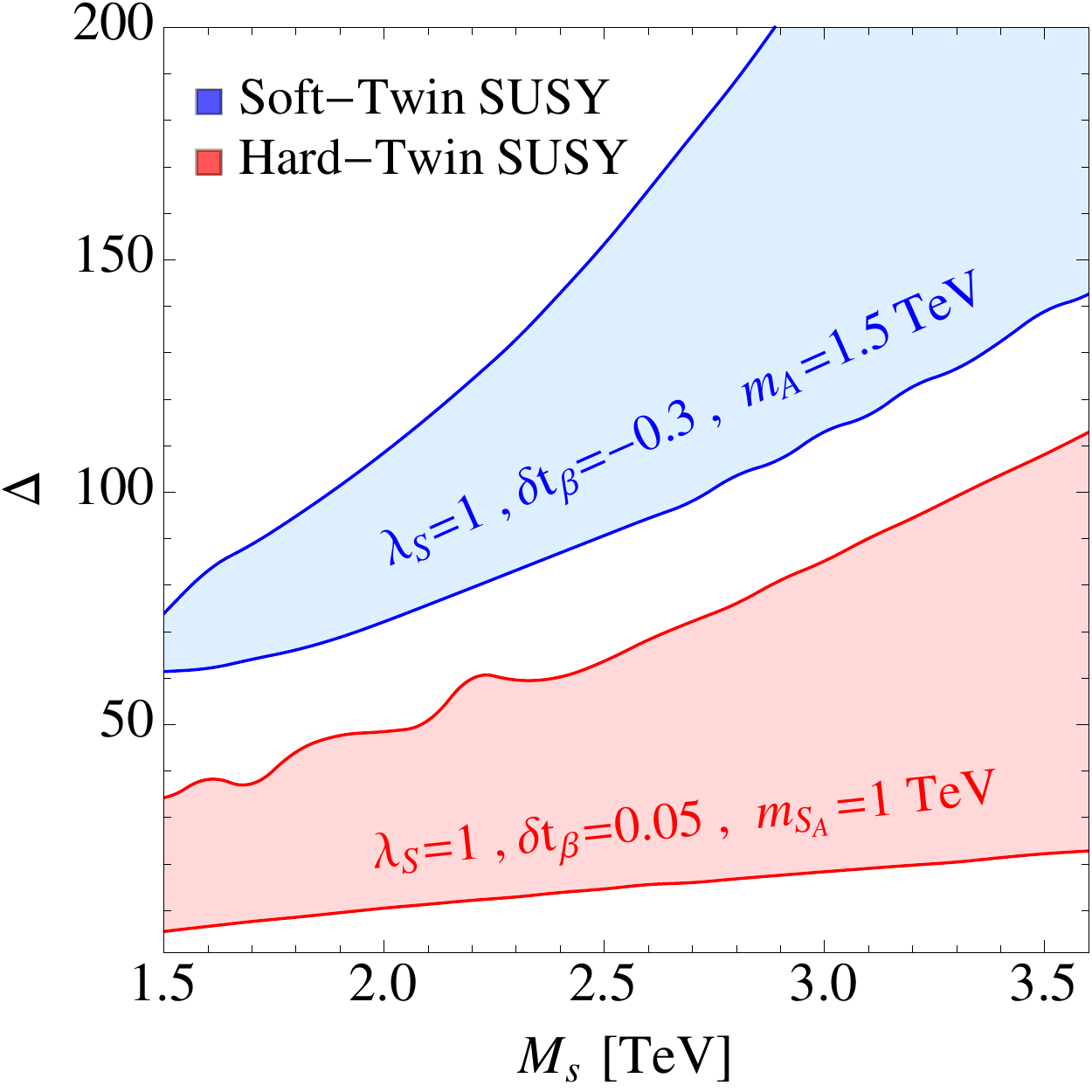}
\caption{Comparison of the fine tuning of the soft Twin SUSY and hard
one. Inside the strips we marginalize over all the parameters of the
model unless specified.}
\label{moneyplot}
\end{figure}

We have studied explicit SUSY UV completions of both soft and hard
$\ztwo$-breaking models. We performed our analysis
both numerically, solving directly the EWSB conditions and the Higgs
mass constraint in the UV theory, and analytically, using a simple
tree level matching to the original Twin Higgs model after the SUSY
states are decoupled.  
Both models 
are  
not saturating the parametric gain in fine-tuning that we na\"ively
expect from Twin SUSY theories with respect to standard SUSY
scenarios, but provide simple existence proofs of both mechanisms of
mirror symmetry breaking. By comparing these two simple SUSY models we
recover the gain in fine tuning of the
hard $\ztwo$-breaking model as obtained in the effective field theory.  As a summary plot we show in Fig.~\ref{moneyplot} the allowed
FT range in the soft and the hard model as a function of the scale of
the colored particles, marginalized over the rest of the
parameters. 

The Twin SUSY constructions discussed here are only valid up to a
scale of roughly $\sim 100\text{ TeV}$. The final goal would be to
build full UV completions up to the GUT scale where both the
SUSY-breaking and the $\ztwo$-breaking are generated
\emph{dynamically} and then mediated to the MSSM and its mirror
copy. Specifying the mechanism of mediation will give a definite
prediction on the SUSY spectrum of both the visible and the mirror
sector.  Further studies in this direction are definitely worthwhile. 

Our analysis has unveiled several important aspects of the parameter space
of Twin SUSY models, highlighting in particular the special role of the
Higgs mass in constraining  the parameter space of both soft and hard
Twin SUSY. To improve the analysis it would be desirable to have a
more precise computation of the SM Higgs mass including at least
2-loop QCD corrections. The implementation of Twin SUSY models in a
package like SARAH~\cite{Staub:2013tta} can bring them to a similar
precision we have for standard SUSY scenarios.  

We hope that our study elucidated the basic building blocks of Twin
SUSY constructions and their connection to the original Twin Higgs
proposal. With these tools at hand it will be easier to construct
``optimal'' models of Twin SUSY, in attempt of maximizing their
parametric gain in fine tuning with respect to standard SUSY
scenarios..   

We also sketched the phenomenology of the Twin SUSY Higgses. 

In particular we showed that LHC searches for a resonance going into
di-bosons can probe a large portion of the parameter space of Twin Higgs
models, independently on how the mirror symmetry is broken. This is a
consequence of the requirement of perturbativity on the
$SU(4)$-invariant quartic coupling which gives an upper bound on the
mass of the twin Higgs at fixed $SU(4)$-breaking VEV $f$, and it is a
distinctive feature of perturbative UV completions of the Twin Higgs
mechanism.  

It would be interesting to further characterize the signatures
associated to the invisible decays of the twin Higgs. These can
potentially distinguish it from a generic singlet that mixes with the
SM Higgs. Indeed, the invisible branching ratio of the twin Higgs is
likely to be enhanced when extra light SUSY states are present in the
mirror sector making it discoverable at the LHC in final states with
large missing transverse energy (MET). 

Interesting signatures can also arise from MSSM-like Higgses which
become light at large $f$. Twin SUSY models then provide an additional
motivation to extend the search program for extra MSSM-like Higgses
in the small $\tan\beta$ regime both at LHC and at future colliders \cite{Craig:2015jba,Hajer:2015gka,Craig:2016ygr}.  
It would be interesting to investigate in detail alternative final
states to probe CP-odd and CP-even MSSM-like scalars, such as the
ones suggested in Ref.~\cite{Craig:2015jba} including for instance $t \bar t A/H$ 
production.  
The possibility of having extra invisible decay channels into the
mirror sector can also give rise to final states with MET, enriching
further the landscape of accessible topologies. 

The phenomenology of the extended Higgs sector is just the tip of an
iceberg of possible phenomenological questions one can ask in the
context of Twin SUSY.  By specifying the detailed structure of both
the SUSY and the mirror spectrum one can explore further the
phenomenological implications of Twin SUSY. For instance it would be
interesting to investigate allowed cosmological scenarios as it has
been done for the original Twin Higgs proposal
\cite{Craig:2015xla,Garcia:2015toa,Garcia:2015loa,Farina:2015uea,
Farina:2016ndq,Barbieri:2016zxn}, identifying possible dark matter
candidates (both symmetric and asymmetric), viable mechanisms for
baryogenesis etc. Of course extra LHC signatures other than the ones
associated to extra Higgses can arise once the full spectrum is
specified (see for example
Refs.~\cite{Craig:2015pha,Burdman:2014zta,Chacko:2015fbc,Curtin:2015fna})
and might require interesting new detecting techniques
\cite{Coccaro:2016lnz,Chou:2016lxi}. All these instances are likely to
have an impact on the fine tuning of Twin SUSY models establishing a
unique interplay between cosmology, collider searches and naturalness
which is a distinctive features of NN scenarios.  

\acknowledgments{We are grateful to Nathaniel Craig, Mark Goodsell, Matthew McCullough, 
Mikolaj Misiak, Filippo Sala, Pietro Slavich and Riccardo Torre for useful discussions and Lorenzo Ubaldi for comments on the manuscript. 
The work of A.K. was performed in part at the Aspen Center for Physics, 
which is supported by National Science Foundation grant PHY-1066293. The work of D.R. and R.Z. was mostly performed 
at the Laboratoire de Physique of Hautes Energies (LPTHE) in Paris. D.R. R.Z. and A.M. also acknowledge the hospitality 
of the Galileo Galilei Institute for Theoretical Physics. A.M. is supported by the Strategic Research Program High Energy Physics 
and the Research Council of the Vrije Universiteit Brussel. A.M. is also supported in part by the Belgian Federal Science Policy 
Office through the Interuniversity Attraction Pole P7/37. 
The  work  of SP is supported by the National Science Centre, Poland, under research grants
DEC-2014/15/B/ST2/02157,
DEC-2012/04/A/ST2/00099 and DEC-2015/18/M/ST2/00054.}

 \appendix
 \section{Appendix: Renormalization Group Equations}\label{app:RGE}
 In this appendix we provide the one-loop RGEs for the Twin SUSY models with a) soft $\ztwo$-breaking, b) hard $\ztwo$-breaking and c) hard $\ztwo$-breaking with bi-doublets. We neglect A-terms, singlet B-terms, electro-weakinos, sbottom and slepton masses, and take into account three families both in the visible and the dark sector (relevant for gauge coupling running and hypercharge D-terms). 
\subsection{Soft $\ztwo$-Breaking Model}\label{sec:RGEsoft}
\subsection*{Superpotential and Soft Terms} 
The superpotential (in the UV) is defined as
\begin{align}
W|_{\rm UV} & = \left( \mu + \lambda_S S \right) \left( h^A_u h^A_d  + h^B_u h^B_d \right)+ y_t \left( Q_3^A U_3^A h_u^A + Q_3^B U_3^B h_u^B \right) + \frac{1}{2} M_S S^2  \, , 
\end{align}
and soft breaking masses (in the UV) are given by
\begin{align}
V_{\rm soft}|_{\rm UV} & = m^2_{H_u} \left( |h_u^A|^2 + |h_u^B|^2 \right) +  m^2_{H_d} \left( |h_d^A|^2 + |h_d^B|^2 \right)   - b \left( h_u^A h_d^A  + h_u^B h_d^B +{\rm h.c.} \right)   \nonumber \\
& + m^2_{Q} \left( |Q_3^A|^2 + |Q_3^B|^2 \right)  + m^2_{U} \left( |U_3^A|^2 +  |U_3^B|^2 \right) + m^2_S |S|^2 \nonumber \\
& + \Delta m^2_{H_u} \left( |h_u^A|^2 - |h_u^B|^2 \right)  + \Delta m^2_{H_d} \left( |h_d^A|^2 - |h_d^B|^2 \right)   \, .
\end{align}
Note that $\ztwo$ is broken in the UV Lagrangian only by soft Higgs masses, which implies that  $\ztwo$ is conserved by the RG flow except for sfermion (stop) masses. Therefore we have to add $\ztwo$-odd stop masses in the IR-potential for consistency (but they are negligible since the $\ztwo$-breaking Higgs masses in the UV are taken to be small in the first place)
 \begin{align}
\delta V_{\rm soft}|_{\rm IR}  & =   \Delta m^2_{Q} \left( |Q_3^A|^2 - |Q_3^B|^2 \right) + \Delta m^2_{U} \left( |U_3^A|^2 -  |U_3^B|^2 \right)  \, .
\end{align}

\subsubsection*{Beta Functions}
The beta function coefficients for the gauge couplings $ \tilde{g}_Y = g_Y ,  \tilde{g}_2 = g_2, \tilde{g}_3 = g_3$ are:
\begin{align}
b_i = \left( 11, 1, -3 \right) \, .
\end{align}
\subsubsection*{Yukawa couplings}

\begin{align}
16 \pi^2 \frac{d}{dt}  \lambda_S & = \lambda_S \left( 6 \lambda_S^2  + 3 y_t^2 - g_Y^2 - 3 g_2^2 \right)  \, , \nonumber \\
16 \pi^2 \frac{d}{dt}  y_t & =  y_t \left( 6 y_t^2 + \lambda_S^2 - \frac{13}{9} g_Y^2 - 3 g_2^2  - \frac{16}{3} g_3^2 \right) \, .
\end{align}
\subsubsection*{SUSY masses}
\begin{align}
16 \pi^2 \frac{d}{dt} \mu  & = \mu \left( 2 \lambda_S^2 + 3 y_t^2   -g_Y^2 - 3 g_2^2 \right) \, , \nonumber \\
16 \pi^2 \frac{d}{dt} M_S  & = 8 M_S   \lambda_S^2 \, .
\end{align}
\subsubsection*{Soft Masses}
Neglecting the  contributions of $\ztwo$-breaking stop masses to the RGEs, one has
\begin{align}
16 \pi^2 \frac{d}{dt} m^2_{H_u}  & = g_Y^2 \xi + 2 \lambda_S^2 X_S   + 6 y_t^2 X_t \, , \label{mhusoft} \nonumber  \\
16 \pi^2 \frac{d}{dt} \Delta m^2_{H_u}  & = g_Y^2 \Delta \xi + 2 \lambda_S^2 \Delta X_S   + 6 y_t^2 \Delta X_t \, , \nonumber  \\
16 \pi^2 \frac{d}{dt} m^2_{H_d}  & = - g_Y^2 \xi + 2 \lambda_S^2 X_S \, , \nonumber  \\
16 \pi^2 \frac{d}{dt} \Delta m^2_{H_d}  & = - g_Y^2 \Delta \xi + 2 \lambda_S^2 \Delta X_S  \, , \nonumber  \\
16 \pi^2 \frac{d}{dt} m^2_{Q}  & = \frac{1}{3} g_Y^2 \xi - \frac{32}{3} g_3^2 M_3^2 + 2 y_t^2 X_t\, , \nonumber  \\
16 \pi^2 \frac{d}{dt} m^2_{U}  & = - \frac{4}{3} g_Y^2 \xi - \frac{32}{3} g_3^2 M_3^2 + 4 y_t^2 X_t\, , \nonumber  \\
16 \pi^2 \frac{d}{dt} m^2_{S}  & =  8 \lambda_S^2 X_S \, . 
\end{align}
with the auxiliary functions
\begin{align}
X_S & \equiv m^2_S + m^2_{H_u} + m^2_{H_d} \, , &  \Delta X_S & \equiv  \Delta m^2_{H_u} + \Delta  m^2_{H_d} \, , \nonumber  \\
X_t & \equiv m^2_{Q} + m^2_{U} + m^2_{H_u}  \, ,  & \Delta X_t & \approx  \Delta  m^2_{H_u} \, , \nonumber  \\
\xi & \equiv 3 m^2_{Q} - 6 m^2_{U} + m^2_{H_u} - m^2_{H_d} \, , &
\Delta \xi & \approx   \Delta m^2_{H_u} - \Delta m^2_{H_d}  \, .
\end{align}
The RGEs for $\ztwo$-breaking stop masses are given by
\begin{align}
16 \pi^2 \frac{d}{dt} \Delta m^2_{Q}  & = \frac{1}{3} g_Y^2 \Delta \xi + 2 y_t^2 \Delta X_t \, , \nonumber  \\
16 \pi^2 \frac{d}{dt} \Delta m^2_{U}  & = - \frac{4}{3} g_Y^2 \Delta \xi + 4 y_t^2 \Delta X_t \, .
\end{align}
Moreover one has
\begin{align}
16 \pi^2 \frac{d}{dt} b & = b \left( 10 \lambda_S^2 + 3 y_t^2  - 3 g_2^2 -  g_Y^2 \right)  \,  , \nonumber  \\
 16 \pi^2 \frac{d}{dt} M_3 & = - 6 g_3^2 M_3 \, . 
\end{align}
\subsection{Hard $\ztwo$-Breaking + Bi-doublets}
For the sake of brevity we give the RGEs only for the general model with bi-doublets. The RGEs of the hard breaking model are obtained by setting to zero the bi-doublet parameters $\lambda_{BD}, M_\Phi, m^2_{B}, m^2_{\bar{B}}$.
\subsubsection*{Superpotential and Soft Terms} 
The superpotential (in the UV) is defined as 
\begin{align}
W|_{\rm UV} & = \left( \mu + \lambda_S S \right) \left( h^A_u h^A_d + h^B_u h^B_d \right) + \lambda_A S_A h^A_u h^A_d  \nonumber \\
& + y_t \left( Q_3^A U_3^A h_u^A + Q_3^B U_3^B h_u^B  \right) +    \frac{1}{2} M_S S^2 + \frac{1}{2} M_{S_A} S_A^2 \nonumber \\ 
& + \lambda_{BD} B h_u^A h_u^B +  M_B \bar{B} B \, ,
\end{align}
where $B$ is a bi-doublet under $SU(2)_A \times SU(2)_B$ and is charged under $U(1)_A \times U(1)_{B}$ as (-1/2, -1/2), while $\bar{B}$ has conjugate quantum numbers. Here we have set to zero the mixed $S$-$S_A$ mass term, a $\ztwo$-breaking 
$\mu$-term and $\ztwo$-breaking top Yukawa and $S$ couplings, as well as the $S_A $ couplings to the $B$-sector . All these operators are generated in the IR as a result of the $\ztwo$-breaking couplings $\lambda_A$, but remain small as far as $\lambda_A$ is not too large:
\begin{align}
\delta W|_{\rm IR} & =   M_{S S_A} S S_A  + \lambda_B S_A  h^B_u h^B_d \nonumber \\
& +  \left( \Delta \mu + \Delta \lambda_S S \right) \left( h^A_u h^A_d - h^B_u h^B_d \right) + \Delta y_t \left( Q_3^A U_3^A h_u^A  - Q_3^B U_3^B h_u^B \right) \, .
\end{align}
Soft breaking masses (in the UV) are defined as
\begin{align}
V_{\rm soft}|_{\rm UV}  & = m^2_{H_u} \left( |h_u^A|^2 +   |h_u^B|^2 \right) + m^2_{H_d} \left( |h_d^A|^2 +   |h_d^B|^2   \right)\nonumber \\
& + m^2_{Q} \left( |Q_3^A|^2 + |Q_3^B|^2 \right) + m^2_{U} \left( |U_3^A|^2 +  |U_3^B|^2  \right) \nonumber \\
& - b \left( h_u^A h_d^A  + h_u^B h_d^B   +{\rm h.c.} \right) 
 + m^2_S |S|^2 + m^2_{S_A} |S_A|^2  \nonumber \\
 & + m_{B}^2 |B|^2 + m_{\bar{B}}^2 |\bar{B}|^2  \, ,
 \end{align}
where we have set to zero the mixed $S$-$S_A$ soft mass, the $\ztwo$-breaking stop masses and soft Higgs masses and the $b$-term. All these operators are generated in the IR as a result of the $\ztwo$-breaking couplings $\lambda_A$, but remain small as long as $\lambda_A$ is not too large:
\begin{align}
\delta V_{\rm soft}|_{\rm IR} & =  m^2_{S S_A} \left( S S_A^* + {\rm h.c.} \right) - \Delta b  \left( h_u^A h_d^A  - h_u^B h_d^B   +{\rm h.c.} \right) \nonumber \\
& + \Delta m^2_{H_u} \left( |h_u^A|^2 -   |h_u^B|^2 \right) + \Delta m^2_{H_d} \left( |h_d^A|^2 -  |h_d^B|^2   \right) \nonumber \\
& + \Delta m^2_{Q} \left( |Q_3^A|^2 - |Q_3^B|^2 \right) + \Delta m^2_{U} \left( |U_3^A|^2 -  |U_3^B|^2  \right) \, .
\end{align}
In the RGEs we can then neglect the contributions of $\ztwo$-breaking couplings except $\lambda_A$ (that is the only $\ztwo$-breaking coupling already present in the UV), as well as the contributions from (SUSY and soft) mixed singlet masses.
\subsubsection*{Beta Functions}
The beta function coefficients for the gauge couplings $ \tilde{g}_Y = g_Y ,  \tilde{g}_2 = g_2, \tilde{g}_3 = g_3$ are:
\begin{align}
b_i^{\rm hard} & = \left( 11, 1, -3 \right) \, , & b_i^{\rm hard+bidoublets} & = \left( 13, 3, -3 \right) \, .
\end{align}
\subsubsection*{Yukawa couplings}
\begin{align}
16 \pi^2 \frac{d}{dt}  \lambda_S & \approx \lambda_S \left[ 6 \lambda_S^2 + 3 y_t^2 + 2 \lambda_A^2   - g_Y^2 - 3 g_2^2 + 2 \lambda_{BD}^2 \right] \, , \nonumber  \\
16 \pi^2 \frac{d}{dt}  \Delta \lambda_S & \approx 2 \lambda_S \lambda_A^2   \, , \nonumber  \\
16 \pi^2 \frac{d}{dt}  \lambda_A & \approx \lambda_A \left[ 4 
\lambda_A^2 + 4 \lambda_S^2    + 3 y_t^2  - g_Y^2 - 3 g_2^2 + 2 \lambda_{BD}^2  \right] \, ,  \nonumber  \\
16 \pi^2 \frac{d}{dt}  \lambda_B & \approx  2 \lambda_A \lambda_S^2   \nonumber  \, , 
\end{align}
\begin{align}
16 \pi^2 \frac{d}{dt} y_t & \approx y_t \left[ 6 y_t^2 + \lambda_S^2 + \frac{1}{2} \lambda_A^2   - \frac{13}{9} g_Y^2 - 3 g_2^2 - \frac{16}{3} g_3^2 + 2 \lambda_{BD}^2 \right]  \, , \nonumber  \\
16 \pi^2 \frac{d}{dt} \Delta y_t & \approx  \frac{1}{2}  y_t \lambda_A^2  \nonumber  \\
16 \pi^2 \frac{d}{dt}  \lambda_{BD} & = \lambda_{BD} \left[ 5 \lambda_{BD}^2  + 2 \lambda_S^2  + \lambda_A^2  - 2 g_Y^2 -6 g_2^2 + 6 y_t^2  \right] \, .
\end{align}
\subsubsection*{SUSY Masses}
\begin{align}
16 \pi^2 \frac{d}{dt} \mu  & \approx \mu \left[ 3 y_t^2  + \lambda_A^2 +  2 \lambda_S^2  - g_Y^2 -3 g_2^2 + 2\lambda_{BD}^2 \right] \ \, , \nonumber  \\ 
16 \pi^2 \frac{d}{dt} \Delta \mu  & \approx   \lambda_A^2  \mu\, , \nonumber  \\ 
16 \pi^2 \frac{d}{dt} M_S  & \approx  8   \lambda_S^2 M_S \, , \nonumber  \\
16 \pi^2 \frac{d}{dt} M_{S_A}  & \approx  4 \lambda_A^2  M_{S_A}   \, , \nonumber  \\
16 \pi^2 \frac{d}{dt} M_{S S_A}  & \approx  2 \lambda_A \lambda_S   \left( M_S + M_{S_A} \right)  \, , \nonumber  \\
16 \pi^2 \frac{d}{dt} M_{B}  & = M_{B} \left[  ( \lambda_{BD})^2 - 6 g_2^2 - 2 g_Y^2 \right]  \, .
\end{align}

\subsubsection*{Soft Masses}
\begin{align}
16 \pi^2 \frac{d}{dt} m^2_{H_u} & \approx g_Y^2 \xi + 2 \lambda_S^2  X_S +    \lambda_A^2 X_{S_A} +  6 y_t^2 X_t + 4\lambda_{BD}^2 \left[ m_{B}^2 + 2 m_{H_u}^2 \right] \, , \nonumber  \\
16 \pi^2 \frac{d}{dt} \Delta m^2_{H_u} & \approx   \lambda_A^2 X_{S_A}  \, , \nonumber  \\
16 \pi^2 \frac{d}{dt} m^2_{H_d} & \approx - g_Y^2 \xi + 2  \lambda_S^2   X_S +    \lambda_A^2  X_{S_A}  \, , \nonumber  \\
16 \pi^2 \frac{d}{dt} \Delta m^2_{H_d} & \approx  \lambda_A^2 X_{S_A} \, , \nonumber  \\
16 \pi^2 \frac{d}{dt} m^2_{Q_3} & = \frac{1}{3} g_Y^2 \xi - \frac{32}{3} g_3^2 M_3^2 + 2  y_t^2  X_t  \, , \nonumber  \\
16 \pi^2 \frac{d}{dt} \Delta m^2_{Q_3} & \approx 0 \, , \nonumber  \\ 
16 \pi^2 \frac{d}{dt} m^2_{U_3} & = - \frac{4}{3} g_Y^2 \xi - \frac{32}{3} g_3^2 M_3^2 + 4 y_t^2  X_t \, , \nonumber  \\
16 \pi^2 \frac{d}{dt} \Delta m^2_{U_3} & \approx 0 \, , \nonumber  \\
16 \pi^2 \frac{d}{dt} m^2_{S} & \approx 8 \lambda_S^2  X_S  \, , \nonumber  \\
16 \pi^2 \frac{d}{dt} m^2_{S_A} & \approx 4  \lambda_A^2 
 X_{S_A} \, , \nonumber  \\
16 \pi^2 \frac{d}{dt} m^2_{S S_A} & \approx  2  \lambda_S   \lambda_A  \left( X_S + X_{S_A} \right) \, , \nonumber  \\
16 \pi^2 \frac{d}{dt} m^2_{B} & =  - 2 g_Y^2 \xi + 2\lambda_{BD}^2 \left[ m_{B}^2 + 2 m^2_{H_u}\right] \, , \nonumber  \\
16 \pi^2 \frac{d}{dt} m^2_{\bar{B}} & =  2 g_Y^2 \xi  \,  .
\end{align}
with the auxiliary functions 
\begin{align}
X_S & \equiv m^2_S + m^2_{H_u} + m^2_{H_d} \, , \nonumber  \\ 
X_{S_A} & \equiv m^2_{S_A} + m^2_{H_u} + m^2_{H_d} \, , \nonumber  \\ 
X_t & \equiv m^2_Q + m^2_U + m^2_{H_u} \, , \nonumber  \\ \xi & \equiv 3 m^2_Q - 6 m^2_U + m^2_{H_u} - m^2_{H_d}  + 2 m^2_{\bar{B}} - 2 m^2_{B} \, . 
\end{align}
Moreover one has
\begin{align}
16 \pi^2 \frac{d}{dt} b & \approx b \left[ 10 \lambda_S^2 + 3 \lambda_A^2  + 3 y_t^2  - g_Y^2 - 3 g_2^2 + 2\lambda_{BD}^2  \right]  \, , \nonumber \\
16 \pi^2 \frac{d}{dt} \Delta b & =  3 \lambda_A^2 b  \, , \nonumber \\
 16 \pi^2 \frac{d}{dt} M_3 & = - 6 g_3^2 M_3 \, . 
\end{align}
 
 \section{Appendix: Higgs sector spectrum}\label{app:2}

In this appendix we give further details about the (SUSY) Twin Higgs model.
First we fully solve the Twin Higgs model as a linear sigma model deriving exact formulas for the VEVs, the masses and the mixing in order to study the validity regime of the PGB approximation.
Then we provide the analytical formulae describing the phenomenology of the Twin SUSY Higgs sector when some hierarchy is present between the twin Higgs and the SUSY Higgses (i.e. for $\lambda_S f \ll m_A$): mass eigenvalues, eigenvectors and decay widths.

\subsection{The Twin Higgs as a linear sigma model}\label{app:twin}
In Sec.~\ref{sec:2} we have explored the parameter space of the Twin Higgs in complete generality, integrating out the radial mode and obtaining a non-linear sigma model description of the SM Higgs as a PGB of the spontaneously broken $SU(4)$-symmetry.  Here we repeat the same exercise working directly at the level of the linear sigma model with both the radial mode/twin Higgs and the SM-like Higgs in the spectrum. We obtain fully general formulas for the mass of the twin Higgs and the SM-like Higgs mass and for their mixing.  We show how the PGB formulas are obtained expanding these expressions at leading order in $k,\sigma \ll \lambda$. 

The Twin potential can be defined in terms of five parameters $\{ \lambda, \rho, \kappa, m^2, \tilde{\mu}^2 \}$
\begin{align}
V_T & = \lambda \left( |H_A|^2 + |H_B|^2 \right)^2 + m^2 \left( |H_A|^2 + |H_B|^2 \right) \nonumber \\ 
& + \kappa \left( |H_A|^4 + |H_B|^4 \right) + \tilde{\mu}^2 |H_A|^2  + \rho |H_A|^4 \, .\label{Twin_potential}
\end{align}
The minimization conditions set the VEVs of $H_A$ and $H_B$ ($v_A$ and $v_B$) as a function of these parameters.
Defining $v_A^2= v^2$ and $v_B^2 = f^2 - v^2$ in analogy with the PGB approximation,  one obtains with the shorthand  notation $\sigma \equiv - 2 \lambda \tilde{\mu}^2/m^2$
\begin{align}
v^2 & = - \frac{m^2}{4} \frac{- \sigma + \kappa \left( 2 - \sigma /\lambda \right)  }{ \lambda \rho + \kappa \left( 2 \lambda + \rho + \kappa\right)} \, , 
&
f^2 & = - \frac{m^2}{4} \frac{  2 \rho + \kappa \left( 4   - \sigma/\lambda  \right)  }{ \lambda \rho + \kappa \left( 2 \lambda + \rho + \kappa\right)} \, . 
\label{eqsvf}
\end{align}

Note that in the limit $\kappa \ll \lambda $ one has $m^2 = - 2 \lambda f^2$, thus recovering the definition of $\sigma$ in Sec.~\ref{sec:2}. Moreover, in the limit  $\kappa,\sigma \ll \lambda$  Eq.~(\ref{eqsvf}) reproduces the PGB formula in Eq.~(\ref{vh1}). We can use the above equations to trade the two VEVs $v$ and $f$ for two Lagrangian parameters, for instance the two scales $m^2$ and $\tilde{\mu}^2$, or equivalently $m^2$ and $\sigma$.

We are now interested in the mass spectrum, in particular the mass of the lightest (PGB) Higgs $h$. The full analytic expression for the masses of the two physical Higgs bosons is given by
\begin{align}
\label{higgs_comp}
m_{h,H}^2=
2 \left[ \rho  v^2+ f^2 (\lambda + \kappa)\left( 1 \mp \sqrt{1 - \mathcal{S}} \right) \right] \, , 
\end{align}
with
\begin{align}
\mathcal{S} =  2 \frac{v^2}{f^2} \frac{ \lambda \rho + \kappa \left( 4 \lambda + \rho + 2 \kappa \right) }{\left( \lambda + \kappa \right)^2} - \frac{v^4}{f^4} \frac{4 \lambda \rho + \rho^2 + \kappa \left( 8 \lambda + 4 \rho + 4 \kappa \right)}{\left( \lambda + \kappa \right)^2} \, , 
\end{align}
Expanding in $v \ll f$ and $\kappa \ll \lambda$ the square root in Eq.~(\ref{higgs_comp}) gives
\be
\sqrt{1- \mathcal{S}} \approx   1- \frac{\rho + \kappa \left( 4  - \rho/\lambda \right)}{\lambda} \frac{v^2}{ f^2} + \frac{2 \rho }{\lambda} \frac{v^4}{f^4} \, .
\ee
Another useful approximation is to expand at leading order in $\kappa/\lambda$ and $\rho/\lambda$, but keep all orders in $v/f$. In this way we get for the Higgs mass
\be
\label{approx_higgs}
m_h^2 \approx 
4 v^2  \left(1-\frac{v^2}{f^2}\right) \left[ \left(2 \kappa +\rho \right) -  \frac{\left( f^2 \kappa - v^2 \left( 2 \kappa + \rho \right) \right)^2}{\lambda f^4} \right] \, , 
\ee
which shows that the PGB expression in Eq.~(\ref{vh2}) tends to overestimate the mass of the SM Higgs. In the same approximation the mass of radial mode is
\begin{align}
m_{H}^2 \approx 4 f^2 \lambda \left[ 1 +\frac{ f^4 \kappa -2 f^2 \kappa v^2+v^4 \left(2 \kappa+\rho \right)}{\lambda f^4} \right] \, .
\end{align}
Defining the mixing angle $\theta$ by ($c_\theta \equiv \cos \theta, s_\theta \equiv \sin \theta$)
\begin{align}
h & = c_\theta \, h_A - s_\theta  \, h_B \, ,  & H & = s_\theta \, h_A + c_\theta  \, h_B \, , 
\end{align}
we find in the same approximation
\bea
s^2_\theta \approx 
\frac{v^2}{f^2} \left[ 1 - 2 \left( 1 - \frac{v^2}{f^2} \right) \frac{\kappa f^2 - v^2 \left( 2 \kappa + \rho \right)}{\lambda f^2} \right] \, . 
\eea

\subsection{The SUSY Twin Higgs}
The SUSY Twin Higgs is a double copy of the MSSM. The scalar potential can be schematically divided into three parts, depending if
they break $U(4)$ and/or the $\mathbb{Z}_2$ symmetry
\begin{align}
\label{app:scalarpot}
&
V  = V_{U(4)}  + V_{\slashed{U}(4),Z_2} + V_{\slashed{U}(4),\slashed{Z}_2} \nonumber \, ,  \\
&
V_{U(4)}  = \lambda_S^2 \left| h^A_u h^A_d + h^B_u h^B_d \right|^2 + m_u^2 \left( |h^A_u|^2 + |h^B_u|^2  \right) + m_d^2 \left( |h^A_d|^2 + |h^B_d|^2  \right) \, , \nonumber \\
&
V_{\slashed{U}(4),Z_2}  = \frac{g_{\rm ew}^2}{8} \left[ \left( |h^A_d|^2 - |h^A_u|^2 \right)^2 + \left( |h^B_d|^2 - |h^B_u|^2 \right)^2  \right] - b \left( h^A_u h^A_d + h^B_u h^B_d  +{\rm h.c.} \right) \nonumber \\
& 
\qquad \qquad \qquad + \delta \lambda_u \left( |h^A_u|^4 +  |h^B_u|^4  \right) + \lambda_{BD}^2 |h^A_u|^2 |h^B_u|^2 \, , \nonumber \\
& 
V_{\slashed{U}(4),\slashed{Z}_2}  = \lambda_A^2 \left| h^A_u h^A_d \right|^2 + \Delta m_u^2 \left| h^A_u \right|^2 + \Delta m_d^2 \left| h^A_d \right|^2 + \delta \rho_u |h^A_u|^4 \, . 
\end{align}
where we have defined $g_{\rm ew}^2 = g^2+g'^2$. Apart from the soft $\mathbb{Z}_2$-breaking terms $\Delta m_{u,d}^2$, hard breaking terms $\lambda_A^2$ and bi-doublet terms $\lambda_{BD}^2$, we have included the terms $\delta \lambda_u$ and $\delta \rho_u$ that are generated at one-loop from the stop/top sector,  with 
\begin{align}
\delta \lambda_u & \approx \frac{3 m_t^4}{16 \pi^2 s_\beta^4 v^4} \log \frac{M_S^2}{m_{t_B}^2} \, , & \delta \rho_u & \approx \frac{3 m_t^4}{16 \pi^2 s_\beta^4 v^4} \log \frac{f^2}{v^2} \, .
\end{align}
We then minimize the potential and trade the parameters $\Delta m_u^2, \Delta m_d^2, m_u^2, m_d^2$ for $v^2, f^2$ and $t_\beta \equiv \tan \beta_A \approx \tan \beta_B$. We do not take into account corrections proportional to $\delta \tan \beta =  \tan \beta_A - \tan \beta_B$, which typically lead only to small corrections. 
We can then compute the mass eigenvalues in the CP-even sector, in the CP-odd sector and for the charged Higgses. These latter results are particularly simple, since the CP-odd and charged Higgs mass matrices are $2\times2$ matrices that can be  diagonalized exactly.
\subsubsection*{CP-odd Higgs Sector}
The CP-odd Higgs eigenvalues and eigenvectors are 
\begin{align}
& m_{A_T}^2  = \frac{2 b}{ s_{2 \beta}} \, ,  \nonumber \\
\label{CPOdd}
& m_{A}^2  = m_{A_T}^2 - \lambda_S^2 f^2 \, , \nonumber \\
& A_T  =  \frac{v}{f} c_\beta  \, a_u^A + \frac{v}{f} s_\beta \, a_d^A + c_\beta \sqrt {1 - \frac{v^2}{f^2}} \, a_u^B +  s_\beta  \sqrt {1 - \frac{v^2}{f^2}} \, a_d^B \, ,  \nonumber \\
& A  = - c_\beta \sqrt {1 - \frac{v^2}{f^2}}  \,   a_u^A -  s_\beta  \sqrt {1 - \frac{v^2}{f^2}}\, a_d^A +  \frac{v}{f} c_\beta \, a_u^B +   \frac{v}{f} s_\beta \, a_d^B \, , 
\end{align}
where we defined $A_T$ as the CP-odd state that is mainly dark for $v^2/f^2 \ll 1$.
\subsubsection*{Charged Higgs Sector}
The charged Higgs eigenvalues and eigenvectors are given by
\begin{align}
& m_{H^\pm_T}^2  = m_A^2  \, , \nonumber \\
&  m_{H^\pm}^2  = m_A^2  - \lambda_A^2 v^2  \, ,  \nonumber \\
& H_T^-  = c_\beta  \, \left( h_u^{+B} \right)^* +  s_\beta \, h_d^{-B}   \, ,  \nonumber \\
&  H^-  =  c_\beta  \, \left( h_u^{+A} \right)^*+  s_\beta \, h_d^{-A} \, ,  
\end{align}
Note that gauge invariance forbids mixing between the visible and dark charged Higgs.
\subsubsection*{CP-even Higgs Sector}

The CP-even sector mass matrix is a $4 \times 4$ matrix whose exact diagonalization expressions are not very illuminating.
In the main text we compute the eigenvalues and eigenvectors numerically to perform the phenomenological study.
In this appendix we provide analytic expressions in some simplifying limits.

As we already discussed in Sec.~\ref{Higgscoupl}, the SUSY Twin Higgs model possesses two different regimes of the Higgs mass parameters which give a different hierarchy in the mass spectrum.
For $m_A^2 \gg \lambda_S^2 f^2$ the next to lightest state after the SM Higgs boson is the twin Higgs, which belongs to the dark sector and develops a VEV of order $f$. In the regime where $m_A^2 \gtrsim \lambda_S^2 f^2$ the lightest scalar above the SM-like higgs is instead
the MSSM-like heavy scalar. In this limit the model resembles the MSSM with a CP-odd mass scale set by the combination $m_A^2 = m_{A_T}^2 -\lambda^2 f^2$.

\begin{figure}[t]
  \centering
 \includegraphics[width=0.45\textwidth]{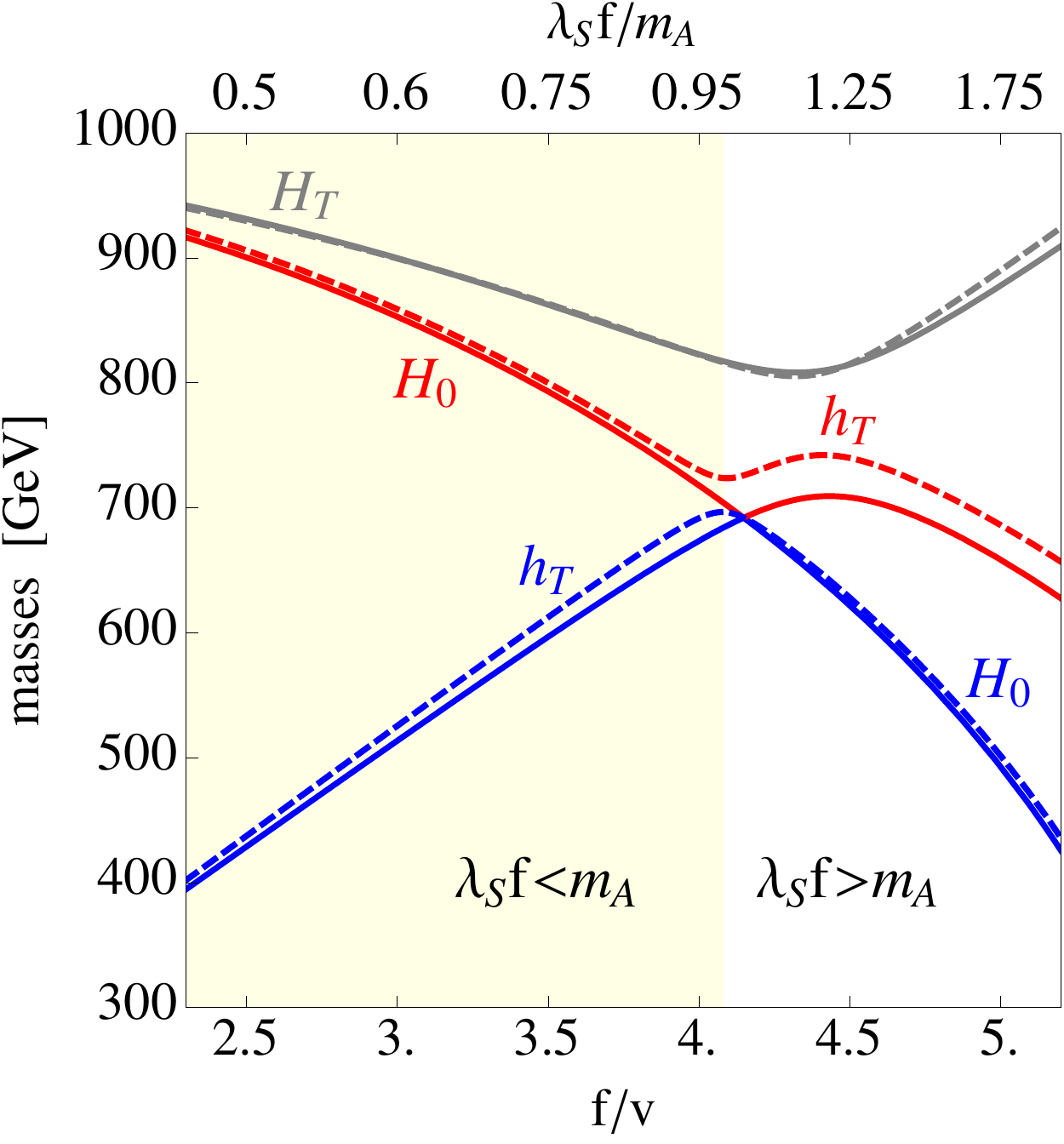}
  \caption{\label{fig:trans_2}
Left: Mass eigenvalues as a function of $f/v$. We fix $\tan \beta=1.2$, $m_{A_T}=1$~TeV and $\lambda_S=1$. The solid lines are obtained using the analytic expressions in Eq.~\eqref{approxmasses}  in the $v\to0$ limit. The dashed lines are obtained numerically including $v/f$ corrections. The light yellow/white area correspond to the region where $\lambda_S f \lessgtr m_A$ as indicated in the plot.}
\end{figure}

In order to obtain expressions for the masses that can describe the transition between these two different regimes, we consider the simplifying limit $v\to 0$ and keep only leading order terms in $g_{\rm ew}^2$ and $\delta \lambda_u$.
In this approximation the mass eigenvalues are
{\small
\begin{align}
&
\label{approxmasses}
m_{h}^2= 0  \, , \nonumber \\
&
m_{H}^2 = m_A^2 =  m_{A_T}^2 - \lambda_S^2 f^2 \, ,  \nonumber \\
&
m_{h_T}^2= \frac{1}{4}\left(2 m_{A_T}^2+ g_{\rm ew}^2 f^2 + 8 \delta \lambda_u f^2 s_\beta^2 -2 m_{A_T}^2 \mathcal{R}
\right) \, ,  \nonumber \\
&
m_{H_T}^2= \frac{1}{4}\left(2 m_{A_T}^2+ g_{\rm ew}^2 f^2 + 8 \delta \lambda_u f^2 s_\beta^2 + 2 m_{A_T}^2 \mathcal{R}
\right)\, ,  \nonumber \\
&
\mathcal{R}\equiv
\sqrt{1- \frac{f^2}{m_{A_T}^2} \left( 4 \lambda_S^2 s_{2 \beta}^2 +  g^2_{\rm ew} c_{4 \beta} - 8 \delta \lambda_u c_{2 \beta} s_\beta^2 \right) + \frac{2 \lambda_S^2 f^4}{m_{A_T}^4} \left( 2 \lambda_S^2 - g_{\rm ew}^2 \right) s_{2 \beta}^2 } \, .
\end{align}
}
We can expand these expressions in the two different regimes. For $\lambda_S^2 f^2 \ll m_{A_T}^2$ we get
\begin{align}
&
m_{h_T}^2 = f^2 (\lambda_S^2  s_{2 \beta}^2 + \frac{1}{2} g_{\rm ew}^2 c_{2 \beta}^2 + 4 \delta \lambda_u s_\beta^4  )\nonumber \\
&
m_{H_T}^2 =m_{A_T}^2 - f^2 s_{2\beta}^2 (\lambda_S^2 -\frac{1}{2} g_{\rm ew}^2 -  \delta \lambda_u  )  \, .
\end{align}
Here the twin Higgs is lighter than the MSSM-like state. 
On the other hand, for small $m_{A_T}^2$ but still with $m_{A_T}^2 \gtrsim \lambda_S^2 f^2$ we get
\begin{align}
&
m_{h_T}^2 =  \frac{1}{2} f^2 g_{\rm ew}^2 + s_{2 \beta}^2 \left( m_{A_T}^2 - \lambda_S^2 f^2 \right) + 4 \delta \lambda_u f^2 s_\beta^4  \, , \nonumber  \\
&
m_{H_T}^2 =\lambda_S^2 f^2 + f^2 \left[ \lambda_S^2 c_{2 \beta}^2 \left( 1 - \frac{\lambda_S^2 f^2}{m_{A_T}^2} \right) +  \delta \lambda_u s_{2 \beta}^2 \right]  \, .
\end{align}
Contrary to the previous regime, 
here the Twin Higgs is always heavier than the MSSM-like Higgs, and their mass splitting is set by
\begin{align}
m_{h_T}^2 - m^2_H & = f^2 \left( \frac{1}{2} g_{\rm{ew}}^2 - \lambda_S^2 c_{2 \beta}^2 \left( 1 - \frac{\lambda_S^2 f^2}{m_{A_T}^2} \right) + 4 \delta \lambda_u s_{\beta}^4  \right) \, .
\end{align}
To make the transition manifest, we plot in Fig.~\ref{fig:trans_2} the eigenvalues as a function of $f/v$ fixing the other parameters
\footnote{As explained in the text, $v$ is set to zero for this plot so the x-axis should be understood as $f$ in units of $174$ GeV.}. 
The transition between the two regimes happens approximately at $m_A^2 \approx  \lambda_S^2 f^2$.
For $\lambda_S^2 f^2$ smaller than this critical value the lightest eigenstate is the twin Higgs, while for larger $\lambda_S^2 f^2$ 
it is MSSM-like.
For completeness we compare the analytic expressions with 
the numerical results where $v$ is turned on.
On the same plot in Fig.~\ref{fig:trans_2} we show in dashed the numerical eigenvalues as a function of $f/v$ on the same benchmark,
showing that $v \neq0 $ only results in a slight separation of the eigenvalues in the transition regime.

\begin{figure}[t]
  \centering
 \includegraphics[width=0.45\textwidth]{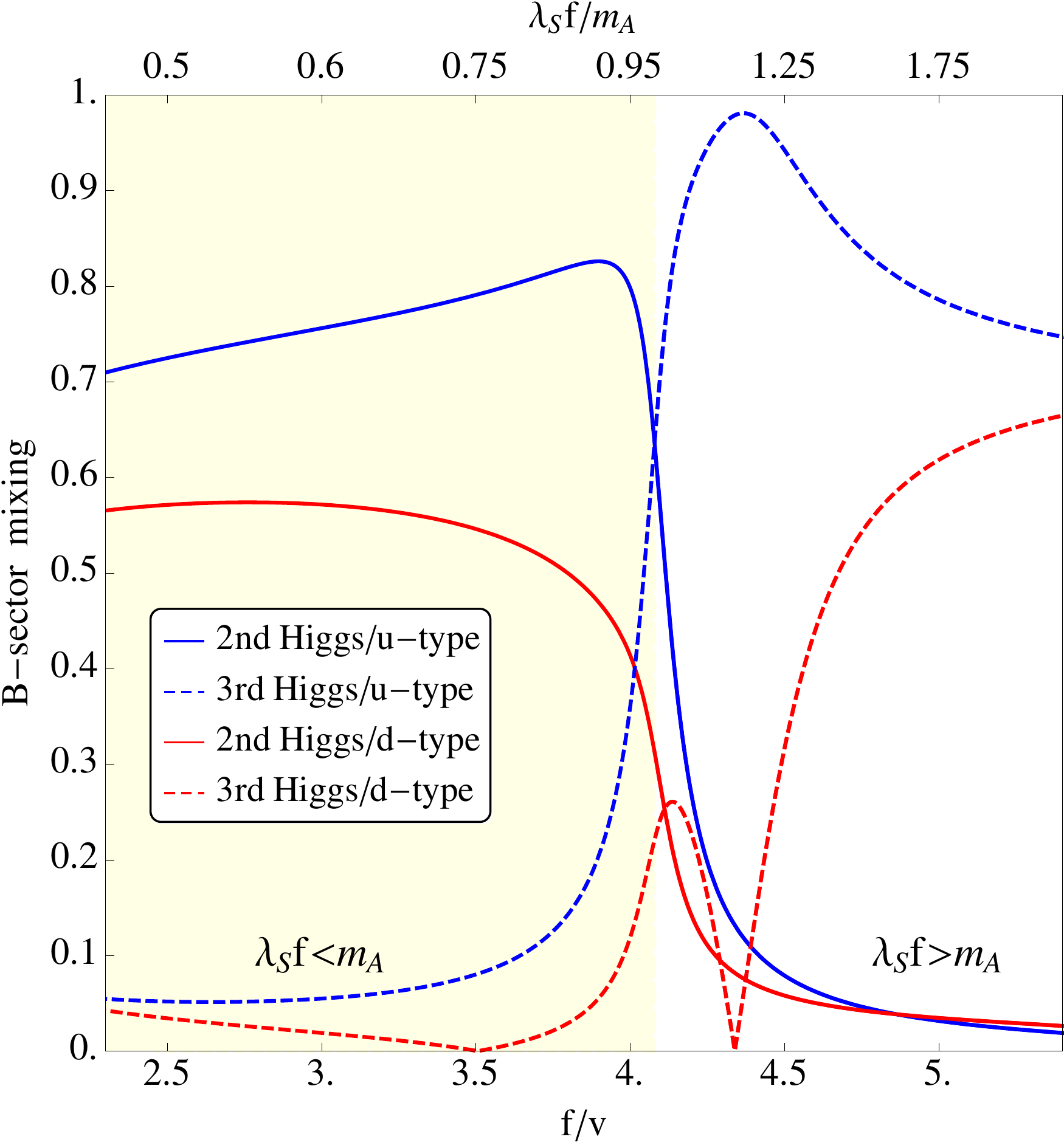}\hfill
  \includegraphics[width=0.45\textwidth]{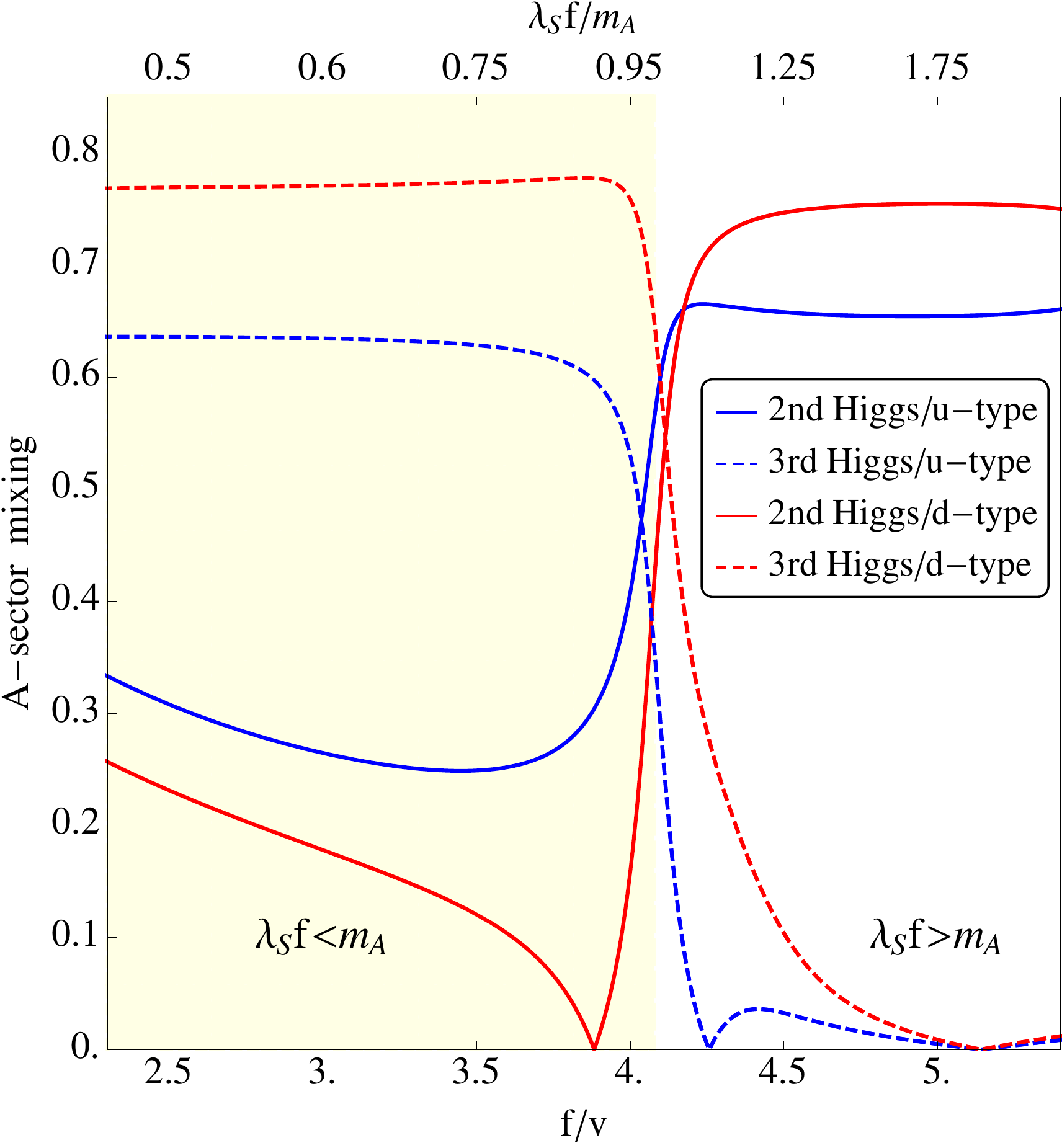}
\caption{\label{fig:mixings}
Mixing as a function of $f/v$. We fix $\tan \beta=1.2$, $m_{A_T}=1$~TeV, $\lambda_S=1$. The light yellow/white area correspond to the region where $\lambda_S f \lessgtr m_A$ as indicated in the plot.
{\bf Left:} $B$-sector components of the next to lightest (thick) and next to next to lightest (dashed) CP-even Higgs. The blue is the $h_u^B$
component while the red is the $h_d^B$ one. 
{\bf Right:} $A$-sector components of the next to lightest (thick) and next to next to lightest (dashed) eigenstate. The blue is the $h_u^A$
component while the red is the $h_d^A$ one.}
\end{figure}

In order to better understand the phenomenology in Sec.~\ref{Higgscoupl} we plot in Fig.~\ref{fig:mixings} the mixing angles of the two mass eigenstates $\{H, h_T \}$ with respect to the gauge eigenbasis
$\{h_u^A,h_d^A,h_u^B,h_d^B \}$. In the left panel of Fig.~\ref{fig:mixings} we focus on the $B$-components into the dark sector and denote with solid blue/red lines the $u/d$-components the next-to-lightest eigenvalues. The dashed lines refer instead to the third lightest eigenvalue with the same color coding (blue for $u$- and red for $d$- components). In the right panel of Fig.~\ref{fig:mixings} we plot with the same color coding the $A$-components into the visible sector  (blue for $u$- and red for $d$- components).

From these plots it is clear that the role of the eigenvalues is inverted in the transition region around $\lambda_S f \approx m_A$ . 
For $\lambda_S f\lesssim m_A$ the next-to-lightest eigenstate is the twin-Higgs, while for $\lambda_S f\gtrsim m_A$ it is MSSM-like. In the transition region with $\lambda_S f\approx m_A$ all states are sizeably mixed one each other.

The visible sector content of the extra Higgs states determines their production cross section at the LHC and their visible branching ratio. 

For small $\lambda_S f\lesssim m_A$ the twin Higgs is the next-to-lightest Higgs and it gets sizeable couplings to the visible sector through its component along $h_u^A$. As a consequence it can be produced at the LHC in gluon fusion and it gives rise to interesting signals in di-boson channels.
In the transition region, when $\lambda_S f\approx m_A$, there is an enhancement of the $h_u^A$ components as it can be seen from the solid blue line in the right panel of Fig.~\ref{fig:mixings}. We then expect a slight improvement in the sensitivity on the twin Higgs searches at LHC into di-bosons. For $\lambda_S f\gtrsim m_A$ the next-to-lightest state becomes MSSM-like and the sensitivity of the di-boson searches drops since MSSM-like have very small branching ratio into gauge bosons and at the same time the twin Higgs has a small mixing angle with the visible sector as it can be seen from the dashed lines in the right panel Fig.~\ref{fig:mixings}. This argument explains the horn-like shape of the di-boson exclusion regions in Fig. \ref{fig:pheno}.

\subsection{Approximate expressions for eigenvalues and eigenvectors}
We have seen that in the regime $\lambda_S^2 f^2 \ll m_A^2$ the collider phenomenology of the model is characterised by a light twin Higgs with interesting signatures. In this regime we can expand the expressions for masses and mixings for large $m_A$ (or equivalently large $m_{A_T}$) and obtain approximate analytical results. From the scalar potential (\ref{app:scalarpot})
we get for the spectrum, keeping only the leading order terms in $g_{\rm ew}^2, \lambda_A^2, \lambda_{BD}^2, \delta \lambda_u, \delta \rho_u$ and  $f^2/m_{A_T}^2$:
\begin{align}
m_h^2 & = v^2 \left( 1 -\frac{v^2}{f^2} \right)  \left( g_{\rm ew}^2  c_{2 \beta}^2   +  \lambda_A^2  s_{2 \beta}^2  - 4 \lambda_{BD}^2 s_\beta^4 + 8 \delta \lambda_u  s_{\beta}^4 + 4 \delta \rho_u s_{\beta}^4 \right)  \, , \nonumber \\
m_{h_T}^2 & = \lambda_S^2 f^2 s_{2 \beta}^2 + \frac{f^2}{2}   \left( g_{\rm ew}^2 c_{2 \beta}^2  +  8  \delta \lambda_u s_\beta^4 \right) + 4 \lambda_{BD}^2  v^2 s_\beta^4 + \frac{v^4}{f^2} \left( \lambda_A^2 s_{2 \beta}^2 + 4 \delta \rho_u s_\beta^4 \right) \, , \nonumber \\
m_{H_T}^2 & = m_{A_T}^2 - f^2 s_{2 \beta}^2  \left(    \lambda_S^2  - \frac{1}{2} g_{\rm ew}^2-  \delta \lambda_u \right)  + \lambda_{BD}^2 v^2 s_{2 \beta}^2 -  \frac{v^4}{f^2} s_{2 \beta}^2 \left(  \lambda_A^2      - \delta \rho_u \right) \, , \nonumber \\
m_{H}^2 & = m_{A_T}^2 - \lambda_S^2 f^2  + v^2 s_{2 \beta}^2  \left(g_{\rm ew}^2   - \lambda_A^2  - \lambda_{BD}^2 + 2 \delta \lambda_u   + \delta \rho_u \right)  \, . 
\end{align}
Similarly we can calculate the mixing angles in the same approximation. Defining the rotation matrix $V$ as
\begin{align}
\begin{pmatrix} h \\ h_T \\ H_T \\ H \end{pmatrix} & = V \begin{pmatrix} h^A_u \\ h^A_d \\ h^B_u \\ h^B_d \end{pmatrix} \, , 
\end{align}
 the entries are, keeping only the leading order terms in $g_{\rm ew}^2, \lambda_A^2, \lambda_{BD}^2, \delta \lambda_u, \delta \rho_u$ and  $f^2/m_{A_T}^2$:
\begin{align}
h & = \left[ s_
\beta  \left( 1 - \frac{v^2}{2 f^2}  +\frac{ \lambda_1^2 v^2 }{2 \lambda_S^2 f^2 t_{2 \beta}^2 }  \right)   \right] h^A_u  + \left[ c_\beta \left(  1 - \frac{v^2}{2 f^2} +\frac{\lambda_1^2 v^2}{2 \lambda_S^2 f^2 t_{2 \beta}^2 }  \right)   \right] h^A_d  \nonumber \\
& +\left[  - \frac{v}{f} s_\beta \left( 1  - \frac{ \lambda_1^2 }{2 \lambda_S^2 t_{2 \beta}^2 }   \right)   \right] h^B_u +\left[  - \frac{v}{f} c_\beta \left( 1  - \frac{ \lambda_1^2 }{2 \lambda_S^2  t_{2 \beta}^2 }   \right)  \right] h^B_d \, ,  \nonumber \\
h_T & = \left[ \frac{v}{f} s_\beta \left(  1 - \frac{ \lambda_1^2 }{2 \lambda_S^2  t_{2 \beta}^2 }  \right) \right] h^A_u  + \left[ \frac{v}{f} c_\beta \left( 1 - \frac{ \lambda_1^2 }{2 \lambda_S^2  t_{2 \beta}^2 }    \right)  \right] h^A_d  \nonumber \\
& +\left[ s_\beta \left( 1 - \frac{v^2}{2 f^2}  + \frac{\lambda_1^2  v^2}{2 \lambda_S^2 f^2 t_{2 \beta}^2 }   \right)  \right] h^B_u   +\left[ c_\beta  \left( 1 - \frac{v^2}{2 f^2}  + \frac{\lambda_1^2 v^2 }{2 \lambda_S^2 f^2 t_{2 \beta}^2 }   \right)   \right] h^B_d \, , \nonumber \\
H_T & = \left[ \frac{v}{f} c_\beta \left( 1  - \frac{ \lambda_2^2 t_{2 \beta}^2 }{2 \lambda_S^2 }   \right)   \right] h^A_u  + \left[  - \frac{v}{f}  s_\beta \left( 1 -  \frac{  \lambda_2^2 t_{2 \beta}^2 }{2 \lambda_S^2 }   \right)  \right] h^A_d  \nonumber \\
& +\left[ c_\beta \left(  1 - \frac{v^2}{2 f^2}  +  \frac{   \lambda_2^2 v^2 t_{2 \beta}^2 }{2 \lambda_S^2 f^2}   \right)  \right] h^B_u  +\left[ - s_\beta \left(  1 - \frac{v^2}{2 f^2}  + \frac{   \lambda_2^2 v^2 t_{2 \beta}^2 }{2 \lambda_S^2 f^2}   \right)  \right] h^B_d \, , \nonumber \\
H & = \left[ c_\beta \left( 1 - \frac{v^2}{2 f^2} + \frac{  \lambda_2^2  v^2 t_{2 \beta}^2 }{2 \lambda_S^2 f^2}  \right)    \right] h^A_u  + \left[ - s_\beta \left(  1 - \frac{v^2}{2 f^2}  + \frac{  \lambda_2^2 v^2 t_{2 \beta}^2 }{2 \lambda_S^2 f^2}  \right)   \right] h^A_d  \nonumber \\
& +\left[ - \frac{v}{f} c_\beta  \left( 1 - \frac{  \lambda_2^2   t_{2 \beta}^2 }{2 \lambda_S^2 }  \right)  \right] h^B_u  +\left[   \frac{v}{f}  s_\beta \left( 1  - \frac{   \lambda_2^2  t_{2 \beta}^2 }{2 \lambda_S^2 }  \right)   \right] h^B_d \, , \end{align}
with the shorthand notation
\begin{align}
\lambda_1^2 & = g_{\rm ew}^2 - 2 \lambda_A^2 \frac{v^2}{f^2} t_{2 \beta}^2 - t_\beta^2 t_{2 \beta}^2 \left( \lambda_{BD}^2 - 2 \delta \lambda_u + 2 \delta \rho_u \frac{ v^2}{f^2} \right) \, , \nonumber \\
 \lambda_2^2 & =   g_{\rm ew}^2 -  \lambda_{BD}^2 + 2 \delta \lambda_u + 2 \lambda_A^2 \frac{v^2}{f^2}  - 2 \delta \rho_u \frac{ v^2}{f^2}  \, .
\end{align}

\subsection{Decay rates}
In the phenomenological study in the main text we make use of several branching ratio for the Higgs sector.
Here we report the most relevant formulas used in the analysis.

\subsubsection*{Decays into fermions}
The relevant Lagrangian is given by 
\begin{align}
{\cal L} & = \frac{m_t}{\sqrt{2} v s_\beta} \left( \overline{t}_A t_A V_{i1} + \overline{t}_B t_B V_{i3}  \right) H_i + \frac{m_b}{\sqrt{2} v c_\beta} \left( \overline{b}_A b_A V_{i2} + \overline{b}_B b_B V_{i4}  \right) H_i\, , 
\end{align}
with Higgs mass eigenstates $H_i = {h,h_T, H_T, H}$ and the rotation matrix $V$ has been given in the previous section.
If kinematically allowed, the decay rates into tops, dark tops, bottoms and dark bottoms are given by:
\begin{align}
\Gamma (H_i \to  t_A t_A ) & = \frac{3 m_t^2 M_{H_i}}{16  \pi v^2 s_\beta^2} |V_{i1}|^2  \left( 1 - 4 \frac{m_t^2}{M_{H_i}^2} \right)^{3/2} \, , \nonumber \\
\Gamma (H_i \to  t_B  t_B ) & = \frac{3 m_t^2 M_{H_i}}{16  \pi v^2 s_\beta^2} |V_{i3}|^2  \left( 1 - 4 \frac{m_{t_B}^2}{M_{H_i}^2} \right)^{3/2} \, , \nonumber \\
\Gamma (H_i \to  b_A b_A ) & = \frac{3 m_b^2 M_{H_i}}{16  \pi v^2 c_\beta^2} |V_{i2}|^2  \left( 1 - 4 \frac{m_b^2}{M_{H_i}^2} \right)^{3/2} \, , \nonumber \\
\Gamma (H_i \to  b_B b_B ) & = \frac{3 m_b^2 M_{H_i}}{16  \pi v^2 c_\beta^2} |V_{i4}|^2  \left( 1 - 4 \frac{m_{b_B}^2}{M_{H_i}^2} \right)^{3/2} \, , 
\end{align}
\subsubsection*{Decays into Vector Bosons}
The relevant Lagrangian is given by 
\begin{align}
{\cal L} & = H_i \left(  \frac{g}{2 c_W} M_Z  Z^\mu Z_\mu + g M_W W^{+ \mu} W_\mu^- \right) \left( s_\beta V_{i1} +  c_\beta V_{i2}  \right) \nonumber \\
& + H_i \left(  \frac{g}{2 c_W} M_{Z_B}  Z^\mu_B Z_{B \mu} + g M_{W_B} W^{+ \mu}_B W_{B \mu^-} \right) \left( s_\beta V_{i3} +  c_\beta V_{i4}  \right)\, . 
\end{align}
If kinematically allowed, the decay rates into visible and dark gauge bosons are given by
\begin{align}
\Gamma (H_i \to  ZZ ) & = \frac{g^2 M_{H_i}^3}{128 \pi c_W^2 M_Z^2} \left|s_\beta V_{i1} +  c_\beta V_{i2} \right|^2  \left( 1 - 4 \frac{M_Z^2}{M_{H_i}^2} \right)^{1/2}  \left( 1- 4 \frac{M_Z^2}{M_{H_i}^2} +12 \frac{M_Z^4}{M_{H_i}^4} \right) \, , \nonumber \\
\Gamma (H_i \to  WW ) & = \frac{g^2 M_{H_i}^3}{64 \pi M_W^2}   \left|s_\beta V_{i1} +  c_\beta V_{i2} \right|^2   \left( 1 - 4 \frac{M_W^2}{M_{H_i} ^2} \right)^{1/2} \left( 1- 4 \frac{M_W^2}{M_{H_i} ^2} +12 \frac{M_W^4}{M_{H_i} ^4} \right) \, , \nonumber \\
\Gamma (H_i \to  Z_B Z_B ) & = \frac{g^2 M_{H_i}^3}{128 \pi c_W^2 M_{Z_B}^2} \left|s_\beta V_{i3} +  c_\beta V_{i4} \right|^2  \left( 1 - 4 \frac{M_{Z_B}^2}{M_{H_i}^2} \right)^{1/2}  \left( 1- 4 \frac{M_{Z_B}^2}{M_{H_i}^2} +12 \frac{M_{Z_B}^4}{M_{H_i}^4} \right) \, , \nonumber \\
\Gamma (H_i \to  W_B W_B ) & = \frac{g^2 M_{H_i}^3}{64 \pi M_{W_B}^2}   \left|s_\beta V_{i3} +  c_\beta V_{i4} \right|^2  \left( 1 - 4 \frac{M_{W_B}^2}{M_{H_i} ^2} \right)^{1/2} \left( 1- 4 \frac{M_{W_B}^2}{M_{H_i} ^2} +12 \frac{M_{W_B}^4}{M_{H_i} ^4} \right) \, , 
\end{align}
\subsubsection*{Decays into Higgs Bosons}
Restricting to the $U(4)$-preserving quartic coupling, the relevant Lagrangian is given by 
\begin{align}
{\cal L} & = \frac{\lambda_S^2}{\sqrt{2}} \left[ v s_\beta h^A_u h^A_d h^A_d + v c_\beta h^A_d h^A_u h^A_u + \sqrt{f^2-v^2} s_\beta h^B_u h^B_d h^B_d + \sqrt{f^2-v^2} c_\beta h^B_d h^B_u h^B_u \right]  \nonumber \\
& + \frac{\lambda_S^2}{\sqrt{2}} \left[ v s_\beta h^A_d h^B_u h^B_d + v c_\beta h^A_u h^B_u h^B_d + \sqrt{f^2-v^2} s_\beta h^A_u h^A_d h^B_d + \sqrt{f^2-v^2} c_\beta h^A_u h^A_d h^B_u \right] \nonumber \\
& \equiv \frac{\lambda_S^2}{\sqrt{2}}  A_{ijk} H_i H_j H_k \, , 
\end{align}
with
\begin{align}
A_{ijk} & = v s_\beta V_{i1} V_{j2}  V_{k2} + v c_\beta V_{k2}  V_{i1}  V_{j1}  + \sqrt{f^2-v^2} s_\beta V_{i3} V_{j4}  V_{k4}  + \sqrt{f^2-v^2} c_\beta V_{i4} V_{j3}  V_{k3} \nonumber \\ & + v s_\beta V_{i2} V_{j3}  V_{k4}  + v c_\beta V_{i1} V_{j3} V_{k4} + \sqrt{f^2-v^2} s_\beta V_{i1} V_{j2} V_{k4} + \sqrt{f^2-v^2} c_\beta V_{i1} V_{j2}  V_{k3} \, .
\end{align}
To calculate a given coupling all permutations have to be summed over, e.g. the Lagrangian coupling ${\cal L} = A_{h_T h h} h_T h h $ is given by $A_{h_T h h}  = \frac{\lambda_S^2}{\sqrt2} \left( A_{122} + A_{212} + A_{221} \right) $.
From the general expression
\begin{align}
{\cal L} & = A_h H h h + A_{h_1 h_2} H h_1 h_2 \, , 
\end{align}
the decay rates of the scalar boson eigenstate $H$ can then be obtained as
\begin{align}
\Gamma (H \to  hh ) & =  \frac{A_{H hh}^2}{8 \pi M_H} \left( 1 - 4 \frac{m_h^2}{M_H^2} \right)^{1/2} \, , \\
\Gamma (H \to  h_1 h_2 ) & =  \frac{A_{H h_1 h_2}^2}{16 \pi M_H} \left( 1 - 2 \frac{m_{h_1}^2 + m_{h_2}^2}{M_H^2} +  \frac{\left( m_{h_1}^2 - m_{h_2}^2 \right)^2}{M_H^4} \right)^{1/2} \, .
\end{align}
 \bibliography{biblio}
\end{document}